\documentclass[preprint,aps,prd,superscriptaddress,nofootinbib]{revtex4}%

\usepackage{amsmath}
\usepackage{amssymb}
\usepackage{graphicx}
\usepackage{dcolumn}
\usepackage{bm}
\usepackage{hyperref}
\usepackage{epsfig}
\usepackage{mathrsfs}

\begin{document}
\preprint{}

\title{Covariant calculation of a two-loop test of
\\
nonrelativistic QCD
factorization}
\author{Geoffrey~T.~Bodwin}
\email[]{gtb@anl.gov}
\affiliation{High Energy Physics Division, Argonne National Laboratory,
Argonne, Illinois 60439, USA}
\author{Hee~Sok~Chung}
\email[]{heesok.chung@tum.de}
\affiliation{Physik-Department, Technische Universit\"at M\"unchen,
James-Franck-Str.\ 1, 85748 Garching, Germany}
\author{June-Haak~Ee}
\email[]{chodigi@gmail.com}
\affiliation{Department of Physics, Korea University, Seoul 02841, Korea}
\author{U-Rae~Kim}
\email[]{sadafada@korea.ac.kr}
\affiliation{Department of Physics, Korea University, Seoul 02841, Korea}
\author{Jungil~Lee}
\email[]{jungil@korea.ac.kr}
\affiliation{Department of Physics, Korea University, Seoul 02841, Korea}
\date{\today}
\begin{abstract}
We test the nonrelativistic QCD factorization conjecture for inclusive
quarkonium production at two loops by carrying out a covariant
calculation of the nonrelativistic quantum chromodynamics (NRQCD) long-distance matrix element (LDME) for a
heavy-quark pair in an $S$-wave, color-octet state to fragment into a
heavy-quark pair in a color-singlet state of arbitrary orbital angular
momentum. The NRQCD factorization conjecture for the universality of the
LDME requires that infrared divergences that it contains be
independent of the direction of the Wilson line that appears in its
definition. We find this to be the case in our calculation. The results
of our calculation differ in some respects from those of a previous
calculation that was carried out by Nayak, Qiu, and Sterman using
light-cone methods. We have identified the sources of some of these
differences. The results of both calculations are consistent with the NRQCD
factorization conjecture. However, the general principle that underlies
this confirmation of NRQCD factorization at two-loop order has yet to be
revealed.
\end{abstract}
\maketitle

\section{Introduction}

The nonrelativistic quantum chromodynamics (NRQCD) factorization
conjecture for inclusive quarkonium production in hard-scattering
QCD processes \cite{Bodwin:1994jh} postulates that the production rates
for those processes at large momentum transfer can be written as a sum
of products of short-distance coefficients (SDCs) and NRQCD
long-distance matrix elements (LDMEs). The SDCs are
process dependent, but they contain no infrared (IR) divergences, and
can be calculated in perturbation theory. The LDMEs contain all of the IR
sensitivity of the process. They are generally nonperturbative in
nature, but they are conjectured to be universal (process
independent)---a property that gives NRQCD factorization much of its
predictive power.

The NRQCD factorization conjecture has enjoyed considerable
phenomenological success. (See, for example,
Refs.~\cite{Brambilla:2004wf,Brambilla:2010cs,Lansberg:2019adr}.)
However, there are also processes for which theory and experiment are in
considerable tension. (See, for example,
Refs.~\cite{Brambilla:2010cs,Lansberg:2019adr}.)

Although the NRQCD factorization conjecture has been in existence for
many years, there is still no demonstration that it is correct to all
orders in perturbation theory or that it fails in perturbation theory.
Important progress toward an all-orders proof of NRQCD factorization was
presented in Ref.~\cite{Nayak:2005rt}. There, an all-orders argument was
given, for the leading and first subleading powers of $m_H/p_T$, that
quarkonium production rates can be written as a sum of SDCs convolved
with fragmentation functions for one or two partons to fragment into a
quarkonium. Here, $m_H$ and $p_T$ are the mass and transverse momentum,
respectively,
of the quarkonium. 

A proof of NRQCD factorization would require the further factorization
of the fragmentation functions into sums of products of SDCs with NRQCD
LDMEs. (In the remainder of this paper, when we refer to SDCs, we mean
the SDCs that relate the LDMEs to fragmentation functions.) One
difficulty in carrying out this factorization is that the fragmentation
functions at the scale of the heavy-quark mass $m_Q$ contain processes
in which gluons are radiated with energies of order $m_Q$ in the
quarkonium center-of-momentum (c.m.) frame. Such gluons cannot be absorbed
into the NRQCD LDMEs, which contain radiation only at the small scale
$m_Qv$, where $v$ is the typical velocity of the heavy quark $Q$ or
antiquark $\bar Q$ in the quarkonium c.m. frame. On the other hand, soft
gluons with arbitrarily small momenta can connect the order-$m_Q$ gluons
to the $Q$ or $\bar Q$, and these soft gluons must be absorbed into the
NRQCD LDME if NRQCD factorization is to hold. A resolution of this
dilemma was suggested in Ref.~\cite{Nayak:2005rt}. It is based on
modifying the original definition of an LDME to include a Wilson line,
which acts as a proxy for an order-$m_Q$ gluon, in that its interactions
with additional soft gluons are identical to those of the order-$m_Q$
gluon. This so-called ``gauge completion'' of the LDMEs is also required
in order to make the LDMEs manifestly gauge invariant
\cite{Nayak:2005rt}. (In this paper, we will always refer to the
gauge-completion Wilson lines as ``Wilson lines'' in order to
distinguish them from soft approximations to the heavy-quark lines,
which we refer to as ``eikonal lines.'')

There are no apparent obstacles to establishing that all of the soft
singularities in the fragmentation functions can be absorbed into the
gauge-completed NRQCD LDMEs, given that the effective field theory NRQCD
is a valid approximation to QCD in the soft limit and that the
gauge-completed LDMEs account for the soft interactions with the
order-$m_Q$ gluons. The central problem in proving NRQCD factorization
is then to demonstrate that the soft divergences in the LDMEs are
independent of the direction(s) of the gauge-completion Wilson line(s).
Without such a demonstration, the LDMEs would depend on the directions
of the order-$m_Q$ gluons, and so universality of the LDMEs would be
lost, even for a single type of quarkonium production process.

Let us contrast this situation with that in exclusive quarkonium
production. A proof of NRQCD factorization for exclusive quarkonium
production at leading power in the inverse of the hard-scattering
momentum transfer is given in Refs.~\cite{Bodwin:2008nf,Bodwin:2009cb}.
That proof focuses on a demonstration that soft-gluon contributions
that do not decouple from the quarkonium can be absorbed completely into
the quarkonium jet. The proof does not address the further factorization
of the quarkonium jet into NRQCD LDMEs. In the proof, it was assumed, on
general grounds, that this factorization would be valid,  since NRQCD is
the effective field theory that describes the relevant low-energy
degrees of freedom. As we have mentioned, the issue of the dependence of
the LDMEs on the Wilson-line direction arises in inclusive quarkonium
production because of the emission of gluons with energies of order
$m_Q$ in the quarkonium c.m. frame. Such real-gluon emissions do not occur
in exclusive quarkonium production, and so no Wilson line is needed to
account for their couplings to soft gluons. Furthermore, the relevant
vacuum-to-quarkonium matrix elements have color-singlet quantum numbers
and do not require a gauge-completion Wilson line. Therefore, the issue
of dependence on the direction of a Wilson line never arises in this
case.

A proof of factorization for inclusive quarkonium decays is also
qualitatively different from a proof for inclusive quarkonium
production. It is thought that NRQCD factorization for inclusive
quarkonium decays can be established along the lines of the argument
that is given in Ref.~\cite{Bodwin:1994jh}, although no complete proof
has been published. The essential element of this argument is that
contributions of final-state soft or collinear gluons that are radiated
from final-state hard partons cancel in an inclusive process because of
the Kinoshita-Lee-Nauenberg theorem. Because soft gluons decouple from the hard-scattering
subdiagram, no Wilson line is needed in this case to account for
soft-gluon exchanges between hard gluons (with energies of order $m_Q$)
and the heavy quark or antiquark. Furthermore, the decay LDMEs do not
require a gauge-completion Wilson line. Again, the issue of dependence
on the direction of a Wilson line never arises in this case.

Returning to the case of inclusive quarkonium production, we note
that perturbative tests at one and two loops of the proposition that
the soft divergences in the LDMEs are independent of the direction of the
gauge-completion Wilson line have been provided by Nayak, Qiu, and
Sterman \cite{Nayak:2005rw,Nayak:2005rt,Nayak:2006fm}. In
Refs.~\cite{Nayak:2005rw,Nayak:2005rt}, one- and two-loop contributions
to the LDMEs were given at the leading nontrivial order in $v$ (order
$v^2$), with the details of the calculations being given in
Ref.~\cite{Nayak:2005rt}. In Ref.~\cite{Nayak:2006fm}, the one- and
two-loop contributions were given to all orders in $v$, although
explicit expressions were given only for the non-Abelian diagrams, which
are the most complicated to evaluate. The calculations in these papers
confirmed that
the soft divergences in the LDMEs are independent of the Wilson-line
direction through two loops. In the remainder of this paper, we refer
collectively to Refs.~\cite{Nayak:2005rt,Nayak:2006fm} as ``NQS''.

The calculations in Refs.~\cite{Nayak:2005rw,Nayak:2005rt,Nayak:2006fm}
make use of light-cone methods, in which one first uses contour
integration to carry out the integration over the component of a loop
momentum that lies in the minus light-cone direction. The calculations
also make use of hard ultraviolet (UV) cutoffs for the phase-space
integrations for real gluons. Hence, the calculations are not manifestly
Lorentz invariant at intermediate stages, although the final results
are. The calculations are quite complicated, especially for the
non-Abelian diagrams, and the independence of the soft divergences from
the Wilson-line direction emerges only after many complicated
intermediate expressions are summed. Therefore, it seems worthwhile to
check these calculations using manifestly covariant methods. One could
hope that such covariant calculations might give insights into the
independence of the soft divergences from the Wilson-line direction.

In this paper, we present such a manifestly covariant calculation of the
soft divergences in an LDME at two-loop order. The calculation
relies on a new UV regulator for the phase-space integrations that
avoids the use of a hard cutoff. As we will see, our covariant
calculation is actually considerably more complicated than the
calculations in NQS. This is probably because the light-cone methods in
NQS allow one to cancel contributions of real-real diagrams against
contributions of real-virtual diagrams and to cancel certain
real-virtual contributions by symmetry, without having to calculate them
completely. In contrast, in our calculations, we compute all of
the IR contributions of the real-real diagrams and real-virtual
diagrams explicitly and implement cancellations only at the end.
Consequently, we must deal with soft poles in dimensional regularization
in $D=4-2\epsilon$ dimensions of order $1/\epsilon^4$, $1/\epsilon^3$,
$1/\epsilon^2$, and $1/\epsilon$ in order to arrive at a final result of
order $1/\epsilon$. Although we can eliminate the $1/\epsilon^4$
contribution trivially at the start by making a Ward-identity
rearrangement of the numerator structure, we still must work out the
cancellation of the $1/\epsilon^3$ and $1/\epsilon^2$ poles, while the
NQS calculation contains only $1/\epsilon^2$ imaginary poles, which
cancel in the absolute square of the amplitude, and $1/\epsilon$ real
poles.

In NQS, it was found that IR poles in the LDME are independent of the
Wilson-line direction, and we also find that to be the case. Our result
for the non-Abelian diagrams agrees with that in NQS, aside from an
overall sign. Our result for the Abelian diagrams differs from that in
NQS in two respects. First, we find new contributions that have the
form of a one-loop contribution to an SDC times the one-loop
IR-divergent contribution to the LDME. For these contributions,
the IR pole in the LDME is independent of the direction of the Wilson
line. We conclude that some of these contributions are absent in NQS
because a mismatch between virtual-gluon and real-gluon contributions in
the UV region was neglected. We also find a new contribution that has
the form of a two-loop IR-divergent contribution to the LDME. Again,
the IR pole in the LDME is independent of the direction of the Wilson
line. This new contribution can be reconciled with the result in NQS
if we reinterpret a UV contribution in NQS as an IR contribution.

The remainder of this paper is organized as follows. In
Sec.~\ref{sec:preliminaries} we describe the LDME and its Feynman rules,
the classes of diagrams that we calculate, and the kinematics.
Section~\ref{sec:phase-space} contains a description of the covariant
phase-space regulator that we use and also contains convenient formulas
for the phase-space integration in dimensional regularization. The main
body of our work is in Sec.~\ref{sec:diagrammatic-analyses}, in which we
give a description of the calculations for the various diagrams.
Because the extraction of analytic expressions for the coefficients of
the IR poles is not straightforward, we present our calculation in
considerable detail, so that an interested reader would be able to
reproduce our results. We have relegated the technical details of the
calculation to appendices. Our results are summarized in
Sec.~\ref{sec:summary}, and we compare them with the results in NQS in
Sec.\ref{sec:comparison}. Finally, we give our conclusions in
Sec.~\ref{sec:conclusions}.

\section{Preliminaries\label{sec:preliminaries}}

The gauge-completed LDME is given by \cite{Nayak:2006fm}
\begin{eqnarray}
\langle {\mathcal O}^H_n(0) \rangle
=
\langle 0 |
\chi^\dagger(0)
{\mathcal{K}}_{n,e}\psi(0)
\Phi_\ell^{(A)}{}^\dagger(0){}_{eb}
\left(a^\dagger_Ha_H\right)
\Phi_\ell^{(A)} (0)_{ba} 
\psi^\dagger(0) {\mathcal{K}}'_{n,a}\chi(0)
| 0 \rangle,
\end{eqnarray}
where $\psi^{\dagger}$ and $\chi$ are two-component Pauli fields 
that create a heavy quark and a heavy antiquark, respectively, 
$\mathcal{K}_{n,e}$ and $\mathcal{K'}_{n,a}$ are local combinations of 
spin and color matrices and polynomials in covariant derivatives, 
and $a_H^{\dagger}$ is the operator that creates the quarkonium $H$.
The operators $\Phi_\ell^{(A)}(0){}$ are Wilson lines, 
which are defined by 
\begin{eqnarray}
\Phi_\ell^{(A)}(0){}
&=&
\mathcal{P} \text{exp} 
\left[
-ig_{s}\int_0^{\infty} d\lambda \,\ell \cdot A^{(A)} (\ell \lambda)
\right],
\end{eqnarray}
where $\mathcal{P}$ stands for path ordering, 
$g_s=\sqrt{4\pi\alpha_s}$ is the QCD coupling constant,
$A^{(A)\mu}=A^\mu_a t_a$
is the gauge field in the adjoint representation, 
$(t_a)_{bc}=-if_{abc}$, and $\ell^{\mu}$ is the momentum of the Wilson line.

Following NQS, we consider the LDME for a $Q\bar Q$ pair in an $S$-wave, 
color-octet state to evolve into a $Q\bar Q$ pair in a color-singlet 
state of arbitrary orbital angular momentum. That is, we consider the 
matrix elements \cite{Nayak:2006fm}
\begin{eqnarray}
\label{eq:octet-ME}
{\cal M}^{(8\to I)}(P_1,P_2,\ell) 
&=&
\sum_X \langle 0| \chi^\dagger(0) T^{(q)}_e\psi(0)
\Phi_\ell^{(A)}{}^\dagger(0){}_{eb} 
|[Q(P_1)\bar{Q}(P_2)]^{(I)} X\rangle
\nonumber\\ 
&&
\times
\langle X [Q(P_1)\bar{Q}(P_2)]^{(I)} |
\Phi_\ell^{(A)} (0)_{ba} \psi^\dagger(0) T^{(q)}_a\chi(0)
|0\rangle,
\end{eqnarray}
where the superscript $I=1,\,8$ denotes the color of the $Q\bar Q$ pair, 
the $T_i^{(q)}$ are the generators of color SU$(3)$ in the
fundamental representation,
and $P_1$ and $P_2$ are momenta of $Q$ and $\bar{Q}$, respectively.
We do not explicitly consider the spin state of the $Q\bar Q$ pair, as
the soft approximation for the $Q$ and $\bar Q$ lines is independent of
the spin.

In computing the matrix elements in Eq.~(\ref{eq:octet-ME}), we
encounter factors that arise from UV-divergent loop integrals. Since
these factors contain no IR sensitivity, they can be interpreted as
contributions to the SDCs that relate the LDMEs to fragmentation
functions. As we have mentioned, the directions of the Wilson lines
correspond to the directions of emitted gluons that have energies of
order $m_Q$ in the quarkonium c.m. frame. Therefore, the directions of the
Wilson lines are process dependent, and the SDCs, which are process
dependent, can depend on the directions of the Wilson lines. Dependence
of the SDCs on the Wilson-line directions is entirely consistent with
NRQCD factorization, provided that the IR poles, which are absorbed into
the LDMEs, are independent of the Wilson-line directions.

\begin{figure}       
\centering                           
\includegraphics[width=0.35\columnwidth]{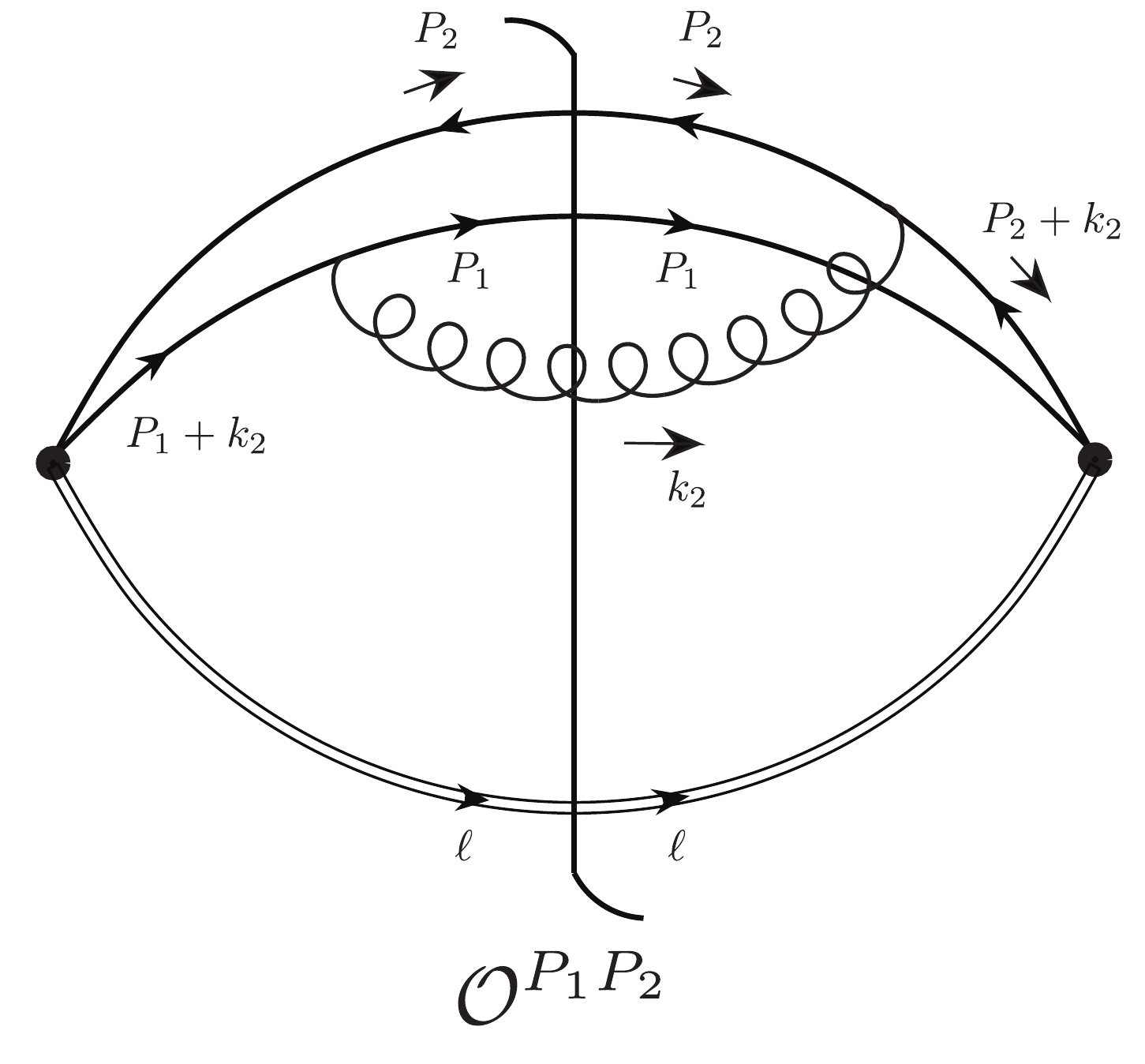}
\caption{\label{figure:O-P1P2}
The one-loop diagram $\mathcal{O}^{P_1P_2}$}
\end{figure}

The only one-loop contribution to the matrix element in 
Eq.~(\ref{eq:octet-ME}) that has a nonvanishing color factor 
is shown in Fig.~\ref{figure:O-P1P2}.
The superscripts $P_jP_k$ specify the gluon connections to the 
$Q$ or $\bar{Q}$ lines on either side of the cut. 

\begin{figure}
\centering                          
\includegraphics[width=0.9\columnwidth]{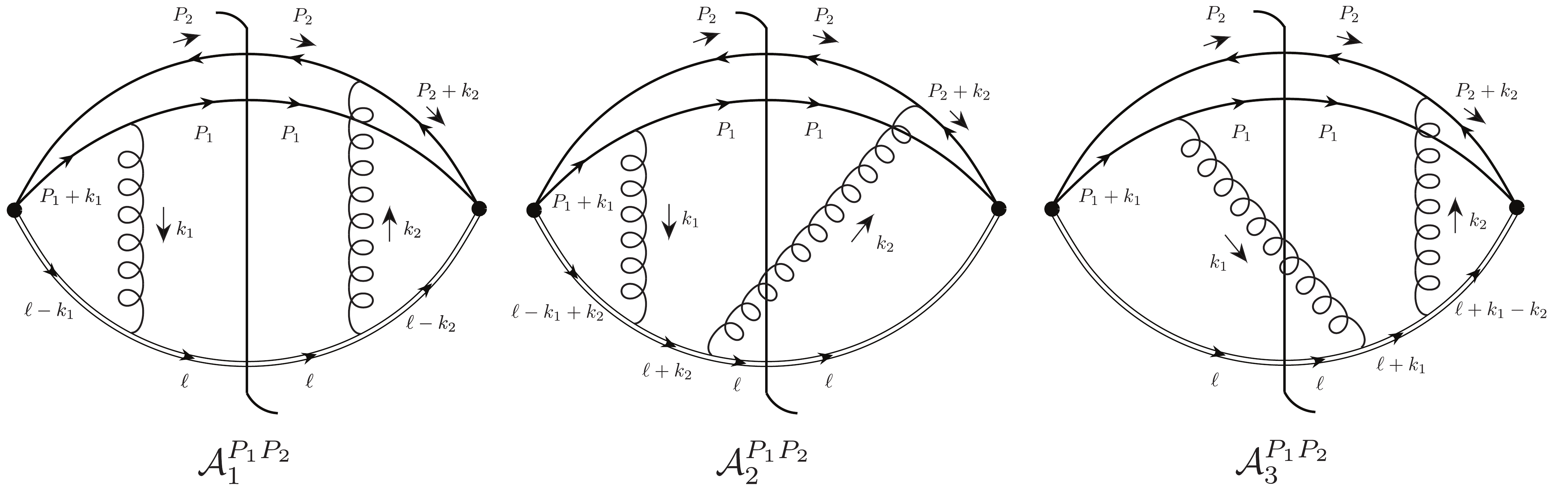}
\caption{\label{figure:A-P1P2}
Ladder diagrams $\mathcal{A}_i^{P_1P_2}$}
\includegraphics[width=0.9\columnwidth]{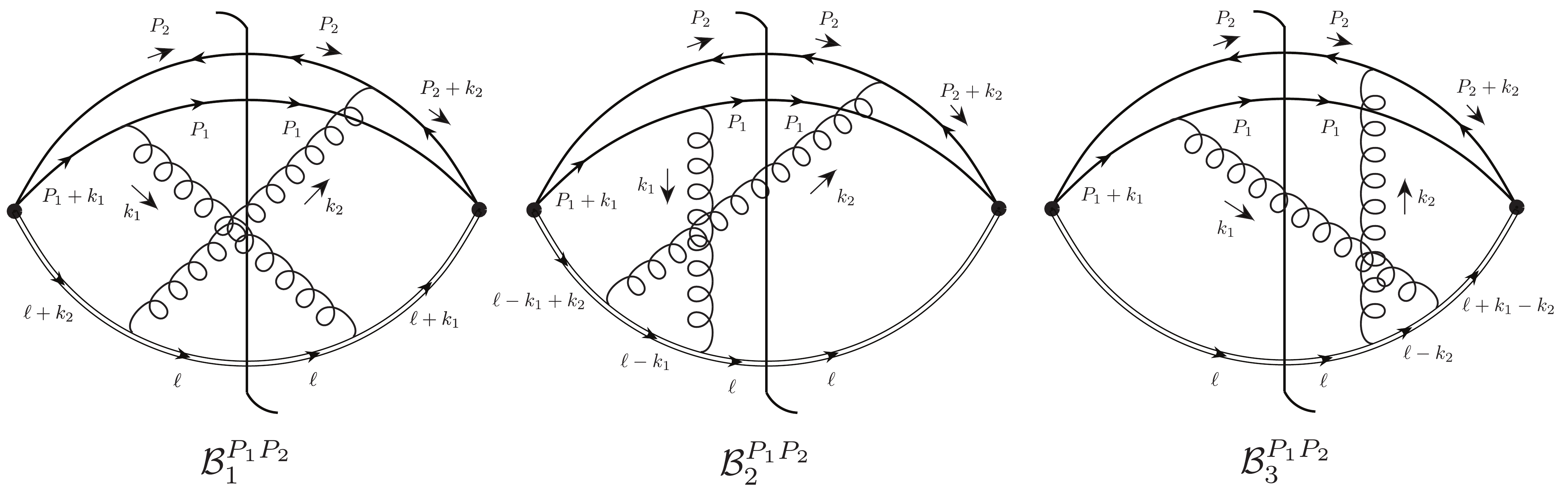}
\caption{\label{figure:B-P1P2}
Crossed ladder diagrams $\mathcal{B}_i^{P_1P_2}$}
\end{figure}

\begin{figure}
\centering                          
\includegraphics[width=0.7\columnwidth]{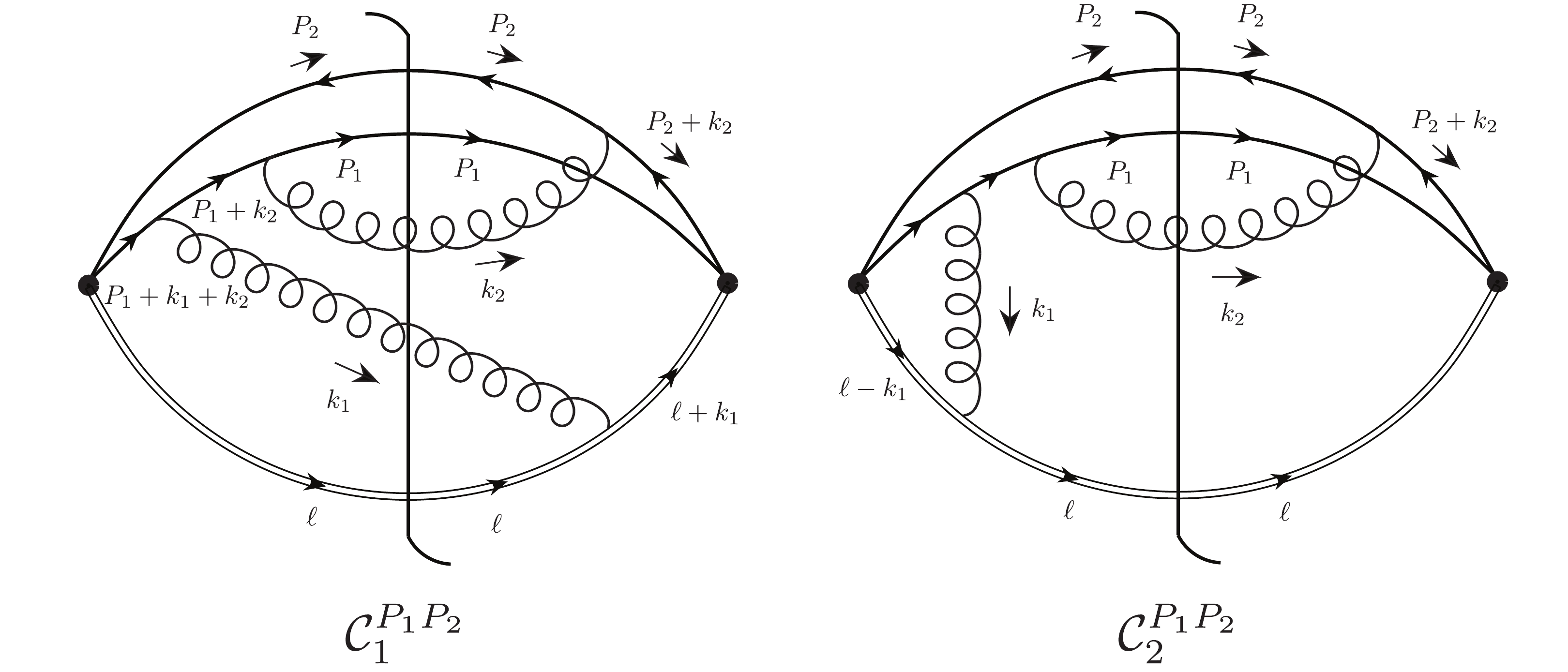}
\caption{\label{figure:C-P1P2}
Diagrams $\mathcal{C}_i^{P_1P_2}$}
\centering                          
\includegraphics[width=0.7\columnwidth]{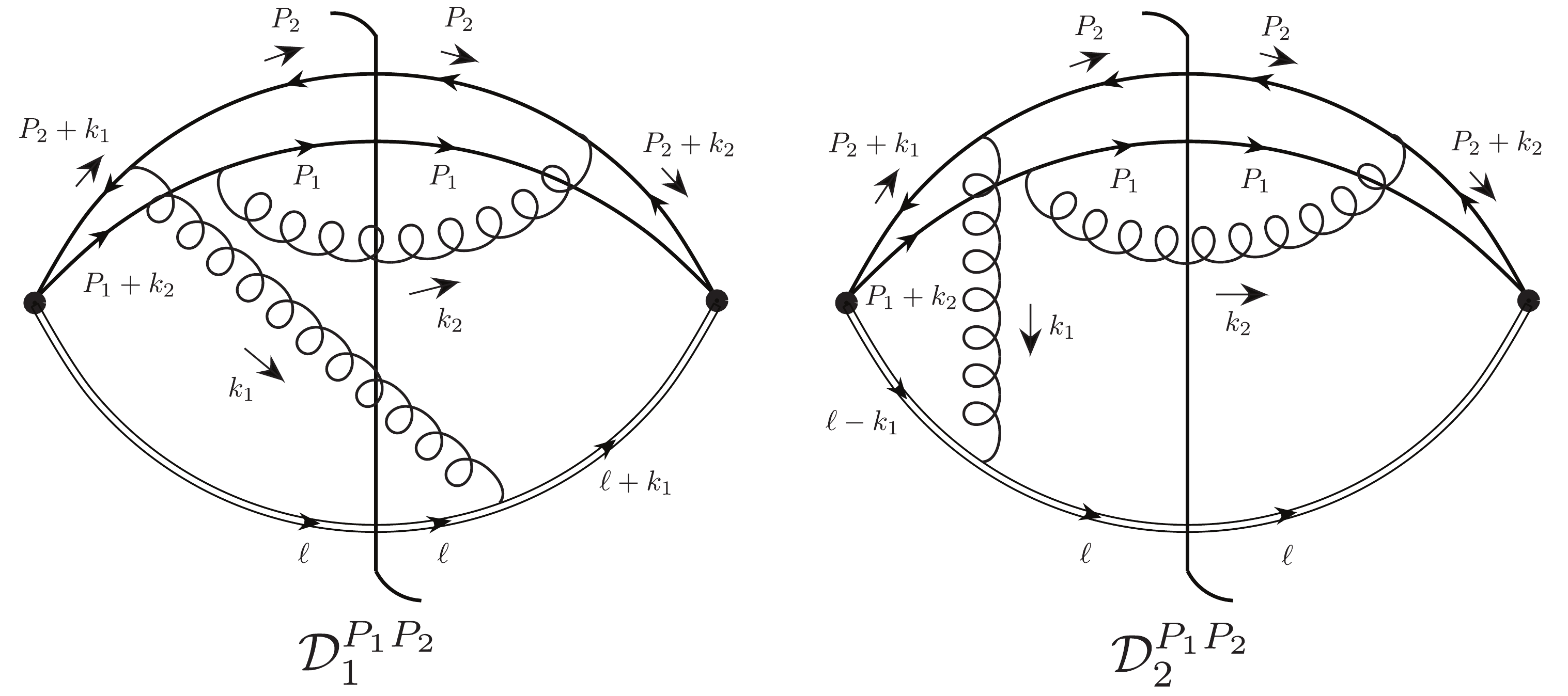}
\caption{\label{figure:D-P1P2}
Diagrams $\mathcal{D}_i^{P_1P_2}$}
\centering 
\includegraphics[width=0.7\columnwidth]{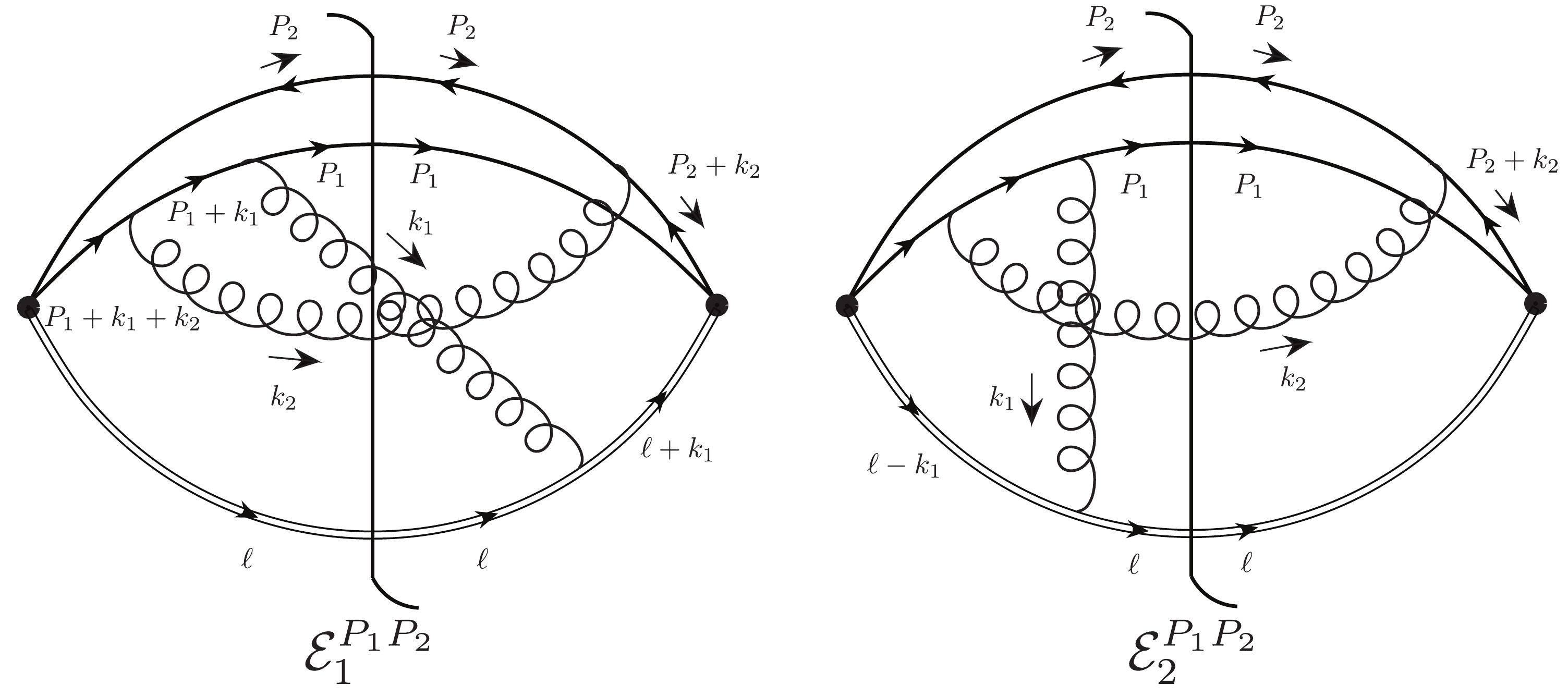}
\caption{\label{figure:E-P1P2}
Diagrams $\mathcal{E}_i^{P_1P_2}$}
\end{figure}

\begin{figure}       
\centering                           
\includegraphics[width=0.7\columnwidth]{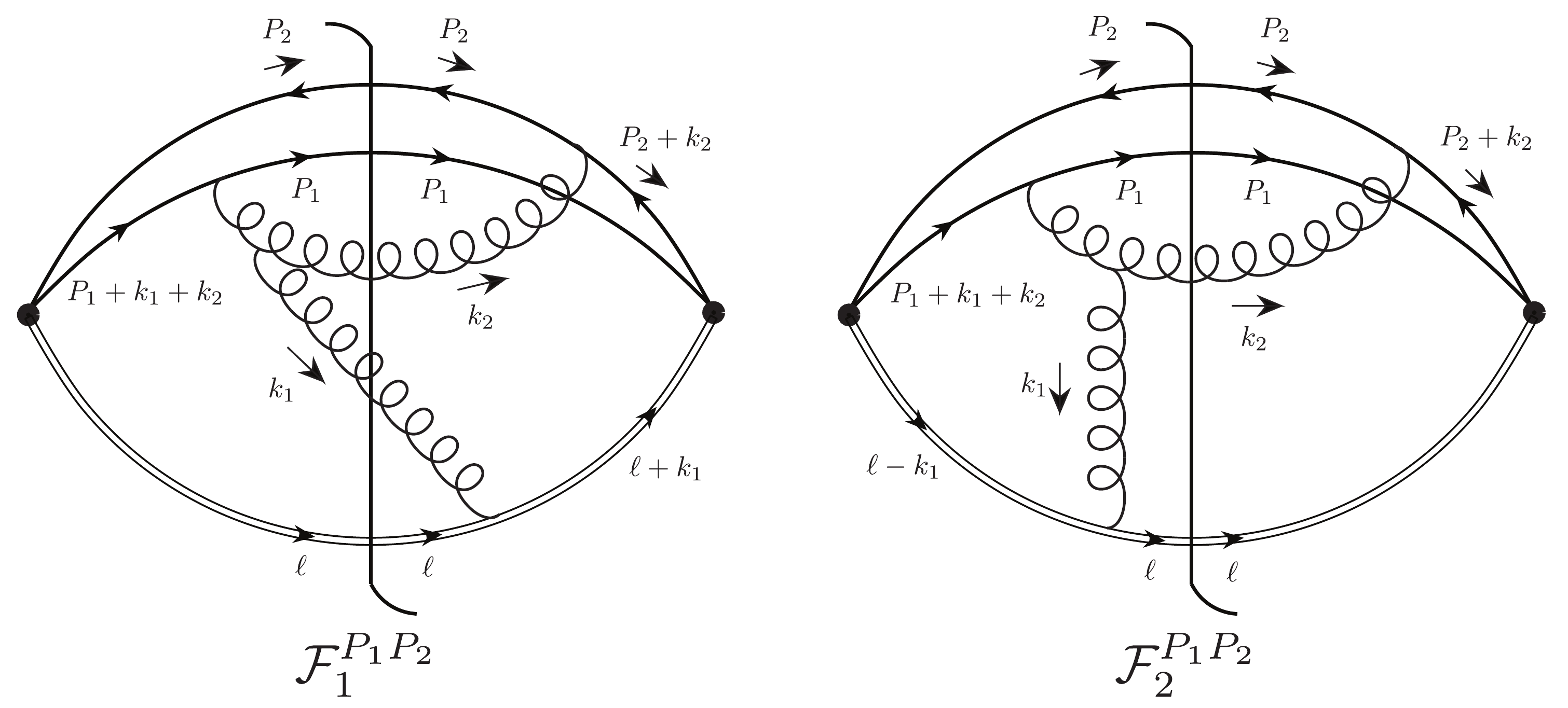}
\caption{\label{figure:F-P1P2}
Non-Abelian diagrams $\mathcal{F}_i^{P_1P_2}$}
\end{figure}

Now let us consider the two-loop diagrams that contribute to the 
matrix element in Eq.~(\ref{eq:octet-ME}).
Since we wish to test for a dependence on the direction of the LDME
Wilson line, we need to consider only those diagrams in which at least
one gluon attaches to the Wilson line. Furthermore, it is clear that a
diagram has a nonzero color factor only if one or more gluons crosses the
final-state cut. The specific classes of diagrams that we compute are
illustrated in Figs.~\ref{figure:A-P1P2}--\ref{figure:F-P1P2}. We use
the notation ${\mathcal X}_i^{P_j P_k}$ for the contributions of these
diagrams. Here, ${\mathcal X}_i$ denotes the class of the diagram. For
each class, the subscript $i$ labels the position of the final-state
cut, and the superscripts $P_j P_k$ specify the gluon connections to the
$Q$ and $\bar Q$ lines, as we will describe below. The classes are
defined as follows.
\begin{itemize}
\item $\mathcal{A}_i^{P_jP_k}$ (Fig.~\ref{figure:A-P1P2}) designates
normal ladder diagrams. $P_j$ indicates the
leftmost gluon connection to the heavy-quark lines; $P_k$ indicates the 
rightmost gluon connection to the heavy-quark lines.

\item $\mathcal{B}_i^{P_jP_k}$ (Fig.~\ref{figure:B-P1P2}) designates
crossed ladder diagrams. $P_j$ indicates the
leftmost gluon connection to the heavy-quark lines; $P_k$ indicates 
the rightmost gluon connection to the heavy-quark lines.

\item $\mathcal{C}_i^{P_jP_k}$
(Fig.~\ref{figure:C-P1P2}) designates Abelian diagrams with two
connections of gluons to the same heavy-quark line to the left of the
cut, in which the gluon that connects to the Wilson line connects to the 
heavy-quark line to the left of the other gluon. 
$P_j$ indicates the two leftmost connections of
gluons to the heavy-quark lines; $P_k$ indicates the rightmost
connection of a gluon to the heavy-quark lines.

\item $\mathcal{D}_i^{P_jP_k}$
(Fig.~\ref{figure:D-P1P2}) designates Abelian diagrams with two
connections of gluons to different heavy-quark lines to the left of the
cut.  $P_j$ indicates the leftmost connection to the 
heavy-quark lines of the gluon that does not attach to the Wilson line; 
$P_k$ indicates the rightmost connection to the 
heavy-quark lines of the gluon that does not attach to the Wilson line.

\item $\mathcal{E}_i^{P_jP_k}$ (Fig.~\ref{figure:E-P1P2})
designates Abelian diagrams with two
connections of gluons to the same heavy-quark line to the left of the
cut, in which the gluon that connects to the Wilson line connects to the 
heavy-quark line to the right of the other gluon. 
$P_j$ indicates the two leftmost connections of
gluons to the heavy-quark lines; $P_k$ indicates the rightmost
connection of a gluon to the heavy-quark lines.

\item $\mathcal{F}_i^{P_jP_k}$ (Fig.~\ref{figure:F-P1P2}) designates the
non-Abelian diagrams. $P_j$ indicates the
leftmost connection of a gluon to the heavy-quark lines; $P_k$
indicates the rightmost connection of a gluon to the heavy-quark lines.

\end{itemize}

The classes of diagrams are constructed such that one can obtain all of
the contributions that we calculate by summing over the indices $i$,
$j$, and $k$ and including Hermitian-conjugate diagrams, except for
the classes $\mathcal{A}$ and $\mathcal{B}$, for which the sums over the
indices $i$, $j$, and $k$ already include the Hermitian-conjugate
contributions. We note that the assignments of our loop momenta $k_1$
and $k_2$ for the diagrams $\mathcal{C}_i$, $\mathcal{D}_i$,
$\mathcal{E}_i$, and $\mathcal{F}_i$ are different from those of NQS.

We define the total momentum of the $Q\bar{Q}$ pair to be $2p$ and the
relative momentum to $Q\bar{Q}$ rest frame to be $q$, and so
\begin{equation}
P_1=p+q,
\quad
P_2=p-q.
\end{equation}
In the rest frame of the $Q\bar{Q}$ pair, $p$ and $q$ are given by
\begin{equation}
\label{eq:p-and-q-in-QQ-rest}
p=(E_Q,\bm{0}),
\quad
q=(0,\bm{q}),
\end{equation}
where $E_Q=\sqrt{m_Q^2+\bm{q}^2}$. The 
relative velocity of the $Q$ and $\bar{Q}$ is 
\begin{equation}
\label{eq:p-and-q-in-QQ-rest-2}
\bm{v}=\frac{2\bm{q}}{E_Q}.
\end{equation}

In comparing with the light-cone calculation in NQS, we need to make use 
of light-cone momentum coordinates
for a $D$-dimensional vector 
$V=(V^0, V^1, \cdots, V^{D-1})$, which we take to be
\begin{eqnarray}
V^+&=&\frac{1}{\sqrt{2}}\left(V^0+V^{D-1}\right),
\nonumber \\
V^-&=&\frac{1}{\sqrt{2}}\left(V^0-V^{D-1}\right),
\nonumber \\
\bm{V}_\perp&=&(V^1, V^2, \cdots, V^{D-2}).
\end{eqnarray}
Then, a scalar product 
of two vectors $V$ and $W$ is given by
\begin{eqnarray}
V \cdot W 
&=& 
V^0 W^0 - V^{D-1} W^{D-1} - \bm{V}_\perp\cdot\bm{W}_\perp
\nonumber \\
&=&
V^+ W^- + V^- W^+ - \bm{V}_\perp\cdot\bm{W}_\perp.
\end{eqnarray}

We note that, in NQS, the momentum of the Wilson line is specified to be
along the minus light-cone direction:
\begin{equation}
\ell=(\ell^+,\ell^-,\bm{\ell}_\perp)=(0,1,\bm{0}_{\perp}).
\end{equation}
However, our covariant calculation does not make use of this assignment.

We define the following Lorentz-invariant quantities, which appear 
throughout our calculations:
\begin{equation}
\label{eq:def-acd}
P_1^2=P_2^2, 
\quad
\ell^2=0,
\quad
a=\frac{P_1\cdot P_2}{P_1^2},
\quad
c=\frac{P_1\cdot \ell}{P_1^2},
\quad
d=\frac{P_2\cdot \ell}{P_1^2}.
\end{equation}

Since we are interested in analyzing the soft singularities, we take
soft approximations for the interactions of the gluons with the $Q$ and
$\bar Q$ lines.\footnote{In the method of regions or threshold
expansion for heavy quarkonium \cite{Beneke:1997zp}, the complete
decomposition of NRQCD amplitudes involves contributions in which gluon
momenta are in the soft, potential, and ultrasoft regions. At two-loop
order, the diagrams that involve the Wilson line do not contain
virtual-gluon exchanges between the heavy quark and the heavy antiquark.
Hence, contributions from the potential region do not arise in our
calculation. The distinction between the soft and ultrasoft regions
affects the approximations that are used for the gluon propagators. The
soft approximation that we have taken for the interactions of gluons
with the heavy quark and heavy antiquark is valid in both the soft and
ultrasoft regions.}  As we have mentioned, we refer to the $Q$ and
$\bar Q$ lines in the soft approximation as ``eikonal lines,'' and we
refer to the gauge-completion Wilson lines as ``Wilson lines,'' in order
to distinguish them from each other. The Feynman rules for the eikonal
lines and Wilson lines are given in
Refs.~\cite{Nayak:2005rt,Nayak:2006fm}. They are summarized below.
\begin{itemize} \item The interaction between a gluon and a $Q$ or $\bar
Q$ line to the left of the final-state cut is shown in
Fig.~\ref{figure:fg}. The propagator for a $Q$ or $\bar Q$ line to the
left of the cut is given by
\begin{equation}
\frac{i}{P_i\cdot k+i\varepsilon},
\end{equation}
and the vertex is given by
\begin{equation}
\quad
-ig_s \mu^{\epsilon} (T_a)_{bc} P_i^{\alpha},
\end{equation}
where $\alpha$ is the Lorentz index of the vertex, $a$ is the color
index of the vertex, $b$ and $c$ are the color indices of the $Q$,
$\mu^\epsilon$ is introduced to account for the dimensionality of the
coupling constant in $D=4-2\epsilon$ dimensions, and the diagrammatic
momentum of the $\bar Q$ line is opposite to the physical momentum.  The
rules for propagators and vertices to the right of the final-state cut
are obtained by taking the Hermitian conjugates of the rules that are
shown. Note that the product of a propagator and a vertex is
invariant with respect to a change of the scale of $P_i$.
\begin{figure}
\centering
\includegraphics[width=0.3\columnwidth]{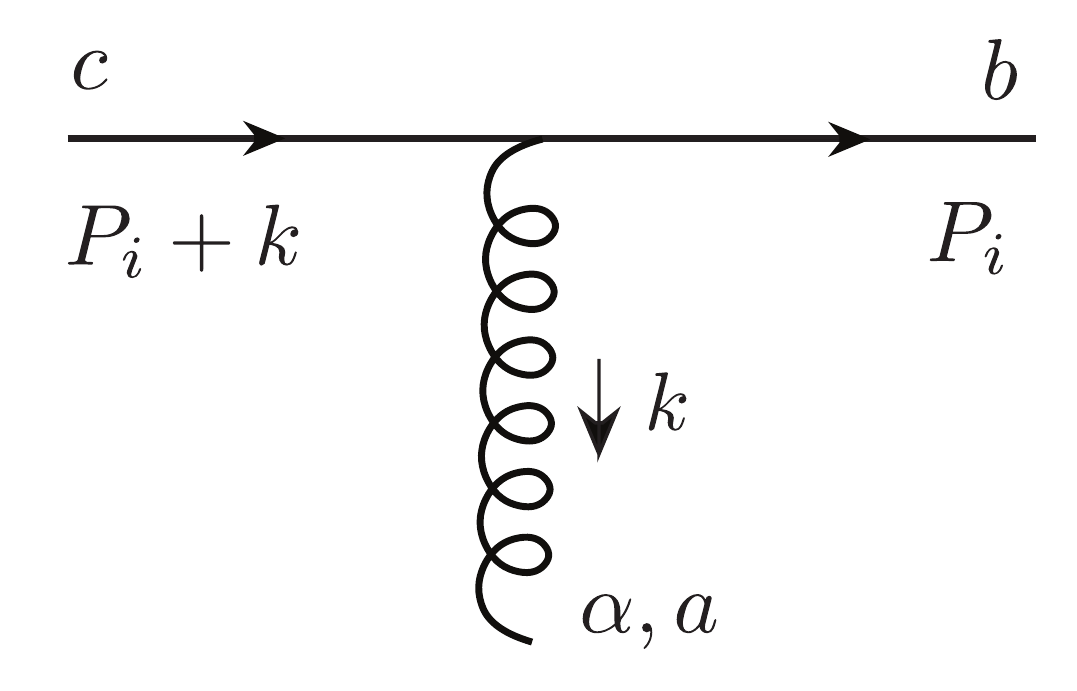}
\caption{\label{figure:fg}Gluon interaction with a $Q$ or $\bar Q$ line.}
\end{figure}

\item The interaction of a 
gluon with a Wilson line is illustrated in Fig.~\ref{figure:LLg}. On 
the left side of the final-state cut, the Wilson-line propagator is given 
by
\begin{equation}
\frac{-i}{\ell\cdot k+i\varepsilon},
\end{equation}
and the vertex is given by
\begin{equation}
-g_s \mu^\epsilon  \ell^{\alpha} f_{abc},
\end{equation}
where $a,\,b,\,c$ are color indices and the $f_{abc}$ are the structure
constants of SU$(3)$. The rules for propagators and vertices to the
right of the final-state cut are obtained by taking the Hermitian
conjugates of the rules that are shown.\footnote{Note that the sign of the 
Wilson-line vertex reverses when the vertex is on the right side of the 
final-state cut, owing to the fact that $f_{abc}$ is anti-Hermitian with
respect to the indices $a$ and $c$. This is also the case for 
the color factor for the
triple-gluon vertex. For both the Wilson-line vertex and the 
triple-gluon vertex, we absorb these sign changes into the expression for the
non-color-factor part of a diagram, so that all of the cuts of a diagram
have the same color factor.}
As in the eikonal-line case, the product of a Wilson-line propagator
and vertex is invariant with respect to a change of the scale of
line momentum $\ell$.
\begin{figure}
\centering
\includegraphics[width=0.3\columnwidth]{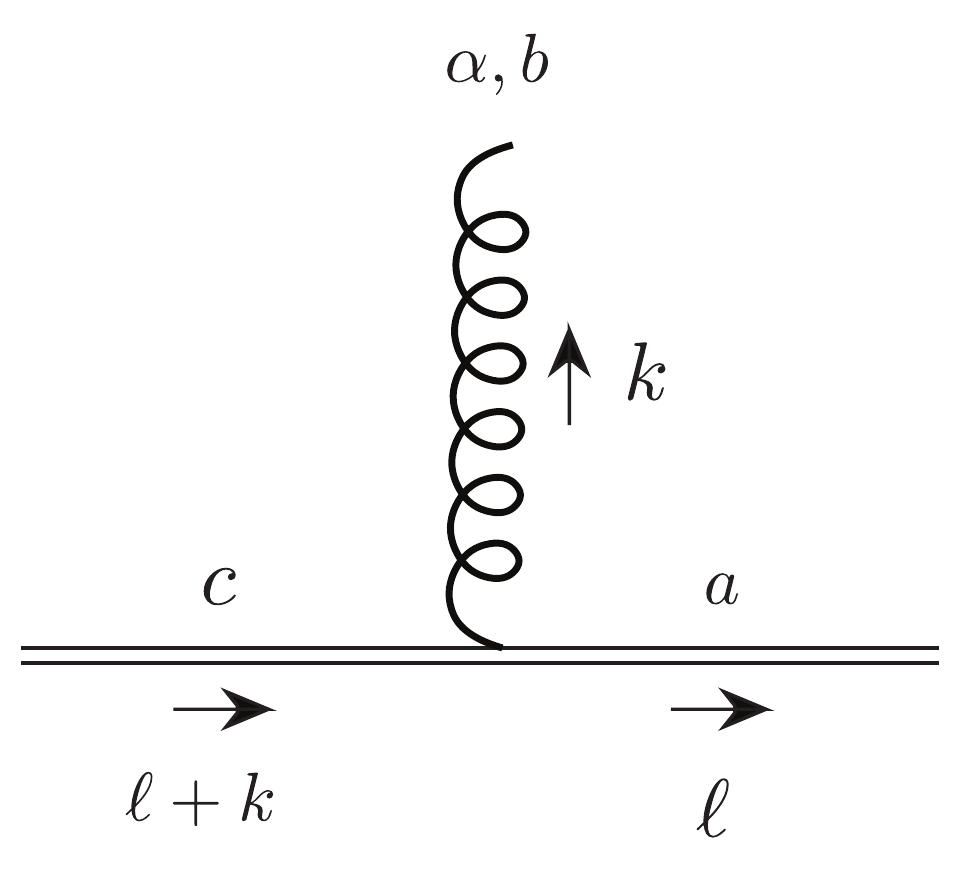}
\caption{\label{figure:LLg}Gluon interaction with a Wilson line.}
\end{figure}
\end{itemize}

The Feynman rules for the non-eikonal and non-Wilson-line 
parts of the diagram are the usual ones. We follow the conventions in 
\cite{Sterman:1994ce}, which are consistent with the conventions that we
have chosen for the eikonal and Wilson lines.
We work in the Feynman 
gauge throughout.

We find that our expressions for the diagrams $\mathcal{C}_i$,
$\mathcal{D}_i$, $\mathcal{E}_i$, and $\mathcal{F}_i$, which contain
one gluon connection to the Wilson line, differ by an overall sign from the
expressions for the corresponding diagrams in NQS.

\section{Phase Space\label{sec:phase-space}}

In this section we discuss the covariant phase-space regulators that we 
employ and also present convenient formulas for carrying out the 
phase-space integrations in dimensional regularization.

\subsection{Phase-space regulators}

Because the eikonal and Wilson propagator denominators are linear in
the loop momenta and the eikonal and Wilson vertices are independent of
the loop momenta, the unregulated expressions for the LDME are invariant
under a simultaneous rescaling of both loop momenta. This scale
invariance would lead to a vanishing of the unregulated loop
integrations in dimensional regularization. That is, UV poles from the
loop integrations would cancel IR poles from the loop integrations.
However, our goal is to isolate and calculate the IR poles. We
accomplish this by imposing an additional UV regulator, which breaks the
scale invariance of the integrals and guarantees that our results
contain only IR poles.

In the fragmentation function in which the LDME is embedded, there are
restrictions on the final-state phase space that follow from
Dirac $\delta$ functions that express conservation of the light-cone energy:
\begin{equation}
\begin{cases}
\delta\left[\ell\cdot(k-P_1-P_2-k_2)\right],
\quad \hbox{when the gluon with momentum $k_2$ is real},
\\
\delta\left[\ell\cdot(k-P_1-P_2-k_1-k_2)\right],
\quad \hbox{when the gluons with momenta $k_1$ and $k_2$ are real}.
\end{cases}
\label{energy-delta-fn}
\end{equation}
Here $k$ is the momentum of the parton (gluon) that fragments into the
$Q\bar Q$ pair. In the fragmentation function at a scale of order $m_Q$,
the light-cone energy that is available to the final-state gluons,
$\ell\cdot k'=\ell\cdot(k-P_1-P_2$), is also of
order $m_Q$.

Motivated by this constraint on the fragmentation-function phase space,
we impose a UV cutoff on the phase space of the LDME. In NQS, there is
also a UV cutoff on the phase space of the LDME. It is imposed as a hard cutoff on $\ell\cdot k'$
(and on the remaining components of $k'$). We find it calculationally 
more convenient to provide a cutoff of order $m_Q$ on $\ell\cdot k'$ by
applying a weight function to the available light-cone energy $\ell\cdot 
k'$:
\begin{equation}
w(\ell\cdot k) \equiv
\frac{\Lambda^2}{2(\ell\cdot k)+\Lambda^2},
\end{equation}
where $\Lambda$ is a cutoff parameter of order $m_Q$. Then, integrating 
over the constraints in Eq.~(\ref{energy-delta-fn}), we obtain
\begin{subequations}%
\label{standard-regulators}%
\begin{equation}
\int_0^\infty d(\ell\cdot k')w(\ell\cdot k')\,
\delta\left[\ell\cdot(k'-k_2)\right]
=
\frac{\Lambda^2}{2\ell\cdot k_2+\Lambda^2},
\end{equation}
when the gluon with momentum $k_2$ is real, and
\begin{equation}
\int_0^\infty d(\ell\cdot k')w(\ell\cdot k')\,
\delta\left[\ell\cdot(k'-k_1-k_2)\right]
=
\frac{\Lambda^2}{2\ell\cdot (k_1+k_2)+\Lambda^2},
\end{equation}
\end{subequations}%
when the gluons with momenta $k_1$ and $k_2$ are real.
We call the factors on the right side of Eq.~(\ref{standard-regulators})
our standard phase-space regulators. For some parts of the calculation,
as we will explain later, we will find it necessary to impose temporarily
additional UV regulators in order to ascertain the IR or UV nature
of poles in $\epsilon$.

\subsection{Phase-space integration}

In computing the phase-space integration for the final-state gluons, it
is convenient to  extend the range of integration to infinity, relying
on UV regulators to remove UV poles. Then, we obtain the following
phase-space integration formulas:
\begin{subequations}%
\label{eq:k1-phase-int-table}%
\begin{eqnarray}
\label{eq:k1-phase-int-table-a}%
\int_k\textrm{PS}
\frac{1}{(2p\cdot k +M^2\pm i\varepsilon)^s} 
&=&
\frac{1}{(4\pi)^{\frac{D}{2}}}
\frac{\Gamma(\frac{D}{2}-1)\Gamma(s-D+2)}
{\Gamma(s)\left(p^0+|\bm{p}|\pm i\varepsilon\right)^{\frac{D}{2}-1}
\left(p^0-|\bm{p}|\pm i\varepsilon\right)^{\frac{D}{2}-1}
(M^2\pm i\varepsilon)^{s-D+2}},\nonumber\\
\\
\label{eq:k1-phase-int-table-b}%
\int_k \textrm{PS}
\frac{k^\alpha}{(2p\cdot k+M^2\pm i\varepsilon)^s}&=&
\frac{p^\alpha}{(4\pi)^{\frac{D}{2}}}
\frac{\Gamma(\frac{D}{2})\Gamma(s-D+1)}
{\Gamma(s)\left(p^0+|\bm{p}|\pm i\varepsilon\right)^{\frac{D}{2}}
\left(p^0-|\bm{p}|\pm i\varepsilon\right)^{\frac{D}{2}}
\left(M^2\pm i\varepsilon\right)^{s-D+1}},\nonumber\\
\end{eqnarray}
\end{subequations}%
where $p^0$ is the temporal component of $p$, $\bm{p}$ is the vector 
of spatial components of $p$, and we
define the measure of the phase-space integration as
\begin{equation}
\int_k\textrm{PS}
\equiv\int\frac{d^Dk}{(2\pi)^D}2\pi\delta(k^2)\theta(k^0).
\end{equation}
In all of the calculations in this paper, we have arranged the parameter
integrations so that $p^0+|\bm{p}|>0$ and $p^0-|\bm{p}|>0$. In this case 
the first two denominator factors in
Eq.~(\ref{eq:k1-phase-int-table-a}) can be written as
$(p^2)^{\frac{D}{2}-1}$, and the first two denominator factors in
Eq.~(\ref{eq:k1-phase-int-table-b}) can be written as
$(p^2)^{\frac{D}{2}}$.
The derivation of the formulas in Eq.~(\ref{eq:k1-phase-int-table}) is given in
Appendix~\ref{app:phase-space-int-formula}.

\section{Analyses of the diagrams\label{sec:diagrammatic-analyses}}

In this section, we outline our calculations for the various classes of 
diagrams. 

\subsection{Method of calculation}

Our general method of calculation is as follows. 

For the Abelian diagrams, we carry out the $k_1$ integration first,
holding $k_2$ fixed. We control UV divergences by imposing our standard
phase-space regulators and, in some cases, temporarily impose 
additional UV regulators in order to determine the IR or UV nature of the
poles in $\epsilon$. We first combine the denominators by using
Feynman parameters that run from 0 to 1 to combine denominators that
have a common term involving $k_1$ and by using Feynman parameters that
run from 0 to $\infty$ to combine the remaining denominators. We then
carry out the integration over $k_1$, using standard formulas of
dimensional regularization if $k_1$ is virtual and using
Eq.~(\ref{eq:k1-phase-int-table}) if $k_1$ is real. We identify the
singular regions of the parameter integrals and isolate these as
integrations over a single parameter, either by rescaling the parameters
that run from 0 to $\infty$ or, in a few cases, by using sector
decomposition. We then expand the integrals around the singularities by
making use of formulas such as
\begin{equation}
\frac{1}{x^{1+a\epsilon}}
=
-
\frac{1}{a\epsilon}\delta(x)
+\left[\frac{1}{x}\right]_+
-a\epsilon
\left[\frac{\log x}{x}\right]_+
+O(\epsilon^2),
\label{plus-expansion}
\end{equation}
which applies when the domain of integration is $0\le x\le 1$.
Here, the plus distribution $[g(x)]_+$ is defined by
\begin{equation}
\int_0^1 dx\,f(x)[g(x)]_+ \equiv \int_0^1 dx[f(x)-f(0)]g(x),
\end{equation}
for a regular function $f(x)$ in $x\in[0,1]$.
However, we avoid expanding 
factors involving $k_2$ in powers of $\epsilon$, as the complete $\epsilon$ 
dependence of these factors is needed to find the coefficient of the 
soft pole from the $k_2$ integration.

We find for the Abelian diagrams that the $k_1$ integrations for the sum
over a class plus Hermitian conjugate are IR finite when $k_2$ is
fixed.\footnote{This is true for a part of the contribution of
the diagrams $\mathcal{B}_i$, while for the remainder, the $k_2$
integration with $k_1$ fixed is IR finite.} In some cases, there
are UV poles from the $k_1$ integration. The IR finiteness of the $k_1$
integration when $k_2$ is fixed is expected on general principles
because the sum over cuts in a class is sufficient to effect a unitarity
cancellation of singularities that occur when $k_1$ is soft relative to
$k_2$ or $k_1$ is collinear to $\ell$.

Some of the contributions from the $k_1$ integration remain finite as
$k_2$ goes to zero, and these can be considered to be SDCs that multiply
possible one-loop IR poles from the $k_2$ integration. In the cases of
diagrams $\mathcal{C}$ and $\mathcal{E}$, there are contributions that
become singular as $k_2$ goes to zero, and these yield two-loop
IR-divergent contributions to the LDME.

Once we have evaluated the $k_1$ integration, 
we can carry out the $k_2$ integration straightforwardly by
combining denominators with Feynman parameters and using
Eq.~(\ref{eq:k1-phase-int-table}) to carry out the phase-space
integration. The remaining parameter integrals are easily reduced,
by using expansions of the type in Eq.~(\ref{plus-expansion}), to
elementary integrals.

In the case of the non-Abelian diagrams, the $k_1$ integration for the
sum over the class plus Hermitian conjugate 
is no longer IR finite when $k_2$ is held fixed. In
this case, as we will explain, the unitarity cancellation requires that
one carry out the $k_2$ integration, as well as the $k_1$ integration.
Therefore, for the non-Abelian diagrams, we combine denominators
involving both $k_1$ and $k_2$ using Feynman parameters that range from
0 to 1 or from 0 to $\infty$, as outlined above. We then carry out the
$k_1$ and $k_2$ integrations, isolate the IR singularities in 
single-parameter 
integrals by using rescaling or sector decomposition, and
evaluate the IR poles by using Eq.~(\ref{plus-expansion}).

In most cases, we have checked the results from direct integration of
the parameter integrals by using the Mellin-Barnes representation to
decompose a denominator so as to obtain parameter integrations that can
be carried out in terms of beta functions. We then evaluate the
Mellin-Barnes integrations by isolating poles in $\epsilon$, using the
methods of Tausk \cite{Tausk:1999vh} or Smirnov \cite{Smirnov:2004ym}, and
computing the remaining finite integrals as a sum over residues in the
complex plane of Mellin-Barnes integration variable.

\subsection{The one-loop IR contribution to the LDME\label{app:1-loop}}

The color factor of the $\mathcal{O}$ diagram 
(Fig.~\ref{figure:O-P1P2}) is given by
\begin{equation}
\label{eq:color-O}
C_{\mathcal{O}^{P_jP_k}}
=\frac{\textrm{Tr}(T_a T_b)}{\sqrt{N_c}} 
\frac{\textrm{Tr}(T_a T_b)}{\sqrt{N_c}}
=\frac{\delta_{aa}}{4N_c} = \frac{N_c^2-1}{4N_c}.
\end{equation}

The expression for the diagram $\mathcal{O}$, with our standard
phase-space regulators in Eq.~(\ref{standard-regulators}), 
is given by
\begin{eqnarray}
\mathcal{O}^{P_1P_2}
&=&
\frac{4g_s^2a}{(P_1^2)^{-1}}
\times
\mu^{2\epsilon}
\int_{k_2}\textrm{PS}
\frac{\Lambda^2}
{\left(2P_1\cdot k_2\right)
\left(2P_2\cdot k_2\right)
\left(2\ell\cdot k_2+\Lambda^2\right)}.
\end{eqnarray}
Combining the denominators by using Feynman parameters and carrying out
the $k_2$ phase-space integration by using Eq.~(\ref{eq:k1-phase-int-table}),
we find that
\begin{eqnarray}
&&
\mu^{2\epsilon}
\int_{k_2} \textrm{PS}\,
\frac{\Lambda^2}
{(2P_1\cdot k_2)
(2P_2\cdot k_2)
\left(2\ell\cdot k_2+\Lambda^2\right)}
\nonumber \\
\,
&=&
\mu^{2\epsilon}
\Lambda^2
\frac{\Gamma(3)}{\Gamma(1)}
\int_0^\infty d\lambda_2
\int_0^\infty d\lambda_3
\int_{k_2} \textrm{PS}\,
\frac{1}{\left[2k_2\cdot(\ell+\lambda_2P_1+\lambda_3 P_2)+\Lambda^2\right]^{3}}
\nonumber \\
&=&
-
\frac{\left(\frac{\tilde\mu^2P_1^2}{\Lambda^4}\right)^{\epsilon}}
{(4\pi)^{2}P_1^2}
\frac{e^{\epsilon\gamma_{{}_{\textrm{E}}}}
\Gamma(1+2\epsilon)\Gamma(1-2\epsilon)\Gamma(1+\epsilon)}
{2\epsilon_\textrm{IR}}
\int_0^\infty d\xi
\frac{1}{A^{1+\epsilon}B^{-2\epsilon}}
\nonumber \\
&=&
-
\frac{1}
{(4\pi)^{2}P_1^2}
\frac{1}{2\epsilon_\textrm{IR}}
\frac{\log\left(a+\sqrt{a^2-1}\right)}{\sqrt{a^2-1}}
+O(\epsilon^0),
\end{eqnarray}
where $\tilde{\mu}^2\equiv 4\pi\mu^2 
e^{-\gamma_{{}_{\textrm{E}}}}$, $\gamma_{{}_{\textrm{E}}}$ 
is the Euler-Mascheroni constant, and we have made the
change of variables $\lambda_3\to\xi\lambda_2$,
carried out $\lambda_2$ integration in terms of the beta function,
introduced the definitions
\begin{eqnarray}
\label{eq:def-of-A-B}
A&\equiv& \xi^2+2a\xi+1,
\nonumber \\
B&\equiv&1+\frac{d}{c}\xi,
\end{eqnarray}
and carried out the $\xi$ integration by
using Eq.~(\ref{eq:xi-integrals}).

Then, taking into account the color factor in 
Eq.~(\ref{eq:color-O}) and summing over the gluon attachments, 
we obtain
\begin{eqnarray}
\label{eq:one-loop-LDME}
\mathcal{O}
&\equiv&
\sum_{j=1}^2\sum_{k=1}^2
C_{\mathcal{O}^{P_jP_k}}
\mathcal{O}^{P_jP_k}
\nonumber \\
&=&
\left[
\frac{\alpha_s}{4\pi}
\frac{N_c^2-1}{4N_c}
\frac{4}{\epsilon_\textrm{IR}}
\left(
1
-
\frac{a\log\left(a+\sqrt{a^2-1}\right)}{\sqrt{a^2-1}}
\right)
+O(\epsilon^0)
\right]_{\textrm{IR}}.
\end{eqnarray}
If we expand Eq.~(\ref{eq:one-loop-LDME}) to leading order in
$\bm{q}^2$, then it agrees with the expression in Eq.~(38) of
Ref.~\cite{Nayak:2005rt}, aside from the color factor, which has been
dropped in Ref.~\cite{Nayak:2005rt}.

\subsection{$\mathcal{A}_i$ diagrams}

We do not need to evaluate the momentum integrations for $\mathcal{A}_i$ 
diagrams, as the color factors vanish:
\begin{eqnarray}
\label{eq:color-A}
C_{\mathcal{A}_i^{P_jP_k}}
&=&
\frac{\textrm{Tr}\left[T^a T^b\right]}{\sqrt{N_c}}
\frac{\textrm{Tr}\left[T^c T^d\right]}{\sqrt{N_c}}
f_{abe}f_{ced}
=
\frac{f_{aae}f_{cec}}{4N_c}=0.
\end{eqnarray}

\subsection{$\mathcal{B}_i$ diagrams}
The color factors of the $\mathcal{B}_i$ diagrams are given by
\begin{eqnarray}
\label{eq:color-B}
C_{\mathcal{B}_i^{P_jP_k}}
&=&
\frac{\textrm{Tr}\left[T^a T^b\right]}{\sqrt{N_c}}
\frac{\textrm{Tr}\left[T^c T^d\right]}{\sqrt{N_c}}
f_{cbe}f_{dae}
=\frac{f_{cae}f_{cae}}{4N_c}=\frac{N_c(N_c^2-1)}{4N_c}.
\end{eqnarray}
The Feynman amplitudes for the 
$\mathcal{B}_i^{P_1P_2}$ diagrams
with our standard phase-space regulators are
given by
\begin{eqnarray}
\mathcal{B}_1^{P_1P_2}
&=&
\frac{16g_s^4 \Lambda^2 cd}{(P_1^2)^{-2}}
\mu^{4\epsilon}
\int_{k_2} \textrm{PS}\,
\frac{1}{(2P_2\cdot k_2)(2\ell\cdot k_2)}
\int_{k_1} \textrm{PS}\,
\frac{1}
{(2P_1\cdot k_1)(2\ell\cdot k_1)
\left[2\ell\cdot (k_1+k_2)+\Lambda^2\right]},
\nonumber \\
\mathcal{B}_2^{P_1P_2}
&=&
\frac{16ig_s^4 \Lambda^2cd}{(P_1^2)^{-2}}
\mu^{4\epsilon}
\int_{k_2} \textrm{PS}\,
\frac{1}{(2P_2\cdot k_2)
(2\ell\cdot k_2+\Lambda^2)}
\nonumber \\
&&
\times
\int\frac{d^Dk_1}{(2\pi)^D}
\frac{1}{
(2P_1\cdot k_1+i\varepsilon)
(-2\ell\cdot k_1+i\varepsilon)
(k_1^2+i\varepsilon)
\left[
2\ell\cdot (-k_1+k_2)+i\varepsilon
\right]
},
\nonumber \\
\mathcal{B}_3^{P_1P_2}
&=&
-
\frac{16ig_s^4 \Lambda^2cd}{(P_1^2)^{-2}}
\mu^{4\epsilon}
\int_{k_1} \textrm{PS}\,
\frac{1}{\left(2P_1\cdot k_1\right)
\left(2\ell\cdot k_1+\Lambda^2\right)}
\nonumber \\
&&
\times
\int\frac{d^Dk_2}{(2\pi)^D}\,
\frac{1}{
\left(2P_2\cdot k_2-i\varepsilon\right)
\left(-2\ell\cdot k_2-i\varepsilon\right)
\left(k_2^2-i\varepsilon\right)
\left[2\ell\cdot(k_1-k_2)-i\varepsilon\right]},
\end{eqnarray}
where $c$ and $d$ are defined in Eq.~(\ref{eq:def-acd}) and
we have suppressed unnecessary $i\varepsilon$'s in the 
denominators.

\subsubsection{IR finiteness of the $k_1$ or $k_2$ integration}
\label{sec:IR-finite-B}%

We wish to establish that, for a certain combination of the
$\mathcal{B}_i$, the integration over $k_1$ with $k_2$ fixed is IR
finite, while for the remaining part of the $\mathcal{B}_i$ the
integration over $k_2$ with $k_1$ fixed is IR finite. For this purpose,
we temporarily impose additional regulators to control all of the UV
divergences.
 
It can be seen by power counting that our standard UV regulators render
the integrations over $\ell\cdot k_1$ and $\ell\cdot k_2$ UV finite.
However, there are still potential UV divergences that could arise 
from the integrations over the other components of $k_1$ and
$k_2$. In order to control these divergences, we introduce the following
additional UV-regulator factor
\begin{eqnarray}
\label{eq:UV-regulator-Li}
\frac{\Lambda\hspace{-0.1em}'^{\hspace{0.07em}4}}
{(2P_1\cdot k_1+\Lambda\hspace{-0.1em}'^{\hspace{0.07em}2})(2P_2\cdot k_2+\Lambda\hspace{-0.1em}'^{\hspace{0.07em}2})},
\end{eqnarray}
where the cutoff $\Lambda'$ is of order $m_Q$. We denote the 
$\mathcal{B}_i$ into which we have inserted this UV-regulator factor 
by $\mathcal{B}_i\big|_\textrm{Reg}$. 

The $\mathcal{B}_i$ contain rapidity (collinear) divergences that are
associated with the vanishing of the denominators of Wilson-line
propagators.\footnote{With appropriate deformations of the $k_1$
integration contours into the complex plane, these can be considered to
be collinear-to-$\ell$ divergences. See, for example, p.~291 of
Ref.~\cite{Collins:2011zzd}.\\{}} We expect these divergences to
cancel, because of unitarity, in the sum over cuts of the
$\mathcal{B}_i$.\footnote{In general, one must sum over all cuts of a
diagram in order to effect a unitarity cancellation. However, in the
case of divergences that arise from the vanishing of specific propagator
denominators, it is necessary only to sum over all cuts of the singular
denominators. Other propagators are relatively far off shell and can be
contracted to a point for the purpose of evaluating divergences.} We can
separate the parts of $\mathcal{B}_1$ that contribute to the
cancellations of the rapidity divergences in  $\mathcal{B}_2$ and
$\mathcal{B}_3$ by making use of the following partial-fraction identity
in $\mathcal{B}_1^{P_1P_2}\big|_\textrm{Reg}$:
\begin{eqnarray}
\frac{1}{(2\ell\cdot k_1)(2\ell\cdot k_2)}
&=&
\frac{1}{\left[2\ell\cdot(-k_1+k_2)-i\varepsilon\right]}
\left(
\frac{1}{2\ell\cdot k_1}-\frac{1}{2\ell\cdot k_2}
\right).
\end{eqnarray}
Then we can write
\begin{equation}
\mathcal{B}_{1}^{P_1P_2}\big|_\textrm{Reg}
=\mathcal{B}_{1a}^{P_1P_2}\big|_\textrm{Reg}
+\mathcal{B}_{1b}^{P_1P_2}\big|_\textrm{Reg},
\end{equation}
where
\begin{eqnarray}
\mathcal{B}_{1a}^{P_1P_2}\big|_\textrm{Reg}
&=&
\frac{16g_s^4 \Lambda^2\Lambda\hspace{-0.1em}'^{\hspace{0.07em}4} cd}{(P_1^2)^{-2}}
\mu^{2\epsilon}
\int_{k_2} \textrm{PS}\,
\frac{1}
{(2P_2\cdot k_2)(2P_2\cdot k_2+\Lambda\hspace{-0.1em}'^{\hspace{0.07em}2})}
\nonumber \\
&&
\times
\mu^{2\epsilon}
\int_{k_1} \textrm{PS}
\frac{1}
{(2P_1\cdot k_1)(2P_1\cdot k_1+\Lambda\hspace{-0.1em}'^{\hspace{0.07em}2})(2\ell\cdot k_1)
\left[2\ell\cdot(-k_1+k_2)-i\varepsilon\right]
\left[2\ell\cdot (k_1+k_2)+\Lambda^2\right]},
\nonumber \\
\mathcal{B}_{1b}^{P_1P_2}\big|_\textrm{Reg}
&=&
\frac{16g_s^4 \Lambda^2\Lambda\hspace{-0.1em}'^{\hspace{0.07em}4} cd}{(P_1^2)^{-2}}
\mu^{2\epsilon}
\int_{k_1} \textrm{PS}
\frac{1}
{(2P_1\cdot k_1)(2P_1\cdot k_1+\Lambda\hspace{-0.1em}'^{\hspace{0.07em}2})
}
\nonumber \\
&&
\times
\mu^{2\epsilon}
\int_{k_2} \textrm{PS}\,
\frac{1}
{(2P_2\cdot k_2)(2P_2\cdot k_2+\Lambda\hspace{-0.1em}'^{\hspace{0.07em}2})(2\ell\cdot k_2)
\left[2\ell\cdot(k_1-k_2)+i\varepsilon\right]
\left[2\ell\cdot (k_1+k_2)+\Lambda^2\right]}.
\nonumber \\
\label{eq:B1a1b-UV}
\end{eqnarray}
In Appendix~\ref{app:IR-finiteness-of-B}, we show that the $k_1$
integration with $k_2$ fixed in
$\mathcal{B}_{1a}^{P_1P_2}\big|_\textrm{Reg}
+\mathcal{B}_2^{P_1P_2}\big|_\textrm{Reg}$ is IR finite. We also show
that the integrand of $\mathcal{B}_{1a}^{P_1P_2}\big|_\textrm{Reg}
+\mathcal{B}_2^{P_1P_2}\big|_\textrm{Reg}$ is the Hermitian conjugate of
the integrand of $\mathcal{B}_{1b}^{P_2P_1}\big|_\textrm{Reg}
+\mathcal{B}_3^{P_2P_1}\big|_\textrm{Reg}$ with $k_1\leftrightarrow
k_2$. It then follows that the $k_2$ integration with $k_1$ fixed in
$\mathcal{B}_{1b}^{P_1P_2}\big|_\textrm{Reg}
+\mathcal{B}_3^{P_1P_2}\big|_\textrm{Reg}$ is also IR finite. We note,
for use below, that it also follows that  $\mathcal{B}_{1b}^{P_2P_1}
+\mathcal{B}_3^{P_2P_1}$ is the Hermitian conjugate of
$\mathcal{B}_{1a}^{P_1P_2} +\mathcal{B}_2^{P_1P_2}$.

\subsubsection{Computation of the diagrams}

Having established that $k_1$ integration with $k_2$ fixed in
$\mathcal{B}_{1a}^{P_1P_2}\big|_\textrm{Reg}
+\mathcal{B}_2^{P_1P_2}\big|_\textrm{Reg}$ is IR finite and that the
$k_2$ integration with $k_1$ fixed in
$\mathcal{B}_{1b}^{P_1P_2}\big|_\textrm{Reg}
+\mathcal{B}_3^{P_1P_2}\big|_\textrm{Reg}$ is IR finite, we can
remove the extra UV-cutoff factor (\ref{eq:UV-regulator-Li}) without
introducing any ambiguities in dimensional regularization.

Carrying out the $k_1$ integrations, we obtain
\begin{eqnarray}
\label{Eq:B1a-B2-p1p2-before-y}
\mathcal{B}_{1a}^{P_1P_2}
&=&
-
\frac{16g_s^4 \Lambda^2 cd}{(4\pi)^2(P_1^2)^{-1}(2c)^{1-2\epsilon}}
\left(\tilde\mu^2P_1^2\right)^{\epsilon}
e^{\epsilon\gamma_{{}_{\textrm{E}}}}\Gamma(1+2\epsilon)
\Gamma(\epsilon)\Gamma(1-2\epsilon)
\nonumber \\
&&
\times
\mu^{2\epsilon}
\int_{k_2} \textrm{PS}\,
\frac{1}
{(2P_2\cdot k_2)(2\ell\cdot k_2+\Lambda^2)}
\nonumber \\
&&
\times
\int_0^1 dy
\left\{
\frac{e^{-i\pi(1+2\epsilon)}}
{\left[
y(2\ell\cdot k_2)\right]^{1+2\epsilon}}
-
\frac{1}
{\left[
(1-y)(2\ell\cdot k_2+\Lambda^2)
-y(2\ell\cdot k_2)+i\varepsilon\right]^{1+2\epsilon}}
\right\},
\nonumber \\
\mathcal{B}_2^{P_1P_2}
&=&
-
\frac{16g_s^4 \Lambda^2 cd}{(4\pi)^2(P_1^2)^{-1}(2c)^{1-2\epsilon}}
\left(\tilde\mu^2P_1^2\right)^{\epsilon}
e^{\epsilon\gamma_{{}_{\textrm{E}}}}\Gamma(1+2\epsilon)\Gamma(\epsilon)\Gamma(1-2\epsilon)
\nonumber \\
&&
\times
\mu^{2\epsilon}
\int_{k_2} \textrm{PS}\,
\frac{1}{(2P_2\cdot k_2)
(2\ell\cdot k_2+\Lambda^2)}
\int_0^1 dy
\frac{1}{
\left[y(2\ell\cdot k_2)\right]^{1+2\epsilon}}.
\end{eqnarray}
Then, combining $\mathcal{B}_{1a}^{P_1P_2}$
and $\mathcal{B}_2^{P_1P_2}$ in Eq.~(\ref{Eq:B1a-B2-p1p2-before-y}) and 
carrying out the $y$ integration, we find that
\begin{eqnarray}
\mathcal{B}_{1a}^{P_1P_2}
+
\mathcal{B}_2^{P_1P_2}
&=&
\frac{16g_s^4 \Lambda^2 cd}{(4\pi)^2(P_1^2)^{-1}(2c)^{1-2\epsilon}}
\left(\tilde\mu^2P_1^2\right)^{\epsilon}
\frac{e^{\epsilon\gamma_{{}_{\textrm{E}}}}\Gamma(1+2\epsilon)
\Gamma(\epsilon)\Gamma(1-2\epsilon)}{2\epsilon}
\nonumber \\
&&
\times
\mu^{2\epsilon}
\int_{k_2} \textrm{PS}\,
\frac{1}
{(2P_2\cdot k_2)(2\ell\cdot k_2+\Lambda^2)}
\nonumber \\
&&
\times
\left\{
\frac{1-e^{-2i\pi\epsilon}}
{(2\ell\cdot k_2)^{1+2\epsilon}}
+
\frac{1}{(4\ell\cdot k_2+\Lambda^2)}
\left[
\frac{e^{-2i\pi\epsilon}}{(2\ell\cdot k_2)^{2\epsilon}}
-\frac{1}{(2\ell\cdot k_2+\Lambda^2)^{2\epsilon}}
\right]
\right\}.
\end{eqnarray}
Carrying out the $k_2$ integrations and expanding the resulting
expression through $O(\epsilon^{-1})$, we have
\begin{eqnarray}
\mathcal{B}_{1a}^{P_1P_2}
+
\mathcal{B}_2^{P_1P_2}
&=&
\frac{4g_s^4}{(4\pi)^4}
\left(\frac{\tilde\mu^{2}P_1^2}{\Lambda^4}\right)^{2\epsilon}
(4cd)^{2\epsilon}
\frac{e^{2\epsilon\gamma_{{}_{\textrm{E}}}}\Gamma(1+2\epsilon)
\Gamma^2(\epsilon)\Gamma(1-2\epsilon)
\Gamma(1+4\epsilon)
\Gamma(-4\epsilon)
\left(1-e^{-2i\pi\epsilon}\right)
}{2\epsilon}
\nonumber \\
&&
+
\frac{4g_s^4}
{(4\pi)^4}
\left(\frac{\tilde\mu^2P_1^2}{\Lambda^4}\right)^{2\epsilon}
(4cd)^{2\epsilon}
\frac{e^{2\epsilon\gamma_{{}_{\textrm{E}}}}\Gamma(1+2\epsilon)
\Gamma^2(\epsilon)\Gamma(1-2\epsilon)
\Gamma(1+4\epsilon)}{4\epsilon}
\nonumber \\
&&
\times
\bigg\{
e^{-2i\pi\epsilon}\Gamma(1-4\epsilon)
\int_0^1 dx
\frac{1}{\left(\frac{1+x}{2}\right)^{1+4\epsilon}}
-
\frac{\Gamma(1-2\epsilon)}
{\Gamma(1+2\epsilon)}
\int_0^1 dx
\frac{x^{2\epsilon}}
{\left(\frac{1+x}{2}\right)^{1+4\epsilon}}\bigg\}
\nonumber \\
&=&
-
\frac{g_s^4}
{(4\pi)^4}
\left(\frac{\tilde\mu^2P_1^2}{\Lambda^4}\right)^{2\epsilon}
\bigg\{
\frac{i\pi}{\epsilon^3}
+\frac{\frac{2\pi^2}{3}+2i\pi \log(c d)
+8i\pi\log 2}{\epsilon^2}
\nonumber \\
&&\quad
\quad\quad\quad\quad\quad\quad\quad\quad
+
\frac{\frac{4\pi^2}{3}\log(4cd)+8\zeta(3)
+2i\pi\log^2(16cd)
+\frac{17i\pi^3}{6}}{\epsilon}
\bigg\},
\end{eqnarray}
where $\zeta(z)\equiv\sum_{n=1}^\infty n^{-z}$ is the Riemann zeta function.
It is easily seen, from the Feynman rules, that 
\begin{eqnarray}
\mathcal{B}_{1a}^{P_2P_1}
+
\mathcal{B}_2^{P_2P_1}
&=&
\left(
\mathcal{B}_{1a}^{P_1P_2}
+
\mathcal{B}_2^{P_1P_2}
\right)
\bigg|_{d\leftrightarrow c},
\nonumber \\
\mathcal{B}_{1a}^{P_1P_1}
+
\mathcal{B}_2^{P_1P_1}
&=&
-
\left(
\mathcal{B}_{1a}^{P_1P_2}
+
\mathcal{B}_2^{P_1P_2}
\right)
\bigg|_{d\to c},
\nonumber \\
\mathcal{B}_{1a}^{P_2P_2}
+
\mathcal{B}_2^{P_2P_2}
&=&
-
\left(
\mathcal{B}_{1a}^{P_1P_2}
+
\mathcal{B}_2^{P_1P_2}
\right)
\bigg|_{c\to d}.
\end{eqnarray}
Then, we have
\begin{eqnarray}
\sum_{j=1}^2\sum_{k=1}^2
\left(
\mathcal{B}_{1a}^{P_jP_k}
+
\mathcal{B}_2^{P_jP_k}
\right)
=
\frac{g_s^4\left(\frac{\tilde\mu^2P_1^2}{\Lambda^4}\right)^{2\epsilon}}
{(4\pi)^4}\frac{4i\pi\log^2\frac{c}{d}}{\epsilon}
+O(\epsilon^0).
\end{eqnarray}
Since, as we have noted at the end of Sec.~\ref{sec:IR-finite-B},
$\mathcal{B}_{1b}^{P_2P_1}+ \mathcal{B}_3^{P_2P_1}$ is the Hermitian
conjugate of $\mathcal{B}_{1a}^{P_1P_2}
+
\mathcal{B}_2^{P_1P_2}$,
we find that
\begin{eqnarray}
\label{eq:B-final-sum-before-real}
\sum_{j=1}^2\sum_{k=1}^2
\left(
\mathcal{B}_{1a}^{P_jP_k}
+
\mathcal{B}_{1b}^{P_jP_k}
+
\mathcal{B}_2^{P_jP_k}
+
\mathcal{B}_3^{P_jP_k}
\right)
=
O(\epsilon^0).
\end{eqnarray}
That is, there are no poles in the sum over the $\mathcal{B}_i$
diagrams.

\subsection{$\mathcal{C}_i$ diagrams}
The color factors of the $\mathcal{C}_i$ diagrams
are given by
\label{eq:color-C}
\begin{eqnarray}
\label{eq:color-fac-Ci}
C_{\mathcal{C}_i^{P_1P_2}}
=
C_{\mathcal{C}_i^{P_1P_1}}
&=&
\frac{\textrm{Tr}\left[T_a T_b T_c\right]}{\sqrt{N_c}}
\frac{\textrm{Tr}\left[T_a T_d\right]}{\sqrt{N_c}}
f_{bcd}
=
\frac{\left(d_{abc}+if_{abc}\right)f_{abc}}{8N_c}
=+\frac{iN_c(N_c^2-1)}{8N_c},
\nonumber \\
C_{\mathcal{C}_i^{P_2P_1}}
=
C_{\mathcal{C}_i^{P_2P_2}}
&=&
\frac{\textrm{Tr}\left[T_a T_b T_c\right]}{\sqrt{N_c}}
\frac{\textrm{Tr}\left[T_d T_c\right]}{\sqrt{N_c}}
f_{adb}
=
-
\frac{\left(d_{abc}+if_{abc}\right)f_{abc}}{8N_c}
=-\frac{iN_c(N_c^2-1)}{8N_c}.
\end{eqnarray}
The expressions for the $\mathcal{C}_i$ diagrams, with our standard
phase-space regulators 
in Eq.~(\ref{standard-regulators}), are given by 
\begin{eqnarray}
\label{eq:Ci-P1P2-organized}
\mathcal{C}_1^{P_1P_2}
&=&
-
\frac{16i
g_s^4
\mu^{4\epsilon}
\Lambda^2
ac}
{(P_1^2)^{-2}}
\int_{k_2} \textrm{PS}\,
\frac{1}
{(2P_1\cdot k_2)
(2P_2\cdot k_2)
}
\nonumber \\
&&
\times
\int_{k_1} \textrm{PS}
\frac{1}{\left[2P_1\cdot (k_1+k_2)\right]
\left(2\ell\cdot k_1\right)\left[2\ell\cdot (k_1+k_2)+\Lambda^2\right]},
\nonumber \\
\mathcal{C}_2^{P_1P_2}
&=&
\frac{16g_s^4
\mu^{4\epsilon}\Lambda^2
ac}
{(P_1^2)^{-2}}
\int_{k_2} \textrm{PS}\,
\frac{1}
{(2P_1\cdot k_2)
(2P_2\cdot k_2)
(2\ell\cdot k_2+\Lambda^2)}
\nonumber \\
&&
\times
\int\frac{d^Dk_1}{(2\pi)^D}
\frac{1}
{\left[2P_1\cdot (k_1+k_2)+i\varepsilon\right]
\left[2\ell\cdot (-k_1)+i\varepsilon\right]
\left(k_1^2+i\varepsilon\right)},
\end{eqnarray}
where $a$ and $c$ are defined in Eq.~(\ref{eq:def-acd}).

\subsubsection{IR finiteness of the $k_1$ integration
in $\mathcal{C}_1$ and $\mathcal{C}_2$}

It can be seen by power counting that our standard UV regulators render
the integration over $\ell\cdot k_1$ UV finite. However, there are still
potential UV divergences that could arise from the integrations over
the other components of $k_1$. In order to control these divergences, we
introduce the following additional UV-regulator factor
\begin{equation}
\label{eq:Ci-addtional-UV-reg}
\frac{\Lambda\hspace{-0.1em}'^{\hspace{0.07em}2}}
{2P_1\cdot (k_1+k_2) + \Lambda\hspace{-0.1em}'^{\hspace{0.07em}2}}.
\end{equation} 
We denote the $\mathcal{C}_{i}$ into which we have inserted 
this UV-regulator factor by 
$\mathcal{C}_{i}\big|_\textrm{Reg}$.

The $k_1$ integrations of the individual $\mathcal{C}_i$ diagrams
contain rapidity (collinear) divergences that arise when the
denominators of the Wilson-line propagators vanish. We expect these
divergences to cancel, by unitarity, in the sum over cuts in the
$\mathcal{C}_i$.

In Appendix~\ref{app:IR-finiteness-C12}, we have carried out the 
$k_1$ integrations of $\mathcal{C}_1^{P_1P_2}\big|_\textrm{Reg}$
and $\mathcal{C}_2^{P_1P_2}\big|_\textrm{Reg}$. The results are
\begin{eqnarray}
\mathcal{C}_1^{P_1 P_2}+\mathcal{C}_2^{P_1 P_2}\big|_\textrm{Reg}
&=&
\frac{8i
g_s^4
\mu^{2\epsilon}
\Lambda^2
a}
{(4\pi)^{2}(P_1^2)^{-1}}
\int_{k_2} \textrm{PS}\,
\frac{\frac{1}{\epsilon_\textrm{IR}}
\log\left(\frac{2P_1\cdot k_2+\Lambda\hspace{-0.1em}'^{\hspace{0.07em}2}}{2P_1\cdot k_2}\right)
+O(\epsilon^0)}
{(2P_1\cdot k_2)
(2P_2\cdot k_2)
(2\ell\cdot k_2+\Lambda^2)
}
\nonumber\\
&&
-
\frac{8ig_s^4
\mu^{2\epsilon}\Lambda^2
a}
{(4\pi)^{2}(P_1^2)^{-1}}
\int_{k_2} \textrm{PS}\,
\frac{\frac{1}{\epsilon_\textrm{IR}}
\log\left(\frac{2P_1\cdot k_2+\Lambda\hspace{-0.1em}'^{\hspace{0.07em}2}}{2P_1\cdot k_2}\right)
+O(\epsilon^0)}
{(2P_1\cdot k_2)
(2P_2\cdot k_2)
(2\ell\cdot k_2+\Lambda^2)},
\end{eqnarray}
where the first term comes from $\mathcal{C}_1^{P_1
P_2}\big|_\textrm{Reg}$ and the second term comes from 
$\mathcal{C}_2^{P_1 P_2}\big|_\textrm{Reg}$. We see the explicit
cancellation of the poles in $\epsilon$ that arise from the rapidity
(collinear) divergences.

\subsubsection{Computation of $\mathcal{C}_1$ and $\mathcal{C}_2$}

Having established that the $k_1$ integration with $k_2$ fixed in
$\mathcal{C}_1^{P_1 P_2}\big|_\textrm{Reg} +\mathcal{C}_2^{P_1
P_2}\big|_\textrm{Reg}$ plus Hermitian conjugate is IR finite, we can
remove the extra UV-regulator factor in
Eq.~(\ref{eq:Ci-addtional-UV-reg}) without introducing any ambiguities
in dimensional regularization. The expressions for the diagrams are
given by
\begin{eqnarray}
\label{eq:C1C2-p1p2-before-int}
\mathcal{C}_1^{P_1P_2}
&=&
-
\frac{16i
g_s^4
\Lambda^2
ac}
{(P_1^2)^{-2}}
\mu^{2\epsilon}
\int_{k_2} \textrm{PS}\,
\frac{1}
{(2P_1\cdot k_2)
(2P_2\cdot k_2)
}
\times C_1^{P_1P_2}
,
\nonumber \\
\mathcal{C}_2^{P_1P_2}
&=&
\frac{16g_s^4\Lambda^2
ac}
{(P_1^2)^{-2}}
\mu^{2\epsilon}
\int_{k_2} \textrm{PS}\,
\frac{1}
{(2P_1\cdot k_2)
(2P_2\cdot k_2)
(2\ell\cdot k_2+\Lambda^2)}
\times C_2^{P_1P_2},
\end{eqnarray}
where the $k_1$ integrations of 
$\mathcal{C}_1^{P_1P_2}$ and $\mathcal{C}_2^{P_1P_2}$ 
are defined by
\begin{eqnarray}
\label{eq:C1C2-Lorentz-cov}
C_1^{P_1P_2}
&\equiv&
\mu^{2\epsilon}
\int_{k_1} \textrm{PS}\,
\frac{1}
{\left[2P_1\cdot (k_1+k_2)\right]
\left(2\ell\cdot k_1\right)
\left[2\ell\cdot (k_1+k_2)+\Lambda^2\right]},
\nonumber \\
C_2^{P_1P_2}
&\equiv&
\mu^{2\epsilon}
\int\frac{d^Dk_1}{(2\pi)^D}
\frac{1}
{\left[2P_1\cdot (k_1+k_2)+i\varepsilon\right]
\left[2\ell\cdot (-k_1)+i\varepsilon\right]
\left(k_1^2+i\varepsilon\right)}.
\end{eqnarray}
Introducing Feynman parameters and performing the $k_1$ integrations,
we obtain
\begin{eqnarray}
\label{C1C2-k1-int-final-before}
C_1^{P_1P_2}
&=&
-\frac{1}{2c}
\frac{\left(
\tilde\mu^2P_1^2
\right)^{\epsilon}}
{(4\pi)^{2}(P_1^2)}
\frac{e^{\epsilon\gamma_{{}_{\textrm{E}}}}\Gamma(1+2\epsilon)\Gamma(1-\epsilon)}
{\epsilon_\textrm{UV}}
\int_0^\infty d\lambda_1
\frac{1}
{\lambda_1^{1-\epsilon}
(\lambda_1+2c)^{-\epsilon}
\left(\lambda_1 K_1
+K_2\right)^{1+2\epsilon}},
\nonumber \\
C_2^{P_1P_2}
&=&
-\frac{i}{4c}
\frac{\left(\tilde\mu^2P_1^2\right)^\epsilon}
{(4\pi)^{2}(P_1^2)}
\frac{e^{\epsilon\gamma_{{}_{\textrm{E}}}}\Gamma(1+2\epsilon)\Gamma(1-\epsilon)}
{e^{-2i\pi\epsilon}\epsilon^2_\textrm{UV}}
\frac{1}{K_1^{2\epsilon}},
\end{eqnarray}
where we define the $k_2$-dependent denominators $K_1$ and $K_2$ as
\begin{eqnarray}
\label{eq:K1K2-def}
K_1
&\equiv&
2P_1\cdot k_2,
\nonumber \\
K_2
&\equiv&
2\ell\cdot k_2+\Lambda^2.
\end{eqnarray}

Then, carrying out the $\lambda_1$ integration in 
Eq.~(\ref{C1C2-k1-int-final-before}) by using 
the formula in Eq.~(\ref{eq:lam-1-int-result-K1-K2}) and inserting 
our results for $C_1^{P_1P_2}$ and $C_2^{P_1P_2}$ into 
Eq.~(\ref{eq:C1C2-p1p2-before-int}), we find that
\begin{eqnarray}
\label{eq:C1C2-p1p2-after-k1-int}%
\mathcal{C}_{1}^{P_1P_2}
+
\mathcal{C}_2^{P_1P_2}
&=&
\frac{8i
g_s^4 a
}
{(4\pi)^{2}(P_1^2)^{-1}}
\left[
-\frac{i\pi}{\epsilon_\textrm{UV}}
+\pi^2
-i\pi\log\left(
\frac{\tilde\mu^2P_1^2}{\Lambda^4}
\right)
+O(\epsilon)
\right]
\nonumber \\
&&
\times
\mu^{2\epsilon}
\int_{k_2} \textrm{PS}\,
\frac{(\Lambda^2)^{1+2\epsilon}}
{(2P_1\cdot k_2)^{2\epsilon} (2P_1\cdot k_2)
(2P_2\cdot k_2)
(2\ell\cdot k_2+\Lambda^2)}
\nonumber \\
&&
+
\frac{8i
g_s^4 a
}
{(4\pi)^{2}(P_1^2)^{-1}}
\left[
\frac{1}{2\epsilon_\textrm{UV}^2}
+\frac{3\pi^2}{8}
+\frac{\log\left[\frac{
\tilde\mu^2 P_1^2(2c)^2}{\Lambda^4}
\right]}
{2\epsilon_\textrm{UV}}
+\frac{\log^2\left[\frac{
\tilde\mu^2 P_1^2(2c)^2}{\Lambda^4}
\right]}{4}
+O(\epsilon)
\right]
\nonumber \\
&&
\times
\mu^{2\epsilon}
\int_{k_2} \textrm{PS}\,
\frac{(\Lambda^2)^{1+2\epsilon}}
{(2P_1\cdot k_2)
(2P_2\cdot k_2)
(2\ell\cdot k_2+\Lambda^2)^{1+2\epsilon}}.
\end{eqnarray}
In Eq.~(\ref{eq:C1C2-p1p2-after-k1-int}), the factor $(2P_1\cdot
k_2)^{2\epsilon}$ in the denominator of the first integrand arises from
the $k_1$ integration. It has an IR sensitivity, in that it becomes
singular when $k_2$ goes to zero. It affects the strength of the IR pole
that will appear in the $k_2$ integration. Consequently, all of the
contributions from the first set of brackets in
Eq.~(\ref{eq:C1C2-p1p2-after-k1-int}) should be regarded as IR in
nature, except for the UV pole.
However, this UV pole has an imaginary coefficient, and 
cancels when we add the Hermitian-conjugate contribution.
The factor
$(2\ell\cdot k_2+\Lambda^2)^{1+2\epsilon}$ in the denominator of the
second integrand in Eq.~(\ref{eq:C1C2-p1p2-after-k1-int}) also arises
from the $k_1$ integration. However, in this case there is no IR
sensitivity because this factor is finite as $k_2$ goes to zero.
Consequently, all of the contributions from the second set of brackets
in Eq.~(\ref{eq:C1C2-p1p2-after-k1-int}) are UV in nature.

Combining denominators in Eq.~(\ref{eq:C1C2-p1p2-after-k1-int}) by
using Feynman parameters and carrying out the $k_2$ phase-space
integration by using Eq.~(\ref{eq:k1-phase-int-table}), we find that
\begin{subequations}
\label{eq:Ci-k2-int-formula}
\begin{eqnarray}
&&
\mu^{2\epsilon}
\int_{k_2} \textrm{PS}\,
\frac{(\Lambda^2)^{1+2\epsilon}}
{(2P_1\cdot k_2)^{1+2\epsilon}
(2P_2\cdot k_2)
(2\ell\cdot k_2+\Lambda^2)}
\nonumber \\
\,
&=&
\mu^{2\epsilon}
(\Lambda^2)^{1+2\epsilon}
\frac{\Gamma(3+2\epsilon)}{\Gamma(1+2\epsilon)}
\int_0^\infty d\lambda_2
\int_0^\infty d\lambda_3
\int_{k_2} \textrm{PS}\,
\frac{\lambda_2^{2\epsilon}}
{
\left[2k_2\cdot(\ell+\lambda_2 P_1+\lambda_3P_2)+\Lambda^2\right]
^{3+2\epsilon}}
\nonumber \\
&=&
-\frac{1}{4}
\frac{(2c)^{4\epsilon}}{(4\pi)^{2}P_1^2}
\left(
\frac{\tilde\mu^2P_1^2}{\Lambda^{4}}
\right)^{\epsilon}
\frac{e^{\epsilon\gamma_{{}_{\textrm{E}}}}
\Gamma(1+4\epsilon)\Gamma(1-4\epsilon)\Gamma(1+3\epsilon)}
{\epsilon_\textrm{IR}\Gamma(1+2\epsilon)}
\int_0^\infty d\xi
\frac{1}
{A^{1+3\epsilon}B^{-4\epsilon}}
\nonumber \\
&=&
-\frac{1}{(4\pi)^2P_1^2}
\frac{1}{4\epsilon_\textrm{IR}}
\int_0^\infty d\xi
\frac{1}{A}+O(\epsilon^0),
\end{eqnarray}
and
\begin{eqnarray}
&&
\mu^{2\epsilon}
\int_{k_2} \textrm{PS}\,
\frac{(\Lambda^2)^{1+2\epsilon}}
{(2P_1\cdot k_2)
(2P_2\cdot k_2)
\left(2\ell\cdot k_2+\Lambda^2\right)^{1+2\epsilon}}
\nonumber \\
\,
&=&
\mu^{2\epsilon}
(\Lambda^2)^{1+2\epsilon}
\frac{\Gamma(3+2\epsilon)}{\Gamma(1+2\epsilon)}
\int_0^\infty d\lambda_2
\int_0^\infty d\lambda_3
\int_{k_2} \textrm{PS}\,
\frac{1}{\left[2k_2\cdot(\ell+\lambda_2P_1+\lambda_3 P_2)+\Lambda^2\right]^{3+2\epsilon}}
\nonumber \\
&=&
-\frac{1}{2}
\frac{(2c)^{2\epsilon}}{(4\pi)^2P_1^2}
\left(\frac{\tilde\mu^2P_1^2}{\Lambda^{4}}\right)^\epsilon
\frac{e^{\epsilon\gamma_{{}_{\textrm{E}}}}
\Gamma(1+4\epsilon)\Gamma(1+\epsilon)\Gamma(1-2\epsilon)}
{\epsilon_\textrm{IR}\Gamma(1+2\epsilon)}
\int_0^\infty d\xi
\frac{1}
{A^{1+\epsilon}
B^{-2\epsilon}}
\nonumber \\
&=&
-\frac{1}{(4\pi)^2P_1^2}
\frac{1}{2\epsilon_\textrm{IR}}
\int_0^\infty d\xi
\frac{1}{A}+O(\epsilon^0),
\end{eqnarray}
\end{subequations}
where we have made change of variables $\lambda_3\to \xi\lambda_2$,
carried out $\lambda_2$ integrations in terms of the beta function, and 
used the definitions of $A$ and $B$ that are given in
Eq.~(\ref{eq:def-of-A-B}).
It can be seen from Eq.~(\ref{eq:def-acd}) that $a\ge1$. Then, 
the $\xi$ integrations in Eq.~(\ref{eq:Ci-k2-int-formula}) can be 
carried out by making use of the formulas in
Eq.~(\ref{eq:xi-integrals}). The result 
is
\begin{eqnarray}
\label{eq:C12-k1-first-result-organized}
\mathcal{C}_1^{P_1 P_2}+\mathcal{C}_2^{P_1 P_2}
&=&
-
\left[\frac{ig_s^4}{(4\pi)^4}
\frac{2\pi^2}{\epsilon_\text{IR}}
\frac{a\log\left(a+\sqrt{a^2-1}\right)}{\sqrt{a^2-1}}
+O(\epsilon^0)\right]_\textrm{IR}
\nonumber \\
&&
+
\left[
\frac{ig_s^2}{(4\pi)^2}
\left(\frac{1}{\epsilon^2_\textrm{UV}}
+\frac{3\pi^2}{4}
+\frac{\log\left[\frac{
\tilde\mu^2 P_1^2(2c)^2}{\Lambda^4}\right]}
{\epsilon_\textrm{UV}}
+\frac{\log^2\left[\frac{
\tilde\mu^2 P_1^2(2c)^2}{\Lambda^4}
\right]}
{2}
\right)
+O(\epsilon)
\right]_\textrm{UV}
\nonumber \\
&&
\times
\left[
-\frac{g_s^2}{(4\pi)^2}
\frac{2}{\epsilon_\text{IR}}
\frac{a\log\left(a+\sqrt{a^2-1}\right)}{\sqrt{a^2-1}}
+O(\epsilon^0)\right]_\textrm{IR},
\end{eqnarray}
where we have omitted imaginary contributions, which cancel when we add
the Hermitian conjugate. The subscripts ``IR'' and ``UV'' are reminders
of the origins of the contributions. The factor labeled IR in the
first term is a two-loop contribution to the LDME. The factor
labeled IR in the second term is the one-loop contribution to the
LDME (absent its color factor), which is computed in
Sec.~\ref{app:1-loop}. The category UV includes all IR-finite
contributions, as well as UV-divergent contributions.

We can find all the other $\mathcal{C}_i^{P_j P_k}$ 
contributions from the relations
\begin{eqnarray}
\label{eq:C1C2-PjPk-symm}
\mathcal{C}_1^{P_2 P_1}+\mathcal{C}_2^{P_2 P_1}
&=&
-
\left(
\mathcal{C}_1^{P_1 P_2}+\mathcal{C}_2^{P_1 P_2}
\right)
\bigg|_{c\leftrightarrow d},
\nonumber \\
\mathcal{C}_1^{P_1 P_1}+\mathcal{C}_2^{P_1 P_1}
&=&
-
\left(
\mathcal{C}_1^{P_1 P_2}+\mathcal{C}_2^{P_1 P_2}
\right)
\bigg|_{d\to c,\,a\to1},
\nonumber \\
\mathcal{C}_1^{P_2 P_2}+\mathcal{C}_2^{P_2 P_2}
&=&
\left(
\mathcal{C}_1^{P_1 P_2}+\mathcal{C}_2^{P_1 P_2}
\right)
\bigg|_{c\to d,\,a\to1}.
\end{eqnarray}
Taking into account the color factors 
in Eq.~(\ref{eq:color-fac-Ci}), we obtain 
\begin{eqnarray}
\label{eq:C12-final-with-color}
&&
\textrm{Re}
\left(
\sum_{i=1}^2
\sum_{j=1}^2\sum_{k=1}^2
C_{\mathcal{C}_i^{P_jP_k}}\mathcal{C}_i^{P_j P_k}
\right)
\nonumber \\
&=&
-
\left[
\frac{g_s^4}{(4\pi)^4}
\frac{N_c(N_c^2-1)}{8N_c}
\frac{4\pi^2}{\epsilon_\text{IR}}
\left(
1-\frac{a\log\left(a+\sqrt{a^2-1}\right)}{\sqrt{a^2-1}}
\right)
+O(\epsilon^0)
\right]_\textrm{IR}
\nonumber \\
&&
-
\left[
\frac{g_s^2}{(4\pi)^2}\frac{N_c}{2}
\left(
\frac{1}{\epsilon^2_\textrm{UV}}
+\frac{3\pi^2}{4}
+\frac{\log\left[\frac{\tilde\mu^2P_1^2(4cd)}{\Lambda^4}\right]}
{\epsilon_\textrm{UV}}
+\frac{\log^2\left[\frac{\tilde\mu^2P_1^2(2c)^2}{\Lambda^4}\right]
+\log^2\left[\frac{\tilde\mu^2P_1^2(2d)^2}{\Lambda^4}\right]}{4}
\right)
+{O}(\epsilon)
\right]_\textrm{UV}
\nonumber \\
&&
\times
\left[\frac{g_s^2}{(4\pi)^2}
\frac{N_c^2-1}{4N_c}
\frac{4}{\epsilon_\text{IR}}
\left(
1-\frac{a\log\left(a+\sqrt{a^2-1}\right)}{\sqrt{a^2-1}}
\right)
+O(\epsilon^0)
\right]_{\text{IR}},
\end{eqnarray}
where we have kept only the real part because the imaginary contributions
cancel when we add the Hermitian-conjugate contribution.

\subsection{$\mathcal{D}_i$ diagrams}
The color factors of the $\mathcal{D}_i$ diagrams are given by
\begin{eqnarray}
\label{eq:color-fac-Di}
C_{\mathcal{D}_i^{P_1P_2}}
=
C_{\mathcal{D}_i^{P_1P_1}}
&=&
\frac{\textrm{Tr}\left[T_a T_b T_c\right]}{\sqrt{N_c}}
\frac{\textrm{Tr}\left[T_a T_d\right]}{\sqrt{N_c}}
f_{cbd}
=
-
\frac{\left(d_{abc}+if_{abc}\right)f_{abc}}{8N_c}
=-\frac{iN_c(N_c^2-1)}{8N_c},
\nonumber \\
C_{\mathcal{D}_i^{P_2P_1}}
=
C_{\mathcal{D}_i^{P_2P_2}}
&=&
\frac{\textrm{Tr}\left[T_a T_b T_c\right]}{\sqrt{N_c}}
\frac{\textrm{Tr}\left[T_d T_c\right]}{\sqrt{N_c}}
f_{abd}
=
\frac{\left(d_{abc}+if_{abc}\right)f_{abc}}{8N_c}
=+\frac{iN_c(N_c^2-1)}{8N_c}.
\end{eqnarray}

The expressions for the $\mathcal{D}_i$ diagrams, with our standard
phase-space regulators in Eq.~(\ref{standard-regulators}), are given by
\begin{eqnarray}
\label{eq:Di-P1P2-organized}
\mathcal{D}_1^{P_1P_2}
&=&
\frac{16ig_s^4\mu^{4\epsilon}\Lambda^2ad}{(P_1^2)^{-2}}
\int_{k_2} \textrm{PS}\,
\frac{1}
{(2P_1\cdot k_2)
(2P_2\cdot k_2)}
\int_{k_1} \textrm{PS}
\frac{1}{
(2P_2\cdot k_1)
(2\ell\cdot k_1)
\left[2\ell\cdot (k_1+k_2)+\Lambda^2\right]},
\nonumber \\
\mathcal{D}_2^{P_1P_2}
&=&
-\frac{16g_s^4\mu^{4\epsilon}\Lambda^2 ad}{(P_1^2)^{-2}}
\int_{k_2} \textrm{PS}\,
\frac{1}
{(2P_1\cdot k_2)
(2P_2\cdot k_2)
(2\ell\cdot k_2+\Lambda^2)}
\nonumber \\
&&
\times
\int\frac{d^Dk_1}{(2\pi)^D}
\frac{1}{
(2P_2\cdot k_1+i\varepsilon)
[2\ell\cdot (-k_1)+i\varepsilon]
(k_1^2+i\varepsilon)}.
\end{eqnarray}

\subsubsection{IR finiteness of the $k_1$ integration 
in $\mathcal{D}_1$ and $\mathcal{D}_2$}

As in the case of the $\mathcal{C}_i$ diagrams, it can be seen by power
counting that our standard UV regulators render the integration over
$k_1$ UV finite. However, there are still potential UV divergences that
could arise from the integrations over the other components of $k_1$.
In order to control these divergences, we introduce the following
additional UV-regulator factor
\begin{equation}
\label{eq:Di-addtional-UV-reg-2}
\frac{\Lambda\hspace{-0.1em}'^{\hspace{0.07em}2}}{2P_2\cdot k_1 + \Lambda\hspace{-0.1em}'^{\hspace{0.07em}2}}.
\end{equation} 
We denote the $\mathcal{D}_i$ into which we have inserted
this UV regulator factor by $\mathcal{D}_i\big|_\textrm{Reg}$.

The $k_1$ integrations of the individual $\mathcal{D}_i$ diagrams
contain rapidity (collinear) divergences, which arise when the
denominators of the Wilson-line propagators vanish, and also contain soft
divergences. We expect these divergences to cancel, by unitarity, in the
sum over cuts in the $\mathcal{D}_i$.

In Appendix~\ref{app:IR-finiteness-D12}, we have carried out
the $k_1$ integrations in $\mathcal{D}_1\big|_\textrm{Reg}$
and $\mathcal{D}_2\big|_\textrm{Reg}$. The results are
\begin{eqnarray}
\label{eq:D1-D2-UV}%
\mathcal{D}_1^{P_1P_2}\big|_\textrm{Reg}
&=&
\frac{8i
g_s^4
\mu^{2\epsilon}
\Lambda^2
a}
{(4\pi)^2(P_1^2)^{-1}}
\int_{k_2} \textrm{PS}\,
\frac{\frac{1}{2\epsilon_\textrm{IR}^2}
+\frac{1}{2\epsilon_\textrm{IR}}
\log\left(\frac{\tilde\mu^2P_1^2}{\Lambda\hspace{-0.1em}'^{\hspace{0.07em}4}}\right)
+O(\epsilon^0)}
{(2P_1\cdot k_2)
(2P_2\cdot k_2)
(2\ell\cdot k_2+\Lambda^2)},
\nonumber \\
\mathcal{D}_2^{P_1P_2}\big|_\textrm{Reg}
&=&
-
\frac{8ig_s^4
\mu^{2\epsilon}\Lambda^2
a}
{(4\pi)^{2}(P_1^2)^{-1}}
\int_{k_2} \textrm{PS}\,
\frac{\frac{1}{2\epsilon_\textrm{IR}^2}
+\frac{1}
{2\epsilon_\textrm{IR}}
\left[2i\pi+\log\left(\frac{\tilde\mu^2P_1^2}{\Lambda\hspace{-0.1em}'^{\hspace{0.07em}4}}\right)\right]
+O(\epsilon^0)}
{(2P_1\cdot k_2)
(2P_2\cdot k_2)
(2\ell\cdot k_2+\Lambda^2)},
\end{eqnarray}
Combining the results for the $k_1$ integrations 
of $\mathcal{D}_1^{P_1 P_2}$ and $\mathcal{D}_2^{P_1 P_2}$, 
we find that
\begin{eqnarray}
\label{eq:IR-finiteness-of-D12}
\mathcal{D}_1^{P_1 P_2}+\mathcal{D}_2^{P_1 P_2}
\big|_\textrm{Reg}
&=&
-
\frac{8i
g_s^4
\mu^{2\epsilon}
a\Lambda^2}
{(4\pi)^2(P_1^2)^{-1}}
\left[
\frac{i\pi}{\epsilon_{\textrm{IR}}}
+O(\epsilon^0)
\right]
\int_{k_2} \textrm{PS}\,
\frac{1}
{(2P_1\cdot k_2)
(2P_2\cdot k_2)
(2\ell\cdot k_2+\Lambda^2)}.
\nonumber \\
\end{eqnarray}
We see that the real double and single poles cancel in the sum over
cuts, leaving only an imaginary pole that cancels when we add the
Hermitian conjugate. 
Hence, we find that the $k_1$ integration with $k_2$ fixed
in $\mathcal{D}_1^{P_1 P_2}+\mathcal{D}_2^{P_1 P_2}$ plus 
Hermitian conjugate is IR finite.

\subsubsection{Calculation of $\mathcal{D}_1$ and $\mathcal{D}_2$}

Having established that the $k_1$ integration with $k_2$ fixed in
$\mathcal{D}_1^{P_1 P_2}+\mathcal{D}_2^{P_1 P_2}$ plus Hermitian
conjugate is IR finite, we can remove the extra UV-cutoff factor
(\ref{eq:UV-regulator-Li}) without introducing any ambiguities in
dimensional regularization. That is, we can assume that any poles that
we encounter in the final result for the sum over diagrams are UV in
origin. We note that, as can be seen from
Eq.~(\ref{eq:Di-P1P2-organized}), the $k_1$ integration in
$\mathcal{D}_2^{P_1P_2}$, without the additional UV regulator in
Eq.~(\ref{eq:Di-addtional-UV-reg-2}), is scaleless and vanishes in
dimensional regularization. That is, the IR and UV poles cancel.
Therefore,
\begin{eqnarray}
\mathcal{D}_1^{P_1P_2}
+
\mathcal{D}_2^{P_1P_2}
=
\mathcal{D}_1^{P_1P_2}
=
\frac{16ig_s^4\Lambda^2ad}{(P_1^2)^{-2}}
\mu^{2\epsilon}
\int_{k_2} \textrm{PS}\,
\frac{1}
{(2P_1\cdot k_2)
(2P_2\cdot k_2)}
D_1,
\end{eqnarray}
where $D_1$ is given by
\begin{eqnarray}
D_1
&\equiv&
\mu^{2\epsilon}
\int_{k_1} \textrm{PS}
\frac{1}{
(2P_2\cdot k_1)
(2\ell\cdot k_1)
\left[2\ell\cdot (k_1+k_2)+\Lambda^2\right]}
\nonumber \\
&=&
\mu^{2\epsilon}
\Gamma(3)
\int_0^{\infty}d\lambda_1 
\int_0^{\infty}d\lambda_2
\int_{k_1} \textrm{PS}
\frac{1}
{\left[2\lambda_1 P_2\cdot k_1
+2\lambda_2\ell\cdot k_1
+2\ell\cdot (k_1+k_2)+\Lambda^2
\right]^3}
\nonumber \\
&=&
-
\frac{1}{2d}
\frac{(\tilde\mu^2P_1^2)^{\epsilon}}{(4\pi)^2P_1^2}
\frac{e^{\epsilon\gamma_{{}_{\textrm{E}}}}\Gamma(1+2\epsilon)\Gamma(1-\epsilon)}
{\epsilon_\textrm{UV}}
\frac{1}{\left(2\ell\cdot k_2+\Lambda^2\right)^{1+2\epsilon}}
\int_0^{\infty}d\lambda_1 
\frac{1}
{\lambda_1^{1-\epsilon}(\lambda_1 + 2d)^{-\epsilon}}.
\phantom{XX}
\end{eqnarray}
The $\lambda_1$ integration can be carried out exactly to obtain
\begin{eqnarray}
\mathcal{D}_1^{P_1P_2}
+
\mathcal{D}_2^{P_1P_2}
&=&
-
\frac{8ig_s^4 a}
{(4\pi)^2(P_1^2)^{-1}}
\left[\frac{\tilde\mu^2P_1^2(2d)^{2}}{\Lambda^4}\right]^{\epsilon}
\frac{e^{\epsilon\gamma_{{}_{\textrm{E}}}}
\Gamma(1+2\epsilon)\Gamma(1-2\epsilon)\Gamma(1+\epsilon)}
{2\epsilon_\textrm{UV}^2}
\nonumber \\
&&
\times
\mu^{2\epsilon}
\int_{k_2} \textrm{PS}\,
\frac{(\Lambda^2)^{1+2\epsilon}}
{(2P_1\cdot k_2)
(2P_2\cdot k_2)\left(2\ell\cdot k_2+\Lambda^2\right)^{1+2\epsilon}}
\nonumber \\
&=&
-
\frac{8ig_s^4 a}
{(4\pi)^2(P_1^2)^{-1}}
\left[
\frac{1}{2\epsilon_\textrm{UV}^2}
+\frac{3\pi^2}{8}
+\frac{\log\left[\frac{\tilde\mu^2P_1^2(2d)^{2}}{\Lambda^4}\right]}{2\epsilon_\textrm{UV}}
+\frac{\log^2\left[\frac{\tilde\mu^2P_1^2(2d)^{2}}{\Lambda^4}\right]}{4}
+O(\epsilon)
\right]
\nonumber \\
&&
\times
\mu^{2\epsilon}
\int_{k_2} \textrm{PS}\,
\frac{(\Lambda^2)^{1+2\epsilon}}
{(2P_1\cdot k_2)
(2P_2\cdot k_2)\left(2\ell\cdot k_2+\Lambda^2\right)^{1+2\epsilon}}.
\end{eqnarray}
The $k_2$-dependent denominator factor that arises from the $k_1$
integration, $(2\ell\cdot k_2+\Lambda^2)^{1+2\epsilon}$, is
nonsingular as $k_2$ goes to zero. Therefore, we conclude that the
entire contribution from the $k_1$ integration is UV in nature.

Making use of the formula for the $k_2$ integration 
in Eq.~(\ref{eq:Ci-k2-int-formula}), we obtain
\begin{eqnarray}
\label{eq:D1-D2-P1P2-final}
\mathcal{D}_1^{P_1P_2}
+
\mathcal{D}_2^{P_1P_2}
&=&
-
\left[
\frac{ig_s^2}{(4\pi)^2}
\left(
\frac{1}{\epsilon_\textrm{UV}^2}
+\frac{3\pi^2}{4}
+\frac{\log\left[\frac{\tilde\mu^2P_1^2(2d)^{2}}{\Lambda^4}\right]}{\epsilon_\textrm{UV}}
+
\frac{\log^2\left[\frac{\tilde\mu^2P_1^2(2d)^{2}}{\Lambda^4}\right]}{2}
\right)
+O(\epsilon)
\right]_{\textrm{UV}}
\nonumber \\
&&
\times
\left[
-\frac{g_s^2}
{(4\pi)^2}
\frac{2}{\epsilon_\textrm{IR}}
\frac{a\log\left(a+\sqrt{a^2-1}\right)}{\sqrt{a^2-1}}
+O(\epsilon^0)
\right]_{\textrm{IR}}.
\end{eqnarray}
Here, as we have mentioned, the subscripts IR and UV
on the brackets indicate the origins of the contributions, and the factor 
labeled IR is the one-loop contribution to the LDME, absent its 
color factor.
At this point we should, in
principle, discard any imaginary contributions, since our proof of the
IR finiteness of the $k_1$ integration with $k_2$ fixed was valid only
for $\mathcal{D}_1^{P_1P_2}+\mathcal{D}_2^{P_1P_2}$ plus Hermitian
conjugate. However, the expression in Eq.~(\ref{eq:D1-D2-P1P2-final})
contains no imaginary parts. [The imaginary IR pole in
$\mathcal{D}_2^{P_1P_2}\big|_\textrm{Reg}$ in Eq.~(\ref{eq:D1-D2-UV}) is 
also present in $\mathcal{D}_2^{P_1P_2}$, but it cancels against an 
imaginary UV pole.]

The relations in Eq.~(\ref{eq:C1C2-PjPk-symm}) also hold for the 
$\mathcal{D}_i^{P_jP_k}$. Taking these relations into account, along 
with the color factors in Eq.~(\ref{eq:color-fac-Di}), we obtain 
\begin{eqnarray}
\label{eq:D12-final-with-color}
&&
\textrm{Re}
\left(
\sum_{i=1}^2
\sum_{j=1}^2\sum_{k=1}^2
C_{\mathcal{D}_i^{P_jP_k}}
\mathcal{D}_i^{P_j P_k}
\right)
\nonumber \\
&=&
-
\left[
\frac{g_s^2}{(4\pi)^2}\frac{N_c}{2}
\left(
\frac{1}{\epsilon^2_\textrm{UV}}
+\frac{3\pi^2}{4}
+\frac{\log\left[\frac{\tilde\mu^2P_1^2(4cd)}{\Lambda^4}\right]}
{\epsilon_\textrm{UV}}
+\frac{\log^2\left[\frac{\tilde\mu^2P_1^2(2c)^{2}}{\Lambda^4}\right]
+\log^2\left[\frac{\tilde\mu^2P_1^2(2d)^{2}}{\Lambda^4}\right]}
{4}
\right)
+O(\epsilon)
\right]_{\textrm{UV}}
\nonumber \\
&&
\times
\left[\frac{g_s^2}{(4\pi)^2}
\frac{N_c^2-1}{4N_c}
\frac{4}{\epsilon_\text{IR}}
\left(
1-\frac{a\log\left(a+\sqrt{a^2-1}\right)}{\sqrt{a^2-1}}
\right)
+O(\epsilon^0)
\right]_{\textrm{IR}}.
\end{eqnarray}

\subsection{$\mathcal{E}_i$ diagrams}
The color factors of the $\mathcal{E}_i^{P_jP_k}$ diagrams are given by
\begin{eqnarray}
\label{eq:colof-fac-Ei}
C_{\mathcal{E}_i^{P_1P_2}}
=
C_{\mathcal{E}_i^{P_1P_1}}
&=&
\frac{\textrm{Tr}\left[T_a T_b T_c\right]}{\sqrt{N_c}}
\frac{\textrm{Tr}\left[T_b T_d\right]}{\sqrt{N_c}}
f_{acd}
=
-\frac{iN_c(N_c^2-1)}{8N_c},
\nonumber \\
C_{\mathcal{E}_i^{P_2P_1}}
=
C_{\mathcal{E}_i^{P_2P_2}}
&=&
\frac{\textrm{Tr}\left[T_a T_b T_c\right]}{\sqrt{N_c}}
\frac{\textrm{Tr}\left[T_b T_d\right]}{\sqrt{N_c}}
f_{adc}
=
+\frac{iN_c(N_c^2-1)}{8N_c}.
\end{eqnarray}

The expressions for the $\mathcal{E}_i$ diagrams with our standard
phase-space regulators in Eq.~(\ref{standard-regulators}) are given by
\begin{eqnarray}
\label{eq:E1E2-sum-ep-1}
\mathcal{E}_1^{P_1P_2}
&=&
-
\frac{16ig_s^4\Lambda^2 ac}{(P_1^2)^{-2}}
\mu^{2\epsilon}
\int_{k_2} \textrm{PS}\,
\frac{1}
{(2P_2\cdot k_2)}
E_1
\nonumber \\
\mathcal{E}_2^{P_1P_2}
&=&
\frac{16g_s^4\Lambda^2 ac}{(P_1^2)^{-2}}
\mu^{2\epsilon}
\int_{k_2} \textrm{PS}\,
\frac{1}{(2P_2\cdot k_2)(2\ell\cdot k_2+\Lambda^2)}
E_2,
\end{eqnarray}
where 
\begin{eqnarray}
\label{eq:E1E2-definition}
E_1
&\equiv&
\mu^{2\epsilon}
\int_{k_1} \textrm{PS}
\frac{1}
{(2P_1\cdot k_1)(2\ell\cdot k_1)\left[2P_1\cdot (k_1+k_2)\right]
\left[2\ell\cdot (k_1+k_2)+\Lambda^2\right]},
\nonumber \\
E_2
&\equiv&
\mu^{2\epsilon}
\int\frac{d^Dk_1}{(2\pi)^D}
\frac{1}
{(2P_1\cdot k_1+i\varepsilon)\left[2\ell\cdot (-k_1)+i\varepsilon\right]
\left[2P_1\cdot (k_1+k_2)+i\varepsilon\right](k_1^2+i\varepsilon)}.
\end{eqnarray}
It can be seen by power counting that the $k_1$ integrations of the 
individual diagrams are rendered UV finite by our 
standard phase-space regulators. There is no need in this case to 
introduce additional UV regulators.

The $k_1$ integrations of the individual $\mathcal{E}_i$ diagrams
contain rapidity (collinear) divergences that arise when the
denominators of the Wilson-line propagators vanish.
We expect these divergences to cancel, by unitarity, in the
sum over cuts in the $\mathcal{E}_i$.

Applying Feynman parametrization 
to Eq.~(\ref{eq:E1E2-definition}),
we obtain
\begin{eqnarray}
E_1
&=&
\mu^{2\epsilon}\Gamma(4)
\int_0^\infty d\lambda_1
\int_0^1 dx
\int_0^1 dy
\int_{k_1} \textrm{PS}
\frac{\lambda_1}
{\left[2k_1\cdot(\ell+\lambda_1 P_1)
+2\lambda_1 y P_1\cdot k_2
+2x\ell\cdot k_2+x\Lambda^2\right]^4},
\nonumber \\
E_2
&=&
\mu^{2\epsilon}\Gamma(4)
\int_0^\infty d\lambda_1
\int_0^\infty d\lambda_2
\int_0^1 dx
\int\frac{d^{D}k_1}{(2\pi)^{D}}
\frac{\lambda_1}
{\left[
k_1^2+2k_1\cdot(\lambda_1P_1-\lambda_2\ell)+2\lambda_1 x P_1\cdot k_2
+i\varepsilon
\right]^4}.
\nonumber \\
\end{eqnarray}
Then, carrying out the $k_1$ integration and the
parameter integrations, except for $\lambda_1$ integration in $E_1$,
we obtain
\begin{eqnarray}
E_1
&=&
-\frac{1}{(2c)}
\frac{\left(\tilde\mu^2P_1^2\right)^\epsilon}
{(4\pi)^{2}(P_1^2)}
\frac{e^{\epsilon\gamma_{{}_{\textrm{E}}}}\Gamma(1+2\epsilon)\Gamma(1-\epsilon)}
{\epsilon_\textrm{IR}}
\frac{1}{K_1}
\nonumber \\
&&
\times
\int_0^\infty d\lambda_1
\frac{1}{\lambda_1^{1-\epsilon}
(\lambda_1+2c)^{-\epsilon}}
\left[
\frac{1}{K_2^{1+2\epsilon}}
-\frac{1}{(\lambda_1 K_1+K_2)^{1+2\epsilon}}
\right],
\nonumber \\
E_2
&=&
\frac{i}{4c}
\frac{\left(\tilde\mu^2P_1^2\right)^{\epsilon}}{(4\pi)^{2}(P_1^2)}
\frac{e^{\epsilon\gamma_{{}_{\textrm{E}}}}\Gamma(1+2\epsilon)\Gamma(1-\epsilon)}{e^{-2i\pi\epsilon}\epsilon_\textrm{IR}^2}
\frac{1}{K_1^{1+2\epsilon}}
,
\end{eqnarray}
where $K_1$ and $K_2$ are given in Eq.~(\ref{eq:K1K2-def}). 
The first $\lambda_1$ integration yields
\begin{eqnarray}
\int_0^\infty d\lambda_1
\frac{1}{\lambda_1^{1-\epsilon}
(\lambda_1+2c)^{-\epsilon}}
\frac{1}{K_2^{1+2\epsilon}}
&=&
\frac{(2c)^{2\epsilon}}{K_2^{1+2\epsilon}}
\left[
\frac{\Gamma(-2\epsilon_\textrm{IR})\Gamma(1+\epsilon)}
{\Gamma(1-\epsilon)}
+
\frac{\Gamma(\epsilon_\textrm{UV})\Gamma(1-2\epsilon)}
{\Gamma(1-\epsilon)}
\right].
\end{eqnarray}
The second $\lambda_1$ integration is given 
in Eqs.~(\ref{eq:lam-1-int-result-K1-K2}).
Then, we obtain 
\begin{eqnarray}
E_1
&=&
-\frac{1}{4c}
\frac{\left(\tilde\mu^2P_1^2\right)^\epsilon}
{(4\pi)^{2}(P_1^2)}
\frac{1}{K_1^{1+2\epsilon}K_2}
\left[
-
\frac{1}{\epsilon_\textrm{IR}^2}
-\frac{5\pi^2}{12}
+O(\epsilon,K_1/\Lambda^2)
\right],
\nonumber \\
E_2
&=&
\frac{i}{4c}
\frac{\left(\tilde\mu^2P_1^2\right)^{\epsilon}}
{(4\pi)^{2}(P_1^2)}
\frac{1}{K_1^{1+2\epsilon}}
\left[
\frac{1}{\epsilon_\textrm{IR}^2}
+\frac{2i\pi}{\epsilon_\textrm{IR}}
-\frac{19\pi^2}{12}
+O(\epsilon)
\right].
\end{eqnarray}
The denominator factors $K_1^{1+2\epsilon}=(2P_1\cdot 
k_2)^{1+2\epsilon}$ become singular as $k_2$ goes to zero, and so all 
of the terms in this expression yield IR contributions.
Inserting these results into Eq.~(\ref{eq:E1E2-sum-ep-1}), we find that
\begin{eqnarray}
\mathcal{E}_1^{P_1P_2}
+
\mathcal{E}_2^{P_1P_2}
&=&
\frac{8ig_s^4 a}
{(4\pi)^{2}(P_1^2)^{-1}}
\left(\frac{\tilde\mu^2P_1^2}{\Lambda^4}\right)^\epsilon
\bigg[
\frac{i\pi}{\epsilon_\textrm{IR}}
-\pi^2
+O\left(\epsilon,2P_1\cdot k_2/\Lambda^2\right)
\bigg]
\nonumber\\
&&
\times
\mu^{2\epsilon}
\int_{k_2} \textrm{PS}\,
\frac{(\Lambda^2)^{1+2\epsilon}}{(2P_1\cdot k_2)^{1+2\epsilon}
(2P_2\cdot k_2)(2\ell\cdot k_2+\Lambda^2)}.
\end{eqnarray}
Making use of the formula for the $k_2$ integration 
in Eq.~(\ref{eq:Ci-k2-int-formula}), we obtain
\begin{eqnarray}
\mathcal{E}_1^{P_1P_2}
+
\mathcal{E}_2^{P_1P_2}
&=&
\left[\frac{ig_s^4}{(4\pi)^{4}}
\frac{2\pi^2}{\epsilon_\textrm{IR}}
\frac{a\log\left(a+\sqrt{a^2-1}\right)}{\sqrt{a^2-1}}
+O(\epsilon^0)\right]_{\textrm{IR}},
\end{eqnarray}
where we have dropped imaginary contributions that cancel when we add the 
Hermitian-conjugate contributions.

The relations in Eq.~(\ref{eq:C1C2-PjPk-symm}) also hold for the
$\mathcal{E}_i^{P_jP_k}$. Taking into account these relations and the
color factors in Eq.~(\ref{eq:color-fac-Di}) we obtain 
\begin{eqnarray}
\label{eq:E-final-with-color}
&&
\textrm{Re}
\left(
\sum_{i=1}^2
\sum_{j=1}^2
\sum_{k=1}^2
C_{\mathcal{E}_i^{P_jP_k}}
\mathcal{E}_i^{P_jP_k}
\right)
\nonumber \\
&=&
-
\left[
\frac{g_s^4}
{(4\pi)^{4}}
\frac{N_c(N_c^2-1)}{8N_c}
\frac{4\pi^2}{\epsilon_\textrm{IR}}
\left(
1-
\frac{a\log\left(a+\sqrt{a^2-1}\right)}{\sqrt{a^2-1}}
\right)
+O(\epsilon^0)\right]_{\textrm{IR}}.
\phantom{XX}
\end{eqnarray}

\subsection{$\mathcal{F}_i$ diagrams}

Now we turn to the calculation of the non-Abelian diagrams 
$\mathcal{F}_i^{P_jP_k}$.

The color factor of the $\mathcal{F}_i^{P_jP_k}$ diagrams is given by
\begin{eqnarray}
\label{eq:color-F}
C_{\mathcal{F}_i^{P_jP_k}}
&=&
\frac{\textrm{Tr}\left[T_a T_b\right]}{\sqrt{N_c}}
\frac{\textrm{Tr}\left[T_c T_d\right]}{\sqrt{N_c}}
f_{aec}f_{bde}
=-\frac{f_{abc}f_{abc}}{4N_c}
=-\frac{N_c(N_c^2-1)}{4N_c}.
\end{eqnarray}
We note that our expression for the color factor of the non-Abelian 
diagrams differs by a factor of $(1/\sqrt{N_c})^2$
from the color factor that is given in 
Eq.~(16) of Ref.~\cite{Nayak:2006fm}
because the factors $1/\sqrt{N_c}$ from the color-singlet projectors were omitted 
in Eq.~(16) of Ref.~\cite{Nayak:2006fm}.

The expressions for the $\mathcal{F}_i^{P_1P_2}$ diagrams, with our 
standard UV  phase-space regulator
in Eq.~(\ref{standard-regulators}) are given by
\begin{subequations}
\begin{eqnarray}
\mathcal{F}_1^{P_1P_2}
&=&
8g_s^4\mu^{4\epsilon}\Lambda^2
\int_{k_1}\textrm{PS}
\int_{k_2}\textrm{PS}
\frac{1}
{(2P_2\cdot k_2)}
\nonumber \\
&&
\times
\frac{N_{\mathcal{F}}}
{[2P_1\cdot(k_1+k_2)]
(2\ell \cdot k_1)
[(k_1+k_2)^2]
\left[
2\ell\cdot (k_1+k_2)+\Lambda^2
\right]},
\label{eq:Fp1p2-1}
\\
\mathcal{F}_2^{P_1P_2}
&=&
8ig_s^4 \mu^{4\epsilon}
\Lambda^2
\int_{k_2}\textrm{PS}
\frac{1}{(2P_2\cdot k_2)\left(2\ell\cdot k_2+\Lambda^2\right)}
\nonumber \\
&&
\times
\int\frac{d^Dk_1}{(2\pi)^D}
\frac{N_{\mathcal{F}}}
{[2P_1\cdot(k_1+k_2)+i\varepsilon]
(-2\ell \cdot k_1+i\varepsilon)(k_1^2+i\varepsilon)
[(k_1+k_2)^2+i\varepsilon]
}.
\phantom{XX}
\label{eq:Fp1p2-2}
\end{eqnarray}
\end{subequations}
The numerator factor $N_{\mathcal{F}}$ is given by 
\begin{equation}
N_{\mathcal{F}}=
P_1^\nu \ell^\mu P_2^\lambda V_{\nu\mu\lambda}(k_1+k_2,-k_1,-k_2),
\end{equation}
where $V_{\mu_1\mu_2\mu_3}$ is the triple-gluon vertex 
\begin{equation}
V_{\mu_1\mu_2\mu_3}
(q_1,q_2,q_3)
=
(q_1-q_2)_{\mu_3} g_{\mu_1\mu_2}
+(q_2-q_3)_{\mu_1} g_{\mu_2\mu_3}
+(q_3-q_1)_{\mu_2} g_{\mu_3\mu_1},
\end{equation}
with all momenta $q_i$ flowing into the vertex. 

The numerator factor can be written as 
\begin{eqnarray}
\label{eq:triple-vertex-Fi}
N_{\mathcal{F}}
&=&
(P_1\cdot \ell)(P_2\cdot k_2)
+(P_2\cdot \ell)[P_1\cdot(k_1+k_2)]
\nonumber\\
&&
-(P_1\cdot P_2)(\ell\cdot k_1)
-2(P_2\cdot \ell)(P_1\cdot k_1) 
+2(P_1\cdot \ell)(P_2\cdot k_1)
-2(P_1\cdot P_2)(\ell\cdot k_2).
\end{eqnarray}
The first two terms on the right side of Eq.~(\ref{eq:triple-vertex-Fi})
cancel the eikonal-propagator denominators $P_2\cdot k_2$ and $P_1\cdot
(k_1+k_2)$, respectively, resulting in expressions that are independent
of $P_1$ or $P_2$. These expressions cancel when we sum over gluon
connections to the quark and antiquark lines. (This is a manifestation of
the graphical Ward identities.) Therefore, we drop these terms in the
numerator and use a modified numerator factor
\begin{equation}
\label{eq:modified-numerator-factor}
\tilde{N}_{\mathcal{F}}=
-(P_1\cdot P_2)(\ell\cdot k_1)
-2(P_2\cdot \ell)(P_1\cdot k_1) 
+2(P_1\cdot \ell)(P_2\cdot k_1)
-2(P_1\cdot P_2)(\ell\cdot k_2).
\end{equation}

It can be seen from power counting that the individual diagrams
$\mathcal{F}_i^{P_1P_2}$ are UV finite with respect to the $k_1$ and
$k_2$ integrations. UV divergences that might occur when the component of
$k_1$ or $k_2$ that is parallel to $\ell$ becomes large cancel because
the terms in the modified numerator factor $\tilde{N}_{\mathcal{F}}$
[Eq.~(\ref{eq:modified-numerator-factor})] that are proportional to
$k_i$ vanish when $k_i$ is proportional to $\ell$.

The $k_1$ and $k_2$ integrations contain IR divergences that arise from 
several sources: (i) the 
rapidity (collinear) divergence that occurs when the denominator 
$\ell\cdot k_1$ vanishes, (ii) the divergence that occurs when 
$k_1$ is collinear to $\ell$ and $k_2$ is collinear to $k_1$, (iii) the 
divergence that occurs when $k_1$ is collinear to $k_2$, (iv) the 
divergence that occurs when $k_1$ is soft relative to $k_2$, and (v) the 
divergence that occurs when both $k_1$ and $k_2$ are soft. These 
divergences can produce poles up to order $1/\epsilon^4$ in the original 
individual diagrams, which occur when $k_1$ and $k_2$ are both soft and 
collinear to $\ell$.  
However, our use of the modified numerator factor $\tilde{N}_{\mathcal{F}}$
[Eq.~(\ref{eq:modified-numerator-factor})] eliminates these 
$1/\epsilon^4$ poles because $\tilde{N}_{\mathcal{F}}$ vanishes when 
$k_i$ and $k_2$ are proportional to $\ell$.

We expect all of the remaining divergences to cancel by unitarity
in the sum over diagrams, except for divergence (v), which is the object
of our calculation. In comparison with the unitarity cancellations that
we found in the Abelian case, the unitarity cancellations in the case of
the $\mathcal{F}_i$ diagrams are rather involved. In particular, when
$k_1$ is soft or collinear to $k_2$, the cancellations involve the
Hermitian conjugate of the original diagram with $k_1\leftrightarrow
k_2$ and $P_1\leftrightarrow P_2$. Rather than attempting to identify
and implement all of the individual unitarity cancellations, we simply
carry out a straightforward evaluation of the diagrams and cancel poles
in the sum over diagrams. This approach leads to rather complicated
intermediate expressions that contain poles of orders $1/\epsilon^3$ and
$1/\epsilon^2$ that cancel to leave a final result of order
$1/\epsilon$.

\subsubsection{$\mathcal{F}_1^{P_1P_2}$ diagram}

Using Eq.~(\ref{eq:Fp1p2-1}) with the modified numerator factor in 
Eq.~(\ref{eq:modified-numerator-factor}), we obtain
\begin{equation}
\mathcal{F}_1^{P_1P_2}=
-8\left[(P_1\cdot P_2)\ell
+2(P_2\cdot \ell)P_1 
-2(P_1\cdot \ell)P_2\right]_\alpha \mathcal{F}_{11}^\alpha
-8\left[2(P_1\cdot P_2)\ell\right]_\alpha 
\mathcal{F}_{12}^\alpha,
\label{eq:F1P1P2-complete-def}
\end{equation}
where
\begin{eqnarray}
\label{eq:Fij-alpha-def}
\mathcal{F}_{11}^\alpha
&=&
\mu^{4\epsilon}\Lambda^2
\int_{k_1}\textrm{PS}
\int_{k_2}\textrm{PS}
\frac{k_1^\alpha}
{(2P_2\cdot k_2)[2P_1\cdot(k_1+k_2)]
(2\ell \cdot k_1)
(k_1+k_2)^2
\left[
2\ell\cdot (k_1+k_2)+\Lambda^2
\right]},
\nonumber \\
\mathcal{F}_{12}^\alpha
&=&
\mu^{4\epsilon}\Lambda^2
\int_{k_1}\textrm{PS}
\int_{k_2}\textrm{PS}
\frac{k_2^\alpha}
{(2P_2\cdot k_2)[2P_1\cdot(k_1+k_2)]
(2\ell \cdot k_1)
(k_1+k_2)^2
\left[
2\ell\cdot (k_1+k_2)+\Lambda^2
\right]}.
\nonumber \\
\end{eqnarray}
Note that any terms in $\mathcal{F}_{11}^\alpha$ and
$\mathcal{F}_{12}^\alpha$ that are proportional to $\ell^\alpha$ vanish
upon contraction with the associated factors in
Eq.~(\ref{eq:F1P1P2-complete-def}). It is this property that 
eliminates the contributions of order $1/\epsilon^4$.

Applying Feynman parametrization to Eq.~(\ref{eq:Fij-alpha-def}), 
we obtain
\begin{eqnarray}
\label{eq:Fij-alpha-def-2}
\mathcal{F}_{11}^\alpha
&=&
\mu^{4\epsilon}
\Lambda^2
\Gamma(5)
\int_0^\infty d\lambda_1
\int_0^\infty d\lambda_2
\int_0^\infty d\lambda_3
\int_0^\infty d\lambda_4
\int_{k_1}\textrm{PS}
\int_{k_2}\textrm{PS}
\nonumber \\
&&
\times
\frac{k_1^\alpha}
{\left[2k_1\cdot(k_2+\lambda_1 P_1+\lambda_2 \ell+\lambda_4\ell)
+2k_2\cdot(\lambda_1 P_1+\lambda_3 P_2+\lambda_4 \ell)
+\lambda_4 \Lambda^2
\right]^5},
\nonumber \\
\mathcal{F}_{12}^\alpha
&=&
\mu^{4\epsilon}
\Lambda^2
\Gamma(5)
\int_0^\infty d\lambda_1
\int_0^\infty d\lambda_2
\int_0^\infty d\lambda_3
\int_0^\infty d\lambda_4
\int_{k_1}\textrm{PS}
\int_{k_2}\textrm{PS}
\nonumber \\
&&
\times
\frac{k_2^\alpha}
{\left[2k_1\cdot(k_2+\lambda_1 P_1+\lambda_2 \ell+\lambda_4\ell)
+2k_2\cdot(\lambda_1 P_1+\lambda_3 P_2+\lambda_4 \ell)
+\lambda_4 \Lambda^2
\right]^5}.
\end{eqnarray}
We first carry out the $k_1$ integration by making use of the phase-space-integration formula in Eq.~(\ref{eq:k1-phase-int-table}). In the result,
we introduce a Feynman parameter $x$ to combine the two $k_2$-dependent
denominators and then carry out the $k_2$ integration by using
Eq.~(\ref{eq:k1-phase-int-table}) once again. The results are
\begin{eqnarray}
\mathcal{F}_{11}^\alpha
&=&
\frac{\mu^{4\epsilon}
\Lambda^2}
{(4\pi)^D}
\Gamma(7-\tfrac{3D}{2})\Gamma(\tfrac{D}{2})
\int_0^\infty [d\lambda_i]
\int_0^1 dx\,x^{\frac{D}{2}-1}(1-x)^{5-D}
\frac{\lambda_1 P_1^\alpha+\lambda_3(1-x)P_2^\alpha}
{(P^2)^{\frac{D}{2}}(M^2)^{7-\frac{3D}{2}}}
\nonumber \\
&&
+
\frac{\mu^{4\epsilon}
\Lambda^2}
{(4\pi)^D}
\Gamma(8-\tfrac{3D}{2})\Gamma(\tfrac{D}{2}-1)
\int_0^\infty [d\lambda_i]
\int_0^1 dx\,x^{\frac{D}{2}-1}(1-x)^{5-D}
\frac{\lambda_1 P_1^\alpha}
{(P^2)^{\frac{D}{2}-1}(M^2)^{8-\frac{3D}{2}}},
\nonumber \\
\mathcal{F}_{12}^\alpha
&=&
\frac{\mu^{4\epsilon}
\Lambda^2}{(4\pi)^D}
\Gamma(7-\tfrac{3D}{2})\Gamma(\tfrac{D}{2})
\int_0^\infty [d\lambda_i]
\int_0^1 dx\,
x^{\frac{D}{2}-2}(1-x)^{6-D}
\frac{\lambda_1 P_1^\alpha+\lambda_3(1-x)P_2^\alpha}
{(P^2)^{\frac{D}{2}}(M^2)^{7-\frac{3D}{2}}}.\phantom{X}
\end{eqnarray}
Here, we have dropped the terms that 
are proportional to $\ell^\alpha$ because they cancel on contraction 
with the other factors in Eq.~(\ref{eq:F1P1P2-complete-def}), and we
have introduced the definitions $[d\lambda_i]\equiv
d\lambda_1d\lambda_2d\lambda_3d\lambda_4$ and
\begin{eqnarray}
P
&\equiv&
\lambda_1 P_1+\lambda_3(1-x)P_2
+\left(\lambda_2x+\lambda_4\right)\ell,
\nonumber \\
M^2
&\equiv&
\lambda_1^2x P_1^2+2\lambda_1(\lambda_2+\lambda_4)x P_1\cdot \ell
+\lambda_4(1-x) \Lambda^2.
\end{eqnarray}

We make a change of variables, replacing $\lambda_2$, $\lambda_3$, and 
$\lambda_4$ with
\begin{eqnarray}
\omega &\equiv& 2c \frac{x\lambda_2}{\lambda_1},
\nonumber \\
\xi &\equiv& \frac{(1-x)\lambda_3}{\lambda_1},
\nonumber \\
\vartheta &\equiv& \frac{\lambda_4}{x\lambda_2}.
\end{eqnarray}
Then, we carry out the $\lambda_1$ integration 
by making use of the integral formula 
in Eq.~(\ref{eq:ablam-general}) and obtain
\begin{eqnarray}
\mathcal{F}_{11}^{\alpha}
&=&
\frac{\left[\frac{\tilde\mu^2 P_1^2(2c)^2}{\Lambda^4}\right]^{2\epsilon}}
{2c(P_1^2)^2(4\pi)^4}
e^{2\epsilon\gamma_{{}_{\textrm{E}}}}
\Gamma(-\epsilon)\Gamma(2-\epsilon)\Gamma(1+4\epsilon)
\int_0^\infty d\xi
\int_0^\infty d\omega\int_0^\infty d\vartheta
\nonumber \\
&&
\times
\int_0^1 dx\,
\frac{\omega^{-4\epsilon}\vartheta^{-1-4\epsilon}
x^{-\epsilon}(1-x)^{-1-2\epsilon}
\left[\omega(1+\vartheta x)+x\right]^{\epsilon}
\left(
P_1+\xi P_2
\right)^\alpha}
{\left[\xi^2+2a\xi+1+\omega(1+\vartheta)(1+\frac{d}{c}\xi)\right]^{2-\epsilon}}
\nonumber \\
&&
+
\frac{\left[\frac{\tilde\mu^2 P_1^2(2c)^2}{\Lambda^4}\right]^{2\epsilon}}
{2c(P_1^2)^2(4\pi)^4}
e^{2\epsilon\gamma_{{}_{\textrm{E}}}}
\Gamma^2(1-\epsilon)\Gamma(1+4\epsilon)
\int_0^\infty d\xi
\int_0^\infty d\omega\int_0^\infty d\vartheta
\nonumber \\
&&
\times
\int_0^1 dx\,
\frac{\omega^{-4\epsilon}\vartheta^{-1-4\epsilon}
x^{-\epsilon}(1-x)^{-1-2\epsilon}
\left[\omega(1+\vartheta x)+x\right]^{-1+\epsilon}
(P_1)^\alpha}
{\left[\xi^2+2a\xi+1+\omega(1+\vartheta)(1+\frac{d}{c}\xi)\right]^{1-\epsilon}},
\nonumber \\
\mathcal{F}_{12}^{\alpha}
&=&
\frac{\left[\frac{\tilde\mu^2 P_1^2(2c)^2}{\Lambda^4}\right]^{2\epsilon}}
{2c(P_1^2)^2(4\pi)^4}
e^{2\epsilon\gamma_{{}_{\textrm{E}}}}
\Gamma(-\epsilon)\Gamma(2-\epsilon)\Gamma(1+4\epsilon)
\int_0^\infty d\xi
\int_0^\infty d\omega\int_0^\infty d\vartheta
\nonumber \\
&&
\times
\int_0^1 dx\,
\frac{\omega^{-4\epsilon}\vartheta^{-1-4\epsilon}
x^{-1-\epsilon}(1-x)^{-2\epsilon}
\left[\omega(1+\vartheta x)+x\right]^{\epsilon}
\left(P_1+\xi P_2\right)^\alpha}
{\left[\xi^2+2a\xi+1+\omega(1+\vartheta)(1+\frac{d}{c}\xi)\right]^{2-\epsilon}}.
\end{eqnarray}
Next, we carry out the parameter integrations, 
except for the $\xi$ integration,
by making use of Eq.~(\ref{plus-expansion}). We find that the 
$1/\epsilon^3$, $1/\epsilon^2$, and $1/\epsilon$ contributions to 
$\mathcal{F}_1^{P_1P_2}$ are given by the following expressions:
\begin{subequations}
\begin{eqnarray}
\label{eq:F1-p1p2-ep3}
\mathcal{F}_1^{P_1P_2}\Big|_{1/\epsilon^3}
&=&
-
\frac{g_s^4\left(\frac{\tilde{\mu}^2P_1^2}{\Lambda^4}\right)^{2\epsilon}}
{(4\pi)^4\epsilon^3}
\int_0^\infty d\xi
\bigg\{
\frac{(3ad-2c)\xi+2d-ac}{2c}
\left(-\frac{1}{A B}\right)
+\frac{(2ad) \xi+2ac}{2c}
\left(-\frac{2}{A B}\right)
\bigg\},
\nonumber \\
\\
\label{eq:F1-p1p2-ep2}
\mathcal{F}_1^{P_1P_2}\Big|_{1/\epsilon^2}
&=&
-
\frac{g_s^4\left(\frac{\tilde{\mu}^2P_1^2}{\Lambda^4}\right)^{2\epsilon}}
{(4\pi)^4\epsilon^2}
\int_0^\infty d\xi
\bigg\{
\frac{(3ad-2c)\xi+2d-ac}{2c}
\nonumber \\
&&
\quad\quad\quad\quad\quad\quad\quad\quad\quad\quad
\times
\left[
\frac{3\log A-4\log B-4\log 2c}{A B}
-
\frac{\log A-\log B}{B(A-B)}
\right]
\nonumber \\
&&
\quad\quad\quad\quad\quad\quad\quad\quad\quad
+\frac{2d-ac}{2c}
\left(\frac{\log A-\log B}{A-B}\right)
\nonumber \\
&&
\quad\quad\quad\quad\quad\quad\quad\quad\quad
+\frac{(2ad) \xi+2ac}{2c}
\left(
\frac{4\log A-6\log B-8\log 2c}{A B}
\right)
\bigg\},
\end{eqnarray}
\begin{eqnarray}
\label{eq:F1-p1p2-ep1}
\mathcal{F}_1^{P_1P_2}\Big|_{1/\epsilon}
&=&
-
\frac{g_s^4\left(\frac{\tilde{\mu}^2P_1^2}{\Lambda^4}\right)^{2\epsilon}}
{(4\pi)^4\epsilon}
\int_0^\infty d\xi
\nonumber \\
&&
\times
\Bigg\{
\frac{(3ad-2c)\xi+2d-ac}{2c}
\nonumber\\
&&
\quad\quad
\times
\Bigg(
\frac{1}{B(A-B)}
\bigg[
-4(\log 2c)
\left(
\log A-\log B
\right)
+\frac{1}{2}\log^2 A
+\frac{3}{2}\log^2 B
-2\log A\log B
\bigg]
\nonumber \\
&&
\quad\quad\quad\quad
+
\frac{1}{AB}
\bigg[
-\frac{13\pi^2}{6}
-8\log^22c
+4(\log 2c)
\left(
3\log A-4\log B
\right)
+4\textrm{Li}_2\left(-\frac{B}{A-B}\right)
\nonumber \\
&&
\quad\quad\quad\quad
-\frac{9}{2}\log^2 A
-6\log^2 B
+2\log^2(A-B)
+12\log A\log B
-4\log B\log(A-B)
\bigg]
\Bigg)
\nonumber \\
&&
\quad\quad
+\frac{2d-ac}{2c}
\Bigg(
\frac{1}{A-B}
\bigg[
4(\log 2c)\left(\log A-\log B\right)
-\frac{1}{2}\log^2 A
-\frac{3}{2}\log^2 B
+2\log A\log B
\bigg]
\Bigg)
\nonumber \\
&&
\quad\quad
+\frac{(2ad) \xi+2ac}{2c}
\Bigg(
\frac{1}{AB}
\bigg[
-\frac{17\pi^2}{3}
-16\log^22c
+8(\log 2c)\left(2\log A-3\log B\right)
\nonumber \\
&&
\quad\quad\quad\quad\quad\quad\quad\quad\quad\quad\quad\quad
-2\textrm{Li}_2\left(-\frac{B}{A-B}\right)
-3\log^2 A
-9\log^2 B
-\log^2(A-B)
\nonumber \\
&&
\quad\quad\quad\quad\quad\quad\quad\quad\quad\quad\quad\quad
+10\log A\log B
+2\log B\log(A-B)
\bigg]
\Bigg)
\Bigg\},
\end{eqnarray}
\end{subequations}
where 
the definitions of $A$ and $B$ are given in
Eq.~(\ref{eq:def-of-A-B}), and
$\textrm{Li}_2(z)$ is the dilogarithm function
\begin{equation}
\textrm{Li}_2(z)=-\int_0^z dt\,\frac{\log(1-t)}{t}.
\end{equation}

\subsubsection{$\mathcal{F}_2$ diagram}

Using Eq.~(\ref{eq:Fp1p2-2}) with the modified numerator factor in 
Eq.~(\ref{eq:modified-numerator-factor}), we have
\begin{equation}
\mathcal{F}_2^{P_1P_2}
=
-8\left[(P_1\cdot P_2)\ell
+2(P_2\cdot \ell)P_1 
-2(P_1\cdot \ell)P_2\right]_\alpha \mathcal{F}_{21}^\alpha
-8\left[2(P_1\cdot P_2)\ell\right]_\alpha 
\mathcal{F}_{22}^\alpha,
\label{eq:F2P1P2-complete-def}
\end{equation}
where
\begin{eqnarray}
\label{eq:F2ij-alpha-def}
\mathcal{F}_{21}^\alpha
&=&
i\mu^{4\epsilon}
\Lambda^2
\int_{k_2}\textrm{PS}
\frac{1}{(2P_2\cdot k_2)(2\ell\cdot k_2+\Lambda^2)}
\nonumber \\
&&
\times
\int\frac{d^Dk_1}{(2\pi)^D}
\frac{k_1^\alpha}
{[2P_1\cdot(k_1+k_2)+i\varepsilon]
(-2\ell \cdot k_1+i\varepsilon)
(k_1^2+i\varepsilon)
[(k_1+k_2)^2+i\varepsilon]
},
\nonumber \\
\mathcal{F}_{22}^\alpha
&=&
i\mu^{4\epsilon}
\Lambda^2
\int_{k_2}\textrm{PS}
\frac{k_2^\alpha}
{(2P_2\cdot k_2)(2\ell\cdot k_2+\Lambda^2)}
\nonumber \\
&&
\times
\int\frac{d^Dk_1}{(2\pi)^D}
\frac{1}
{[2P_1\cdot(k_1+k_2)+i\varepsilon]
(-2\ell \cdot k_1+i\varepsilon)
(k_1^2+i\varepsilon)
[(k_1+k_2)^2+i\varepsilon]}.
\end{eqnarray}

Applying Feynman parametrization to Eq.~(\ref{eq:F2ij-alpha-def}), 
we obtain
\begin{eqnarray}
\label{eq:F2ij-alpha-def-2}
\mathcal{F}_{21}^{\alpha}
&=&
i\mu^{4\epsilon}\Lambda^2\Gamma(6)
\int_0^\infty [d\lambda_i]
\int_0^1 dx
\int_{k_2}\textrm{PS}
\int\frac{d^Dk_1}{(2\pi)^D}
\nonumber \\
&&\times
\frac{-[(1-x)k_2-\lambda_2 \ell+\lambda_1 P_1]^\alpha}
{\{
k_1^2
+2[\lambda_1 x P_1+\lambda_3P_2+\lambda_2(1-x)\ell+\lambda_4\ell]
\cdot k_2
+2\lambda_1\lambda_2 P_1\cdot \ell
-\lambda_1^2 P_1^2
+\lambda_4\Lambda^2
+i\varepsilon
\}^{6}},
\nonumber \\
\mathcal{F}_{22}^\alpha
&=&
i\mu^{4\epsilon}\Lambda^2\Gamma(6)
\int_0^\infty [d\lambda_i]
\int_0^1 dx
\int_{k_2}\textrm{PS}
\int\frac{d^Dk_1}{(2\pi)^D}
\nonumber \\
&&\times
\frac{k_2^\alpha}
{\{
k_1^2
+2[\lambda_1 x P_1+\lambda_3 P_2+\lambda_2(1-x)\ell+\lambda_4\ell]
\cdot k_2
+2\lambda_1\lambda_2 P_1\cdot \ell
-\lambda_1^2 P_1^2
+\lambda_4\Lambda^2
+i\varepsilon
\}^{6}},
\nonumber \\
\end{eqnarray}
where we have made the translation 
$k_1\to k_1-[(1-x)k_2-\lambda'\ell+\lambda P_1]$.

We carry out the $k_1$ and $k_2$ integrations 
in Eq.~(\ref{eq:F2ij-alpha-def-2}) by 
using the standard formulas for the virtual-gluon loop integration and 
the phase-space integration formula in 
Eq.~(\ref{eq:k1-phase-int-table}). The result is
\begin{eqnarray}
\mathcal{F}_{21}^{\alpha}
&=&
-
\frac{\mu^{4\epsilon}\Lambda^2\Gamma(6-\frac{D}{2})}{(4\pi)^\frac{D}{2}}
\int_0^\infty [d\lambda_i]
\int_0^1 dx
\left[-(1-x)F_{22}^{\alpha}+(\lambda_2\ell-\lambda_1 P_1)^\alpha F_{21}\right],
\nonumber \\
\mathcal{F}_{21}^{\alpha}
&=&
-
\frac{\mu^{4\epsilon}\Lambda^2\Gamma(6-\frac{D}{2})}{(4\pi)^\frac{D}{2}}
\int_0^\infty [d\lambda_i]
\int_0^1 dx\,
F_{22}^{\alpha},
\end{eqnarray}
where
\begin{eqnarray}
F_{21}
&=&
\frac{1}{(4\pi)^{\frac{D}{2}}e^{-i\pi(D-2)}}
\frac{\Gamma(8-\frac{3D}{2})\Gamma(\frac{D}{2}-1)}{\Gamma(6-\frac{D}{2})}
\frac{1}{(P'^2)^{\frac{D}{2}-1}}
\frac{1}{(M'^2-i\varepsilon)^{8-\frac{3D}{2}}},
\nonumber \\
F_{22}^{\alpha}
&=&
\frac{1}{(4\pi)^{\frac{D}{2}}e^{-i\pi(D-1)}}
\frac{\Gamma(7-\frac{3D}{2})\Gamma(\frac{D}{2})}{\Gamma(6-\frac{D}{2})}
\frac{P'^\alpha}{(P'^2)^{\frac{D}{2}}}
\frac{1}{(M'^2-i\varepsilon)^{7-\frac{3D}{2}}},
\end{eqnarray}
and
\begin{eqnarray}
P'
&\equiv&\lambda_1 x P_1+\lambda_3P_2
+\left[\lambda_2(1-x)+\lambda_4\right]\ell,
\nonumber \\
M'^2
&\equiv&\lambda_1^2P_1^2-2\lambda_1\lambda_2 P_1\cdot \ell 
-\lambda_4\Lambda^2.
\end{eqnarray}

Now we make a change of variables, replacing $\lambda_2$, $\lambda_3$, 
$\lambda_4$, and $x$ with
\begin{eqnarray}
\omega &\equiv& 2c \frac{\lambda_2}{\lambda_1},
\nonumber \\
\xi &\equiv& \frac{\lambda_3}{x\lambda_1},
\nonumber \\
\vartheta &\equiv& \frac{\lambda_4}{(1-x)\lambda_2},
\nonumber \\
t&\equiv&\frac{1-x}{x}.
\end{eqnarray}
Then, we can express the $\lambda_1$ integrations in terms of the beta
function by making use of the integral formula in
Eq.~(\ref{eq:ablam-general}) to obtain
\begin{eqnarray}
\mathcal{F}_{21}^{\alpha}
&=&
\frac{\left[\frac{\tilde\mu^2 P_1^2(2c)^2}{\Lambda^4}\right]^{2\epsilon}}
{2c(P_1^2)^2(4\pi)^4}
\frac{e^{2\epsilon\gamma_{{}_{\textrm{E}}}}\Gamma(2-\epsilon)\Gamma(-\epsilon)\Gamma(1+4\epsilon)}
{e^{-i\pi\epsilon}}
\int_0^\infty d\vartheta
\int_0^\infty d\xi
\int_0^\infty d\omega 
\int_0^\infty dt
\nonumber \\
&&
\times
\frac{
\omega^{-4\epsilon} t^{1-4\epsilon}\vartheta^{-1-4\epsilon}
\left(P_1+\xi P_2\right)^\alpha}
{(1+t)^{1-2\epsilon}
(\omega-1+i\varepsilon)^{-\epsilon}
\left[
\xi^2+2a\xi+1+\omega t\left(1+\vartheta\right)\left(1+\frac{d}{c}\xi\right)
\right]^{2-\epsilon}}
\nonumber \\
\nonumber \\
&&
+
\frac{\left[\frac{\tilde\mu^2 P_1^2(2c)^2}{\Lambda^4}\right]^{2\epsilon}}
{2c(P_1^2)^2(4\pi)^4}
\frac{e^{2\epsilon\gamma_{{}_{\textrm{E}}}}\Gamma^2(1-\epsilon)\Gamma(1+4\epsilon)}{e^{-i\pi\epsilon}}
\int_0^\infty d\vartheta
\int_0^\infty d\xi
\int_0^\infty d\omega 
\int_0^\infty dt
\nonumber \\
&&
\times
\frac{\omega^{-4\epsilon} t^{-4\epsilon}\vartheta^{-1-4\epsilon}
\left(P_1\right)^\alpha}
{(1+t)^{1-2\epsilon}
(\omega-1+i\varepsilon)^{1-\epsilon}
\left[
\xi^2+2a\xi+1+\omega t\left(1+\vartheta\right)\left(1+\frac{d}{c}\xi\right)
\right]^{1-\epsilon}},
\nonumber \\
\mathcal{F}_{22}^{\alpha}
&=&
-
\frac{\left[\frac{\tilde\mu^2 P_1^2(2c)^2}{\Lambda^4}\right]^{2\epsilon}}
{2c(P_1^2)^2(4\pi)^4}
\frac{e^{2\epsilon\gamma_{{}_{\textrm{E}}}}\Gamma(2-\epsilon)\Gamma(-\epsilon)\Gamma(1+4\epsilon)}
{e^{-i\pi\epsilon}}
\int_0^\infty d\vartheta
\int_0^\infty d\xi
\int_0^\infty d\omega 
\int_0^\infty dt
\nonumber \\
&&
\times
\frac{
\omega^{-4\epsilon} t^{-4\epsilon}\vartheta^{-1-4\epsilon}
\left(P_1+\xi P_2\right)^\alpha}
{(1+t)^{-2\epsilon}
(\omega-1+i\varepsilon)^{-\epsilon}
\left[
\xi^2+2a\xi+1+\omega t\left(1+\vartheta\right)\left(1+\frac{d}{c}\xi\right)
\right]^{2-\epsilon}}.
\end{eqnarray}
Again, we have dropped the terms that are proportional to
$\ell^\alpha$, as they vanish on contraction with the other factors in
Eq.~(\ref{eq:F1P1P2-complete-def}). Then, we carry out the parameter
integrations, except for the $\xi$ integration by making use of
Eq.~(\ref{plus-expansion}). Inserting the results into
Eq.~(\ref{eq:F2P1P2-complete-def}), we find that the $1/\epsilon^3$,
$1/\epsilon^2$, and $1/\epsilon$ contributions to
$\mathcal{F}_2^{P_1P_2}$ are given by the following expressions:
\begin{subequations}
\begin{eqnarray}
\label{eq:F2-p1p2-ep3}
\mathcal{F}_2^{P_1P_2}\Big|_{1/\epsilon^3}
&=&
-
\frac{g_s^4\left(\frac{\tilde{\mu}^2P_1^2}{\Lambda^4}\right)^{2\epsilon}}
{(4\pi)^4\epsilon^3}
\int_0^\infty d\xi
\bigg\{
\frac{(3ad-2c)\xi+2d-ac}{2c}
\left(-\frac{1}{AB}\right)
+\frac{(2ad) \xi+2ac}{2c}
\left(
\frac{3}{AB}
\right)
\bigg\},
\nonumber\\
\\
\label{eq:F2-p1p2-ep2}
\mathcal{F}_2^{P_1P_2}\Big|_{1/\epsilon^2}
&=&
-
\frac{g_s^4\left(\frac{\tilde{\mu}^2P_1^2}{\Lambda^4}\right)^{2\epsilon}}
{(4\pi)^4\epsilon^2}
\int_0^\infty d\xi
\bigg\{
\frac{(3ad-2c)\xi+2d-ac}{2c}
\left(
\frac{-2i\pi -4\log 2c+3\log A-4\log B}{AB}
\right)
\nonumber \\
&&
\quad\quad\quad\quad\quad\quad\quad\quad\quad
+\frac{(2ad) \xi+2ac}{2c}
\left(
\frac{4i\pi+12\log 2c-7\log A+10\log B}{AB}
\right)
\bigg\},
\end{eqnarray}
\begin{eqnarray}
\label{eq:F2-p1p2-ep1}
\mathcal{F}_2^{P_1P_2}\Big|_{1/\epsilon}
&=&
-
\frac{g_s^4\left(\frac{\tilde{\mu}^2P_1^2}{\Lambda^4}\right)^{2\epsilon}}
{(4\pi)^4\epsilon}
\int_0^\infty d\xi
\nonumber \\
&&\times
\Bigg\{
\frac{(3ad-2c)\xi+2d-ac}{2c}
\nonumber \\
&&
\quad\quad
\times
\Bigg(
\frac{1}{B(A-B)}
\bigg[
-2i\pi(\log A-\log B)
+\log^2 A+\log^2 B-2\log A\log B
\bigg]
\nonumber \\
&&
\quad\quad\quad\quad
+
\frac{1}{AB}
\bigg[
-\frac{3\pi^2}{2}
-8i\pi\log 2c
-8\log^22c
+2i\pi(3\log A-4\log B)
\nonumber \\
&&
\quad\quad\quad\quad\quad\quad
+4(\log 2c)(3\log A-4\log B)
-\frac{9}{2}\log^2 A
-8\log^2 B
+12\log A\log B
\bigg]
\Bigg)
\nonumber \\
&&
\quad\quad
+\frac{2d-ac}{2c}
\Bigg(
\frac{1}{A-B}
\bigg[
2i\pi(\log A-\log B)
-\log^2 A-\log^2 B+2\log A\log B
\bigg]
\Bigg)
\nonumber \\
&&
\quad\quad
+\frac{(2ad) \xi+2ac}{2c}
\Bigg(
\frac{1}{AB}
\bigg[
\frac{13\pi^2}{2}
+16i\pi\log 2c
+24\log^2 2c
-4i\pi(2\log A-3\log B)
\nonumber \\
&&
\quad\quad\quad\quad
-4(\log 2c)(7\log A-10\log B)
+\frac{15}{2}\log^2 A
+16\log^2 B
-22\log A\log B
\bigg]
\Bigg)
\Bigg\},
\nonumber \\
\end{eqnarray}
\end{subequations}
where $A$ and $B$ are defined in Eq.~(\ref{eq:def-of-A-B}).

\subsubsection{Result for the IR poles in 
$\mathcal{F}_1+\mathcal{F}_2$}
Now we compute the IR poles of various orders in $\epsilon$ in
$\mathcal{F}_1+\mathcal{F}_2$.

First, let us consider the $1/\epsilon^3$ poles. From
Eqs.~(\ref{eq:F1-p1p2-ep3}) and (\ref{eq:F2-p1p2-ep3}),
we find that 
\begin{equation}
\label{eq:F1-E2-P1P2-ep-3}
\mathcal{F}_1^{P_1P_2}
+
\mathcal{F}_2^{P_1P_2}\Big|_{1/\epsilon^{3}}
=
\frac{g_s^4}
{(4\pi)^4\epsilon^3}
\int_0^\infty   d\xi
\left[
-
\frac{(3ad+2c)\xi+(7ac-2d)}{2(\xi^2+2a\xi+1)(c+d\xi)}
+
\frac{(3ac+2d)\xi+(7ad-2c)}{2(\xi^2+2a\xi+1)(d+c\xi)}
\right],
\end{equation}
where we have made a change of variable $\xi\to1/\xi$ for 
$\mathcal{F}_1^{P_1P_2}\Big|_{1/\epsilon^{3}}$ 
and used the definitions of
$A$ and $B$ that are given in Eq.~(\ref{eq:def-of-A-B}).
Since the integrand in Eq.~(\ref{eq:F1-E2-P1P2-ep-3}) 
is antisymmetric under $P_1\leftrightarrow P_2$, or 
$c\leftrightarrow d$, we find that the triple poles cancel
after symmetrization under $P_1\leftrightarrow P_2$:
\begin{equation}
\mathcal{F}_1^{P_1P_2}
+
\mathcal{F}_2^{P_1P_2}
+
\mathcal{F}_1^{P_2P_1}
+
\mathcal{F}_2^{P_2P_1}\Big|_{1/\epsilon^{3}}
=0.
\end{equation}
The triple poles in $\mathcal{F}_1^{P_1P_1}+\mathcal{F}_2^{P_1P_1}$ and
$\mathcal{F}_1^{P_2P_2}+\mathcal{F}_2^{P_2P_2}$ cancel in a similar
fashion.

Next, let us consider the $1/\epsilon^2$ poles.
From Eqs.~(\ref{eq:F1-p1p2-ep2}) and (\ref{eq:F2-p1p2-ep2}),
we find that
\begin{eqnarray}
\mathcal{F}_1^{P_1P_2}+
\mathcal{F}_2^{P_1P_2}\Big|_{1/\epsilon^2}
&=&
\left(\mathcal{F}_1^{P_1P_2}+
\mathcal{F}_2^{P_1P_2}\Big|_{1/\epsilon^2}\right)_{AB}
+
\left(\mathcal{F}_1^{P_1P_2}+
\mathcal{F}_2^{P_1P_2}\Big|_{1/\epsilon^2}\right)_{A-B},
\end{eqnarray}
where
\begin{eqnarray}
\label{eq:AB-A-B-definition}
\left(\mathcal{F}_1^{P_1P_2}+
\mathcal{F}_2^{P_1P_2}\Big|_{1/\epsilon^2}\right)_{AB}
&\equiv&
-
\frac{g_s^4}
{(4\pi)^4\epsilon^2}
\int_0^\infty   d\xi
\nonumber \\
&&
\times
\bigg\{
\frac{
(3ad-2c)\xi+2d-ac
}{2c}
\left(
\frac{-2i\pi+6\log A-8\log B-8\log 2c}{AB}
\right)
\nonumber \\
&&
\quad
+
\frac{
(2ad) \xi+2ac
}{2c}
\left(
\frac{4i\pi-3\log A+4\log B+4\log 2c}{AB}
\right)
\bigg\},
\nonumber \\
\left(\mathcal{F}_1^{P_1P_2}+
\mathcal{F}_2^{P_1P_2}\Big|_{1/\epsilon^2}\right)_{A-B}
&\equiv&
-
\frac{g_s^4}
{(4\pi)^4\epsilon^2}
\int_0^\infty   d\xi
\bigg\{
\frac{
(3ad-2c)\xi+2d-ac
}{2c}
\left[
-\frac{\log A-\log B}{B(A-B)}
\right]
\nonumber \\
&&
\quad\quad\quad\quad\quad\quad\quad~
+
\frac{2d-ac}{2c}
\left[
\frac{\log A-\log B}{(A-B)}
\right]
\bigg\}.
\end{eqnarray}
We compute the combination $\left(\mathcal{F}_1^{P_1P_2}+
\mathcal{F}_2^{P_1P_2}\Big|_{1/\epsilon^2}\right)_{AB}$ by making use of
the same method that we used to compute the $1/\epsilon^{3}$
contributions. The result is
\begin{eqnarray}
\label{eq:AB-symp-trick}
&&
\left(\mathcal{F}_1^{P_1P_2}+
\mathcal{F}_2^{P_1P_2}\Big|_{1/\epsilon^2}\right)_{AB}
+
\left(\mathcal{F}_1^{P_2P_1}+
\mathcal{F}_2^{P_2P_1}\Big|_{1/\epsilon^2}\right)_{AB}
\nonumber \\
&=&
-
\frac{g_s^4}
{(4\pi)^4\epsilon^2}
\int_0^\infty d\xi
\left[
\frac{6i\pi a}{1+2a\xi+\xi^2}
+(c^2-2acd+d^2)\frac{2(1-\xi^2)\log\xi}
{(d+c\xi)(c+d\xi)(1+2a\xi+\xi^2)}
\right].\nonumber
\\
\end{eqnarray}
Also, one can show that
\begin{eqnarray}
&&
\left(\mathcal{F}_1^{P_1P_2}+
\mathcal{F}_2^{P_1P_2}\Big|_{1/\epsilon^2}\right)_{A-B}
+
\left(\mathcal{F}_1^{P_2P_1}+
\mathcal{F}_2^{P_2P_1}\Big|_{1/\epsilon^2}\right)_{A-B}
\nonumber \\
&=&
-
\frac{g_s^4}
{(4\pi)^4\epsilon^2}
(c^2-2acd+d^2)
\int_0^\infty   d\xi
\bigg[
\frac{\log(1+2a\xi+\xi^2)-\log\left(d+c\xi\right)+\log d}
{(d+c\xi)(2ad-c+d\xi)}
\nonumber \\
&&
\quad\quad\quad\quad\quad\quad\quad\quad\quad\quad\quad\quad\quad\quad
+\frac{\log(1+2a\xi+\xi^2)-\log\left(c+d\xi\right)+\log c}{(c+d\xi)(2ac-d+c\xi)}
\bigg].
\end{eqnarray}
The $\xi$ integrations can be carried out through a straightforward, but
tedious process, by partial-fractioning the denominators and factoring 
the arguments of the logarithms. We obtain a lengthy result, which we do not
reproduce here, that contains logarithms and dilogarithms. This result
can be greatly simplified by making use of the polylogarithm
identities in Eq.~(\ref{eq:dilog-identity}) to obtain
\begin{eqnarray}
\mathcal{F}_1^{P_1P_2}+
\mathcal{F}_2^{P_1P_2}
+\mathcal{F}_1^{P_2P_1}+
\mathcal{F}_2^{P_2P_1}
\Big|_{1/\epsilon^2}
&=&
\frac{g_s^4}
{(4\pi)^4\epsilon^2}
\left[
-6i\pi\times\frac{a\log(a+\sqrt{a^2-1})}{\sqrt{a^2-1}}
\right].
\end{eqnarray}
This expression contains no real IR poles.

Next, let us consider the $1/\epsilon$ contribution.
From Eqs.~(\ref{eq:F1-p1p2-ep1}) and
(\ref{eq:F2-p1p2-ep1}), we find that
\begin{equation}
\label{eq:F1F2-intermsof-AB-A-B-Li2}
\mathcal{F}_1^{P_1P_2}+
\mathcal{F}_2^{P_1P_2}
\Big|_{1/\epsilon}
=
\left(
\mathcal{F}_1^{P_1P_2}+
\mathcal{F}_2^{P_1P_2}
\Big|_{1/\epsilon}
\right)_{AB}
+
\left(
\mathcal{F}_1^{P_1P_2}+
\mathcal{F}_2^{P_1P_2}
\Big|_{1/\epsilon}
\right)_{A-B}
+
\left(
\mathcal{F}_1^{P_1P_2}+
\mathcal{F}_2^{P_1P_2}
\Big|_{1/\epsilon}
\right)_{\textrm{Li}_2},
\end{equation}
where
\begin{eqnarray}
\frac{\left(
\mathcal{F}_1^{P_1P_2}+
\mathcal{F}_2^{P_1P_2}
\Big|_{1/\epsilon}
\right)_{AB}}
{\frac{g_s^4}
{(4\pi)^4\epsilon}}
&=&
\int_0^\infty
d\xi
\frac{a}{A}
\bigg[
\pi^2
-12i\pi\log 2c
+i\pi(5\log A-8\log B)
\bigg]
\nonumber \\
&&
+
\int_0^\infty
d\xi
\left(
\frac{2a-\frac{c}{d}-\frac{d}{c}}{AB}
-
\frac{a-\frac{c}{d}}{A}
\right)
\nonumber \\
&&
\quad\quad\quad\quad
\times
\bigg[
-\frac{11\pi^2}{3}
-8i\pi\log 2c
-16\log^22c
\nonumber \\
&&
\quad\quad\quad\quad\quad~
+(2i\pi+8\log 2c)(3\log A-4\log B)
\nonumber \\
&&
\quad\quad\quad\quad\quad~
-9\log^2 A
-14\log^2 B
+24\log A\log B
\bigg],
\nonumber \\
\frac{\left(
\mathcal{F}_1^{P_1P_2}+
\mathcal{F}_2^{P_1P_2}
\Big|_{1/\epsilon}
\right)_{A-B}}{
\frac{g_s^4}{(4\pi)^4\epsilon}}
&=&
\int_0^\infty d\xi
\frac{\left(2a-\frac{c}{d}-\frac{d}{c}\right)(1-B)}{B(A-B)}
\nonumber \\
&&
\quad\quad\quad
\times
\bigg[
-(2i\pi+4\log2c)(\log A-\log B)
\nonumber\\
&&
\quad\quad\quad\quad~
+\frac{3}{2}\log^2 A+\frac{5}{2}\log^2 B-4\log A\log B
\bigg]
\nonumber \\
&&
+
\int_0^\infty 
d\xi
\left(
\frac{2a-\frac{c}{d}-\frac{d}{c}}{AB}
-
\frac{a-\frac{c}{d}}{A}
\right)
\nonumber \\
&&
\quad\quad\quad\quad
\times
\bigg[
2\log^2(A-B)
-4\log B\log(A-B)
\bigg],
\nonumber \\
\frac{
\left(
\mathcal{F}_1^{P_1P_2}+
\mathcal{F}_2^{P_1P_2}
\Big|_{1/\epsilon}
\right)_{\textrm{Li}_2}}
{\frac{g_s^4}{(4\pi)^4\epsilon}}
&=&
\int_0^\infty d\xi
\left(
\frac{2a-\frac{c}{d}-\frac{d}{c}}{AB}
-
\frac{a-\frac{c}{d}}{A}
\right)
\left[
4\textrm{Li}_2\left(-\frac{B}{A-B}\right)
\right].
\end{eqnarray}

Taking only the real parts, we obtain
\begin{eqnarray}
\frac{\textrm{Re}\left(
\mathcal{F}_1^{P_1P_2}+
\mathcal{F}_2^{P_1P_2}
\Big|_{1/\epsilon}
\right)}
{\frac{g_s^4}
{(4\pi)^4\epsilon}}
&=&
\int_0^\infty
d\xi
\frac{a}{A}
\pi^2
\nonumber \\
&&
+
\int_0^\infty
d\xi
\left(
\frac{2a-\frac{c}{d}-\frac{d}{c}}{AB}
-
\frac{a-\frac{c}{d}}{A}
\right)
\nonumber \\
&&
\quad\quad\quad\quad
\times
\bigg[
-\frac{11\pi^2}{3}
-16\log^22c
+(8\log 2c)(3\log A-4\log B)
\nonumber \\
&&
\quad\quad\quad\quad\quad~
-9\log^2 A
-14\log^2 B
+24\log A\log B
\nonumber \\
&&
\quad\quad\quad\quad\quad~
+2\log^2(A-B)
-4\log B\log(A-B)
+
4\textrm{Li}_2\left(-\frac{B}{A-B}\right)
\bigg]
\nonumber \\
&&
+
\int_0^\infty d\xi
\frac{\left(2a-\frac{c}{d}-\frac{d}{c}\right)(1-B)}{B(A-B)}
\bigg[
-(4\log2c)(\log A-\log B)
\nonumber\\
&&
\quad\quad\quad\quad\quad\quad\quad
\quad\quad\quad\quad\quad\quad\quad
+\frac{3}{2}\log^2 A+\frac{5}{2}\log^2 B-4\log A\log B
\bigg].
\nonumber \\
\end{eqnarray}
For the contribution that contains a dilogarithm in the integrand, we
eliminate the dilogarithm by integrating by parts. Then, the $\xi$
integrations can again be carried out by partial-fractioning the
denominators and factoring the arguments of the logarithms. This leads
to an expression, which we do not reproduce here, that is hundreds of
terms long and contains logarithms, dilogarithms, and trilogarithms. This
result can be reduced, by making use of the polylogarithm identities in
Eq.~(\ref{eq:dilog-identity}) and symmetrizing under $P_1\leftrightarrow
P_2$ to obtain a remarkably simple form:
\begin{eqnarray}
\textrm{Re}\left(
\mathcal{F}_1^{P_1P_2}+
\mathcal{F}_2^{P_1P_2}
+\mathcal{F}_1^{P_2P_1}+
\mathcal{F}_2^{P_2P_1}
\Big|_{1/\epsilon}
\right)
&=&
\frac{g_s^4a}
{(4\pi)^4\epsilon}
\int_0^\infty
d\xi
\frac{2\pi^2}{A}
\nonumber \\
&=&
\frac{g_s^4}
{(4\pi)^4}
\frac{2\pi^2}{\epsilon_\textrm{IR}}
\frac{a\log\left(a+\sqrt{a^2-1}\right)}{\sqrt{a^2-1}}.
\end{eqnarray}
Since this result shows no dependence on the UV cutoff, we conclude
that it is entirely IR in nature. We have confirmed this conclusion
by carrying out a calculation of the $k_1$ integration for
$\mathcal{F}$ using light-cone methods.

We can find all the other $\mathcal{F}_i^{P_j P_k}$ 
contributions from the relations
\begin{eqnarray}
\label{eq:F1F2-PjPk-symm}
\mathcal{F}_1^{P_2 P_1}+\mathcal{F}_2^{P_2 P_1}
&=&
\left(
\mathcal{F}_1^{P_1 P_2}+\mathcal{F}_2^{P_1 P_2}
\right)
\bigg|_{c\leftrightarrow d},
\nonumber \\
\mathcal{F}_1^{P_1 P_1}+\mathcal{F}_2^{P_1 P_1}
&=&
-
\left(
\mathcal{F}_1^{P_1 P_2}+\mathcal{F}_2^{P_1 P_2}
\right)
\bigg|_{d\to c,\,a\to1},
\nonumber \\
\mathcal{F}_1^{P_2 P_2}+\mathcal{F}_2^{P_2 P_2}
&=&
-
\left(
\mathcal{F}_1^{P_1 P_2}+\mathcal{F}_2^{P_1 P_2}
\right)
\bigg|_{c\to d,\,a\to1}.
\end{eqnarray}
Taking into account the color factors 
in Eq.~(\ref{eq:color-F}), we obtain
\begin{eqnarray}
\label{eq:F-final-with-color}
&&
\textrm{Re}\left(\sum_{i=1}^2
\sum_{j=1}^2
\sum_{k=1}^2
C_{\mathcal{F}_i^{P_jP_k}}
\mathcal{F}_i^{P_jP_k}\right)
\nonumber \\
&=&
\left[
\frac{g_s^4}
{(4\pi)^{4}}
\frac{N_c(N_c^2-1)}{4N_c}
\frac{2\pi^2}{\epsilon_\textrm{IR}}
\left(
1-
\frac{a\log\left(a+\sqrt{a^2-1}\right)}{\sqrt{a^2-1}}
\right)
+O(\epsilon^0)\right]_{\textrm{IR}}.
\nonumber \\
\end{eqnarray}

\section{Summary of results\label{sec:summary}}
Now let us summarize the results of our calculations.

From Eq.~(\ref{eq:color-A}) we have
\begin{eqnarray}
\mathcal{A}
\equiv
\sum_{i=1}^{3} \sum_{j=1}^{2} \sum_{k=1}^{2}
C_{\mathcal{A}_i^{P_jP_k}}
\mathcal{A}_{i}^{P_jP_k}
=0,
\end{eqnarray}
and from Eqs.~(\ref{eq:color-B}) and
(\ref{eq:B-final-sum-before-real}) we have
\begin{eqnarray}
\mathcal{B}
\equiv
\sum_{i=1}^{3} \sum_{j=1}^{2} \sum_{k=1}^{2}
C_{\mathcal{B}_i^{P_jP_k}}
\mathcal{B}_{i}^{P_j P_k}
=O(\epsilon^0).
\end{eqnarray} 
That is, there are no IR divergences in the ladder diagrams.

From Eqs.~(\ref{eq:C12-final-with-color})
(\ref{eq:D12-final-with-color}), (\ref{eq:E-final-with-color}), and 
(\ref{eq:F-final-with-color}), we have
\begin{eqnarray}
\mathcal{C}
&\equiv&
2\textrm{Re}
\left(
\sum_{i=1}^2
\sum_{j=1}^2\sum_{k=1}^2
C_{\mathcal{C}_i^{P_jP_k}}\mathcal{C}_i^{P_j P_k}
\right)
\nonumber \\
&=&
-
\left[
\frac{\alpha_s^2}{(4\pi)^2}
\frac{N_c(N_c^2-1)}{4N_c}
\frac{4\pi^2}{\epsilon_\text{IR}}
\left(
1-\frac{a\log\left(a+\sqrt{a^2-1}\right)}{\sqrt{a^2-1}}
\right)
+O(\epsilon^0)
\right]_\textrm{IR}
\nonumber \\
&&
-
\left[
\frac{\alpha_s}{4\pi}N_c
\left(
\frac{1}{\epsilon^2_\textrm{UV}}
+\frac{3\pi^2}{4}
+\frac{\log\left[\frac{\tilde\mu^2P_1^2(4cd)}{\Lambda^4}\right]}
{\epsilon_\textrm{UV}}
+\frac{\log^2\left[\frac{\tilde\mu^2P_1^2(2c)^2}{\Lambda^4}\right]
+\log^2\left[\frac{\tilde\mu^2P_1^2(2d)^2}{\Lambda^4}\right]}{4}
\right)
+{O}(\epsilon)
\right]_\textrm{UV}
\nonumber \\
&&
\times
\left[
\frac{\alpha_s}{4\pi}
\frac{N_c^2-1}{4N_c}
\frac{4}{\epsilon_\text{IR}}
\left(
1-\frac{a\log\left(a+\sqrt{a^2-1}\right)}{\sqrt{a^2-1}}
\right)
+O(\epsilon^0)
\right]_{\text{IR}},
\\
\mathcal{D}
&\equiv&
2\textrm{Re}
\left(
\sum_{i=1}^2
\sum_{j=1}^2\sum_{k=1}^2
C_{\mathcal{D}_i^{P_jP_k}}
\mathcal{D}_i^{P_j P_k}
\right)
\nonumber \\
&=&
-
\left[
\frac{\alpha_s}{4\pi}N_c
\left(
\frac{1}{\epsilon^2_\textrm{UV}}
+\frac{3\pi^2}{4}
+\frac{\log\left[\frac{\tilde\mu^2P_1^2(4cd)}{\Lambda^4}\right]}
{\epsilon_\textrm{UV}}
+\frac{\log^2\left[\frac{\tilde\mu^2P_1^2(2c)^{2}}{\Lambda^4}\right]
+\log^2\left[\frac{\tilde\mu^2P_1^2(2d)^{2}}{\Lambda^4}\right]}{4}
\right)
+O(\epsilon)
\right]_{\textrm{UV}}
\nonumber \\
&&
\times
\left[
\frac{\alpha_s}{4\pi}
\frac{N_c^2-1}{4N_c}
\frac{4}{\epsilon_\text{IR}}
\left(
1-\frac{a\log\left(a+\sqrt{a^2-1}\right)}{\sqrt{a^2-1}}
\right)
+O(\epsilon^0)
\right]_{\textrm{IR}},
\\
\mathcal{E}
&\equiv&
2
\textrm{Re}
\left(
\sum_{i=1}^2
\sum_{j=1}^2
\sum_{k=1}^2
C_{\mathcal{E}_i^{P_jP_k}}
\mathcal{E}_i^{P_jP_k}
\right)
\nonumber \\
&=&
-
\left[
\frac{\alpha_s^2}
{(4\pi)^{2}}
\frac{N_c(N_c^2-1)}{4N_c}
\frac{4\pi^2}{\epsilon_\textrm{IR}}
\left(
1-
\frac{a\log\left(a+\sqrt{a^2-1}\right)}{\sqrt{a^2-1}}
\right)
+O(\epsilon^0)\right]_\textrm{IR},
\phantom{XX}
\\
\mathcal{F}
&\equiv&
2\textrm{Re}
\left(
\sum_{i=1}^2
\sum_{j=1}^2
\sum_{k=1}^2
C_{\mathcal{F}_i^{P_jP_k}}
\mathcal{F}_i^{P_jP_k}
\right)
\nonumber \\
&=&
\left[
\frac{\alpha_s^2}
{(4\pi)^{2}}
\frac{N_c(N_c^2-1)}{4N_c}
\frac{4\pi^2}{\epsilon_\textrm{IR}}
\left(
1-
\frac{a\log\left(a+\sqrt{a^2-1}\right)}{\sqrt{a^2-1}}
\right)
+O(\epsilon^0)\right]_{\textrm{IR}},
\end{eqnarray}
where we have used $g_s^2 = 4\pi\alpha_s$. We remind the reader that
the subscripts IR and UV on the brackets indicate the origins of
the contributions and that the factors labeled IR are
proportional to the one-loop
contribution to the LDME, which is computed in 
Sec.~\ref{app:1-loop}.
The expression for $\mathcal{C}$ contains
two contributions: (i) a UV factor times an IR factor, which is a
one-loop contribution to an SDC times the one-loop IR-divergent
contribution to the LDME; (ii) a two-loop IR-divergent contribution
to the LDME. The expression for $\mathcal{D}$ contains a UV factor times
an IR factor, which is a one-loop contribution to an SDC times the
one-loop IR-divergent contribution to the LDME. The expressions for
$\mathcal{E}$ and for the non-Abelian contribution $\mathcal{F}$ are
both two-loop IR-divergent contributions to the LDME. Note that all of
the IR-divergent contributions to the LDME are independent of the
direction of the Wilson line and, so, are consistent with the NRQCD
factorization conjecture. All of the two-loop IR-divergent contributions
to the LDME are of the same form, and their sum is nonzero.

\section{Comparison with the results of NQS\label{sec:comparison}}

\subsection{The NQS calculation and its correspondence to our 
calculation}

In NQS, the integration over the minus component of 
the loop momentum of a virtual gluon is carried out by closing the 
integration contour in the complex plane and using Cauchy's theorem. 
Some of the pole residues cancel against contributions in which the 
virtual gluon is replaced with a real gluon, and those contributions 
are not calculated. As we will see below, this cancellation is not exact 
because of sensitivity of the real-gluon contribution to the UV regulator. 

In the calculations in Ref.~\cite{Nayak:2005rt}, a velocity expansion is
made for the interactions of the gluons with the $Q$ and $\bar Q$ lines.
This expansion mixes the contributions of some of the diagrams that
appear in our calculations. Consequently, there is not a one-to-one
correspondence between the diagrams in NQS and our diagrams. However,
the following correspondences hold.
\begin{eqnarray}
I+II &\leftrightarrow& 
\mathcal{A}+ \mathcal{B},\nonumber\\
IV+V+VI &
\leftrightarrow& \mathcal{C}+ \mathcal{D}+ \mathcal{E},\nonumber\\
III&\leftrightarrow& \mathcal{F}.
\end{eqnarray}
Here, the Roman numerals on the left sides of the correspondences are 
the notations for the diagrams in  Ref.~\cite{Nayak:2005rt}.

\subsection{Comparison}

Our result for the sum of the ladder and crossed ladder diagrams
$\mathcal{A}+\mathcal{B}$ contains no IR poles, which is in agreement
with the order-$v^2$ results for $\textrm{I}+\textrm{II}$ in
Ref.~\cite{Nayak:2005rt}.

Our result for the non-Abelian diagrams $\mathcal{F}$ is in agreement
with the corresponding result in Ref.~\cite{Nayak:2006fm}, except for an
overall sign. The result for the non-Abelian diagrams in
Ref.~\cite{Nayak:2006fm}, when expanded to order $v^2$, is identical to
the result for $III$ in Ref.~\cite{Nayak:2005rt}.

Our result for the Abelian diagrams that involve one interaction on the
Wilson line is contained in $\mathcal{C}+ \mathcal{D}+ \mathcal{E}$. In
order to compare this result with the calculation of $IV+V+VI$ in
Ref.~\cite{Nayak:2005rt}, we need to complete some of the calculations
in NQS, which were left in the form of integrals in
Ref.~\cite{Nayak:2005rt}. We have carried out these calculations in
Appendix~\ref{app:NQS-calc}, correcting some typographical errors
in the expressions in
Ref.~\cite{Nayak:2005rt}, inserting missing color factors, and
correcting the overall signs.  Our results, from
Eqs.~(\ref{eq:NQS-final-IV}), (\ref{eq:NQS-V-final}), and
(\ref{eq:VIA-Eq-58-final-q3}), are
\begin{eqnarray}
IV^{(k_2^0)}
&=&
-
\left(\frac{\alpha_s}{\pi}\right)^2
(N_c^2-1)
(2\pi)^{2\epsilon}
\Gamma(1+\epsilon)
\left[
-i\pi-\pi^2\epsilon + O(\epsilon^2)
\right]
\int_0^\Lambda \frac{dk_1^+}{(k_1^+)^{1+4\epsilon}}
\nonumber \\
&&
\times
\left[\bm{q}^2
-\epsilon\left(2q_\perp^2+\gamma_{{}_{\textrm{E}}}\bm{q}^2\right)
+O(\epsilon^2)\right],
\nonumber \\
V^{(k_2^0)}
&=&
-
\left(\frac{\alpha_s}{\pi}\right)^2
(N_c^2-1)
(2\pi)^{2\epsilon}\Gamma(1+\epsilon)
\left[
-i\pi-\pi^2\epsilon + O(\epsilon^2)
\right]
\int_0^\Lambda \frac{dk_1^+}{(k_1^+)^{1+4\epsilon}}
\nonumber \\
&&
\times
\left(
\frac{1}{2\epsilon_\textrm{UV}}
+1
\right)
\left[
-\frac{2}{3}\bm{q}^2
+O(\epsilon)
\right],
\nonumber \\
VI^{(k_2^0,k_2^0-k_1^0)}
&=&
-
\left(\frac{\alpha_s}{\pi}\right)^2
(N_c^2-1)
(2\pi)^{2\epsilon}
\Gamma(1+\epsilon)
\left[
-i\pi-\pi^2\epsilon
+O(\epsilon^2)
\right]
\int_0^\Lambda \frac{dk_1^+}{(k_1^+)^{1+4\epsilon}}
\nonumber \\
&&
\times
\left[-2\epsilon q_3^2+O(\epsilon^2)\right].
\end{eqnarray}
The integral over $k_1^+$ yields an infrared pole:
\begin{equation}
\int_0^\Lambda \frac{dk_1^+}{(k_1^+)^{1+4\epsilon}}  
\sim \frac{-1}{4\epsilon_{\textrm{IR}}}.
\end{equation}
Note that the sum of diagrams IV and VI is 
rotationally invariant:
\begin{eqnarray}
IV^{(k_2^0)}
+
VI^{(k_2^0,k_2^0-k_1^0)}
&=&
-
\left(\frac{\alpha_s}{\pi}\right)^2
(N_c^2-1)
(2\pi)^{2\epsilon}
\Gamma(1+\epsilon)
\left[
-i\pi-\pi^2\epsilon + O(\epsilon^2)
\right]
\int_0^\Lambda \frac{dk_1^+}{(k_1^+)^{1+4\epsilon}}
\nonumber \\
&&
\times
\bm{q}^2
\left[1
-\epsilon\left(2+\gamma_{{}_{\textrm{E}}}\right)
+O(\epsilon^2)\right].
\end{eqnarray}
The real IR pole comes only from diagram V:
\begin{equation}
2\textrm{Re}
\left(V^{(k_2^0)}\right)
=
-
\frac{\alpha_s^2
(N_c^2-1)}{\epsilon_\textrm{IR}}
\left(-\frac{\bm{q}^2}{6}\right).
\label{eq:NQS-V}
\end{equation}
We note that, in NQS, this contribution is dropped because it is
regarded as a one-loop correction to an SDC, times the one-loop
correction to the LDME. However, in Appendix~\ref{app:diagram-V}, we
have checked this assignment of the contribution of diagram V and
conclude that it is a two-loop IR-divergent contribution to the LDME.

In order to compare our results with those in NQS, we expand our results
in Sec.~\ref{sec:summary} through order $v^2$ (order $\bm{q}^2$) to
obtain
\begin{eqnarray}
\label{eq:expanded-C-D-E}%
\mathcal{C}&=&
-
\left[
\frac{\alpha_s^2}{(4\pi)^2}
\frac{N_c(N_c^2-1)}{4N_c}
\left(-
\frac{16\pi^2}{3\epsilon_\text{IR}}\bm{q}^2
\right)
+O(\epsilon^0,\bm{q}^4)
\right]_\textrm{IR}
\nonumber \\
&&
-
\left[
\frac{\alpha_s}{4\pi}
N_c
\left(
\frac{1}{\epsilon^2_\textrm{UV}}
+\frac{3\pi^2}{4}
+\frac{\log\left[\frac{\tilde\mu^2P_1^2(4cd)}{\Lambda^4}\right]}
{\epsilon_\textrm{UV}}
+\frac{\log^2\left[\frac{\tilde\mu^2P_1^2(2c)^2}{\Lambda^4}\right]
+\log^2\left[\frac{\tilde\mu^2P_1^2(2d)^2}{\Lambda^4}\right]}{4}
\right)
+{O}(\epsilon)
\right]_\textrm{UV}
\nonumber \\
&&
\times
\left[
\frac{\alpha_s}{4\pi}
\frac{N_c^2-1}{4N_c}
\left(
-\frac{16}{3\epsilon_\text{IR}}\bm{q}^2
\right)
+O(\epsilon^0,\bm{q}^4)
\right]_{\text{IR}},
\nonumber \\
\mathcal{D}
&=&
-
\left[
\frac{\alpha_s}{4\pi}N_c
\left(
\frac{1}{\epsilon^2_\textrm{UV}}
+\frac{3\pi^2}{4}
+\frac{\log\left[\frac{\tilde\mu^2P_1^2(4cd)}{\Lambda^4}\right]}
{\epsilon_\textrm{UV}}
+\frac{\log^2\left[\frac{\tilde\mu^2P_1^2(2c)^{2}}{\Lambda^4}\right]
+\log^2\left[\frac{\tilde\mu^2P_1^2(2d)^{2}}{\Lambda^4}\right]}{4}
\right)
+O(\epsilon)
\right]_{\textrm{UV}}
\nonumber \\
&&
\times
\left[
\frac{\alpha_s}{4\pi}
\frac{N_c^2-1}{4N_c}
\left(
-\frac{16}{3\epsilon_\text{IR}}\bm{q}^2
\right)
+O(\epsilon^0,\bm{q}^4)
\right]_{\textrm{IR}},
\nonumber \\
\mathcal{E}
&=&
-
\left[
\frac{\alpha_s^2}
{(4\pi)^{2}}
\frac{N_c(N_c^2-1)}{4N_c}
\left(
-
\frac{16\pi^2}{3\epsilon_\text{IR}}\bm{q}^2
\right)
+O(\epsilon^0,\bm{q}^4)\right]_\textrm{IR},
\end{eqnarray}
where, following NQS, we have normalized the quark mass $m_Q$ to unity.

There are several differences between our results and those in NQS.

First, there are UV double poles in $\mathcal{C}$ and  $\mathcal{D}$ in
our calculation that are not present in the NQS result. In
Appendix~\ref{app:source-disc}, we demonstrate these UV double poles
appear in an NQS-style calculation because the following two quantities
cancel incompletely: (i) the double-real-gluon contribution $VB$ and (ii)
the part of the real-virtual-gluon contribution $VA$ that comes from the
residue of the pole in the virtual-gluon propagator. These contributions
cancel in the IR, but they yield a net nonzero contribution in the UV
because of a UV-regulator mismatch: Both real gluons have a UV
phase-space regulator in $VB$, while only the real gluon has a UV
phase-space regulator in $VA$. In the calculation in NQS, this 
contribution from the UV-regulator mismatch was discarded.

Of course, the appearance of these UV contributions  from a
UV-regulator mismatch is dependent on the choice of UV
regulator, and the results that we have obtained are specific to our
standard phase-space UV regulator. In any case, such UV-regulator
dependences can always be absorbed into an SDC, and so they do not bear
on the issue of the dependence of the LDME on the Wilson-line 
direction.

The single and double logarithms in $\mathcal{C}$ and  $\mathcal{D}$,
which are also UV in origin, do not appear in the NQS calculation. The
sum of the constant terms $3\pi^2/4$ in $\mathcal{C}$ and $3\pi^2/4$ in
$\mathcal{D}$, which are UV in origin, does not agree with the
coefficient of the IR pole in $V$ [Eq.~(\ref{eq:NQS-V})]. As we have
said, this coefficient is regarded in NQS as being UV in origin, but we
believe it to be IR in origin. It is plausible that these
discrepancies might also be removed once the UV-regulator mismatch has
been taken into account fully in the NQS calculation, but we have not
checked this. In any case, the UV factors, some of which are $\ell$
dependent, are consistent with NRQCD factorization because they can be interpreted as one-loop contributions to
an SDC.

UV contributions from the UV-regulator mismatch could, in principle,
occur in individual contributions from the non-Abelian diagrams in the
calculation in NQS. There could be a residual contribution from the
mismatch between the double-real-gluon contribution and
real-virtual-gluon contribution that comes from the residue of the pole
in the virtual-gluon propagator that is labeled as $k_{1[(k_1-k_2)^2]}$
in NQS. There could also be a residual contribution from the residue of
the pole in the virtual-gluon propagator that is labeled as
$k_{1[k_1^2]}$ in NQS. That residue is discarded in NQS because of the
antisymmetry of the integrand under the interchanges
$P_1\leftrightarrow P_2$, $k_1^+\leftrightarrow k_2^+$, and
$k_{1\perp}\leftrightarrow k_{2\perp}$. However, that symmetry is
violated by the UV phase-space regulator. Evidently, such residual
contributions, if they are present, cancel in the final result, since we
find no UV contributions in the coefficient of the IR pole in our
covariant calculations.

Finally, let us consider the Abelian two-loop IR-divergent contributions,
which come from diagrams $\mathcal{C}$ and $\mathcal{E}$
in our calculation. The sum of these contributions is equal to the
contribution in $V$ in the NQS calculation, and so our result for the
two-loop IR-divergent contribution agrees with that in NQS, provided
that we interpret $V$ as being a two-loop IR-divergent contribution.

\section{Conclusions \label{sec:conclusions}}

The central issue in establishing the NRQCD factorization conjecture for
inclusive quarkonium production is the question of the universality of
the NRQCD LDMEs. Universality of the
LDMEs requires that any IR divergences that they contain be independent
of the direction of the Wilson line which makes the LDMEs gauge
invariant.

In order to test for a possible dependence on the Wilson-line direction,
we have used covariant methods to carry out a two-loop calculation of
the NRQCD LDME for a heavy $Q\bar Q$ pair in an $S$-wave, color-octet
state to evolve into a color-singlet state. Our calculation provides a
check of a previous two-loop calculation by Nayak, Qiu, and Sterman in
Refs.~\cite{Nayak:2005rw,Nayak:2005rt,Nayak:2006fm}, who used light-cone
methods to carry out their calculation. Although our results differ from
those of Nayak, Qiu, and Sterman in several respects, we find, as did
they, that the LDME is independent of the direction of the Wilson line.

One might hope that a covariant calculation would reveal simplifications
in comparison with the rather complicated light-cone calculation. That
did not prove to be the case in our calculation, probably because, for the
non-Abelian diagrams, we did not implement the unitarity cancellations
between real- and virtual-gluon corrections at the integrand level.
Consequently, we had to deal with poles in the
dimensional-regularization parameter $\epsilon$ of orders $1/\epsilon^3$,
$1/\epsilon^2$, which canceled, in order to obtain a final result of
order $1/\epsilon$. Furthermore, in the case of the order-$1/\epsilon$
contribution, our intermediate expressions contained hundreds of terms
involving dilogarithms and trilogarithms, which ultimately canceled to
yield a very simple expression.

Our results for the non-Abelian diagrams $\mathcal{F}$ agree with those
in Refs.~\cite{Nayak:2005rw,Nayak:2005rt,Nayak:2006fm}, up to an overall
sign, and are independent of the Wilson line direction. Our results for
the Abelian diagrams $\mathcal{A}$ and $\mathcal{B}$ that involve two
gluon connections on the Wilson line do not contribute an IR pole, in
agreement with the calculations in Ref.~\cite{Nayak:2005rt}. However,
our results for the Abelian diagrams $\mathcal{C}$, $\mathcal{D}$, and
$\mathcal{E}$ that involve one gluon connection on the Wilson line
differ in some respects from those in Ref.~\cite{Nayak:2005rt}, 
although both our
results and those in Ref.~\cite{Nayak:2005rt} are consistent with the 
NRQCD factorization conjecture.
In our case, we find dependences on the Wilson-line direction in some
contributions. However, these contributions  can be factored into a
one-loop contribution to a short-distance coefficient times the one-loop
contribution to the LDME, the latter of which is independent of the
Wilson-line direction.

We have identified one source of the differences between our calculation 
and that in Ref.~\cite{Nayak:2005rt}. In Ref.~\cite{Nayak:2005rt}, 
a double-real-gluon contribution is canceled against a particular 
real-virtual-gluon contribution that comes from the residue of the 
pole in the virtual-gluon propagator. That cancellation is exact in the 
IR limit, but leaves residual UV contributions from the virtual-gluon 
loop because, unlike the corresponding real-gluon loop, the 
virtual-gluon loop is not constrained by phase space in the UV. These 
residual contributions account for UV double poles that are present in 
our result but not in the result in Ref.~\cite{Nayak:2005rt}. They may 
also account for UV single poles and UV constant terms that are present in 
our results, but not in the results in Ref.~\cite{Nayak:2005rt}, 
although we have not checked this explicitly.

In our result, we find a two-loop IR-divergent contribution to the
LDME that arises from the Abelian diagrams $\mathcal{C}$ and
$\mathcal{E}$. This contribution is not present in the result for the
Abelian diagrams in Ref.~\cite{Nayak:2005rt}. However, if we reinterpret
the contribution from diagram V in Ref.~\cite{Nayak:2005rt} as a
two-loop IR-divergent contribution to the LDME, rather than as a
one-loop contribution to the SDC times the one-loop contribution to the
LDME, as is done in Ref.~\cite{Nayak:2005rt}, then we find agreement
with Ref.~\cite{Nayak:2005rt} for the total two-loop IR-divergent
contribution to the LDME.

While our result confirms the NRQCD factorization conjecture through two 
loops, the principle that underlies that result has not emerged, and its 
elucidation remains a challenge for future work.

\begin{acknowledgments}

We thank Jianwei Qiu and George Sterman for many helpful discussions.
The work of G.T.B.\ is supported by the U.S.\ Department of Energy,
Division of High Energy Physics, under Contract No.\ DE-AC02-06CH11357.
The work of H.S.C.\ at CERN is supported by the Korean Research
Foundation (KRF) through the CERN-Korea fellowship program. 
The work of J.-H.E. and J.L. is supported by the National Research 
Foundation of Korea (NRF) under Contract No. NRF-2017R1E1A1A01074699
and NRF-2020R1A2C3009918.
The work of U-R.K. is supported by the National Research Foundation of 
Korea (NRF) under Contract No. NRF-2019R1A6A3A01096460. The submitted
manuscript has been created in part by UChicago Argonne, LLC, Operator
of Argonne National Laboratory. Argonne, a U.S.\ Department of Energy
Office of Science laboratory, is operated under Contract No.\
DE-AC02-06CH11357. The U.S.\ Government retains for itself, and others
acting on its behalf, a paid-up nonexclusive, irrevocable worldwide
license in said article to reproduce, prepare derivative works,
distribute copies to the public, and perform publicly and display
publicly, by or on behalf of the Government.

\end{acknowledgments}

\appendix

\section{Useful formulas}

In this appendix, we compile some formulas that are useful in our 
calculation. 

The Feynman parametrization that we use
for parameters that run from $0$ to $1$ is given by
\begin{eqnarray}
\label{eq:Feynman-parametrization}
\frac{1}{A_1^{\alpha_1} \cdots A_n^{\alpha_n}}
&=&
\frac{\Gamma(\alpha_1+\cdots +\alpha_n)}
{\Gamma(\alpha_1)\cdots \Gamma(\alpha_n)}
\int_0^1 du_1\cdots \int_0^1 du_n
\frac{\delta(1-\sum_{k=1}^n u_k)u_1^{\alpha_1-1}\cdots u_n^{\alpha_n-1}}
{(\sum_{k=1}^n u_k A_k)^{\sum_{k=1}^n\alpha_k}}.
\end{eqnarray}
The Feynman (Schwinger) parametrization that we use
for parameters that run from $0$ 
to $\infty$ is given by
\begin{eqnarray}
\label{eq:alpha-parametrization}
\frac{1}{A^nB^m}
&=&
\frac{\Gamma(n+m)}{\Gamma(n)\Gamma(m)}
\int_0^\infty d\lambda\frac{\lambda^{m-1}}{(A+B\lambda)^{n+m}}.
\end{eqnarray}

We derive the following integral formula by calculating first with 
real values of the parameters $a$, $b$, $\alpha$, and $\beta$ and then 
constructing an analytic continuation in those parameters that is 
consistent with the original integral:
\begin{equation}
\label{eq:ablam-general}
\int_0^\infty \frac{d\lambda}{\lambda^\alpha(a\lambda+b\pm i\varepsilon)^\beta}
=
\frac{1}{(a\pm i\varepsilon)^{1-\alpha}(b\pm i\varepsilon)^{\alpha+\beta-1}}
\frac{\Gamma(1-\alpha)\Gamma(\alpha+\beta-1)}{\Gamma(\beta)},
\end{equation}
where $a$, $b$, $\alpha$, and $\beta$ are complex numbers. This formula
is used for the phase-space integrations and the parameter integrations
in our calculations.

In computing the $k_1$ integrations of the Abelian diagrams,
we encounter a
parameter integration of the following form:
\begin{eqnarray}
\label{eq:lam-in-terms-of-common-ep}
&&
\int_0^\infty d\lambda_1
\frac{1}
{\lambda_1^{1-\epsilon}
(\lambda_1+2c)^{-\epsilon}
\left(\lambda_1 K_1+K_2\right)^{1+2\epsilon}}
\end{eqnarray}
where $K_1$ and $K_2$ are $k_2$-dependent Lorentz products.
We perform the $\lambda_1$ integration in
Eq.~(\ref{eq:lam-in-terms-of-common-ep}) by making use of the
Mellin-Barnes representation
\begin{eqnarray}
\label{eq:lam-in-terms-of-epi}
&&
\int_0^\infty d\lambda_1
\frac{1}
{\lambda_1^{1-\epsilon_1}
(\lambda_1+2c)^{-\epsilon_2}
\left(\lambda_1 K_1+K_2\right)^{1+2\epsilon_3}}
\nonumber \\
&=&
\frac{1}{K_2^{1+2\epsilon_3}}
\oint
\frac{dz}{2\pi i}
\left(
\frac{K_1}{K_2}
\right)^z
\frac{\Gamma(-z)\Gamma(1+2\epsilon_3+z)}{\Gamma(1+2\epsilon_3)}
\int_0^\infty d\lambda_1
\frac{1}
{\lambda_1^{1-\epsilon_1-z}
(\lambda_1+2c)^{-\epsilon_2}}
\nonumber \\
&=&
\frac{(2c)^{\epsilon_1+\epsilon_2}}{K_2^{1+2\epsilon_3}}
\oint
\frac{dz}{2\pi i}
\left(
\frac{2cK_1}{K_2}
\right)^z
\frac{\Gamma(-z)\Gamma(1+2\epsilon_3+z)
\Gamma(-\epsilon_1-\epsilon_2-z)\Gamma(\epsilon_1+z)}
{\Gamma(1+2\epsilon_3)\Gamma(-\epsilon_2)},
\end{eqnarray}
with
\begin{equation}
\label{eq:ep1-ep2-z0-condition}
\textrm{Re}(\epsilon_1+\epsilon_2+z_0)<0
\textrm{~and~}
\textrm{Re}(\epsilon_1+z_0)>0.
\end{equation}
The initial contour for the $z$ integration runs parallel to the
imaginary axis with $\textrm{Re}(z)=z_0$ and separates the left poles
and right poles of the $\Gamma$ functions, where the left (right) poles
have a positive (negative) sign in the argument of the $\Gamma$
function. We analytically continue the function in
Eq.~(\ref{eq:lam-in-terms-of-epi}) in $\epsilon_1$, $\epsilon_2$, and
$\epsilon_3$ to a common value near zero, deforming the contour for the
$z$ integration so that the poles in the $\Gamma$ functions never cross
the contour. The result is a curved contour, in the style of the method
of Smirnov \cite{Smirnov:2004ym}, that still separates the left and
right poles. If we close the contour at $\infty$ in the right half of
the $z$ plane, we pick up only the right poles. The result is an
asymptotic expansion in powers of $K_1/K_2$:
\begin{eqnarray}
&&
\int_0^\infty d\lambda_1
\frac{1}
{\lambda_1^{1-\epsilon}
\left(\lambda_1+2c\right)^{-\epsilon}
\left(\lambda_1 K_1+K_2\right)^{1+2\epsilon}}
\nonumber \\
&=&
-
\frac{(2c)^{2\epsilon}}{K_2^{1+2\epsilon}}
\sum_{n=0}^\infty
\textrm{Res}\left[
\left(\frac{2cK_1}{K_2}\right)^z
\frac{\Gamma(-z)\Gamma(1+2\epsilon+z)
\Gamma(-2\epsilon-z)\Gamma(\epsilon+z)}
{\Gamma(1+2\epsilon)\Gamma(-\epsilon)}
\right]_{z=n-2\epsilon}
\nonumber \\
&&
-\frac{(2c)^{2\epsilon}}{K_2^{1+2\epsilon}}
\sum_{n=0}^\infty
\textrm{Res}\left[
\left(\frac{2cK_1}{K_2}\right)^z
\frac{\Gamma(-z)\Gamma(1+2\epsilon+z)
\Gamma(-2\epsilon-z)\Gamma(\epsilon+z)}
{\Gamma(1+2\epsilon)\Gamma(-\epsilon)}
\right]_{z=n}.
\end{eqnarray}
In our applications of this parameter integral, $K_1/K_2$ is small,
and we can retain the contribution of leading order in $K_1/K_2$, which
comes from the $n=0$ terms. This leading contribution is given below.
\begin{eqnarray}
\label{eq:lam-1-int-result-K1-K2}
\int_0^\infty d\lambda_1
\frac{1}
{\lambda_1^{1-\epsilon}
(\lambda_1+2c)^{-\epsilon}
\left(\lambda_1 K_1
+K_2\right)^{1+2\epsilon}}
&=&
\frac{1}{K_1^{2\epsilon}K_2}
\frac{1}{2\epsilon_\textrm{UV}}
+\frac{(2c)^{2\epsilon}}{K_2^{1+2\epsilon}}
\frac{\Gamma(1-2\epsilon)\Gamma(1+\epsilon)}
{2\epsilon_\textrm{UV}\Gamma(1-\epsilon)}
+O\left(\frac{K_1}{K_2}\right).
\nonumber \\
\end{eqnarray}

We also make use of the following elementary integrals, which are valid 
for $a\ge 0$:
\begin{eqnarray}
\label{eq:xi-integrals}%
\int_0^\infty d\xi \frac{1}{A}
&=&
\int_0^\infty  \frac{d\xi}{1+2a\xi+\xi^2}
=
\frac{\log\left(a+\sqrt{a^2-1}\right)}{\sqrt{a^2-1}},
\nonumber \\
\int_0^\infty d\xi \frac{1}{A}\bigg|_{a=1}
&=&
\int_0^\infty  \frac{d\xi}{1+2\xi+\xi^2}
=
1.
\end{eqnarray}

We simplify a number of expressions in our calculations by 
applying the following polylogarithm identities:
\begin{subequations}
\label{eq:dilog-identity}
\begin{eqnarray}
\textrm{Li}_2(x)
&=&
-\textrm{Li}_2(1-x)
+\frac{\pi^2}{6}
-\log x\log(1-x),
\nonumber \\
\textrm{Li}_2(x)
&=&
-\textrm{Li}_2\left(-\frac{x}{1-x}\right)
-\frac{1}{2}\log^2(1-x),
\quad\textrm{for $x<1$},
\nonumber \\
\textrm{Li}_2(x)
&=&
-\textrm{Li}_2\left(\frac{x}{x-1}\right)
+\frac{\pi^2}{2}
-\frac{1}{2}\log^2(x-1)
+i\pi\log\left(\frac{x-1}{x^2}\right),
\quad\textrm{for $x>1$},
\nonumber \\
\textrm{Li}_2(x)
&=&
-\textrm{Li}_2\left(\frac{1}{x}\right)
+\frac{\pi^2}{3}
-\frac{1}{2}\log^2x
-i\pi\log x
\quad\textrm{for $x>1$},
\nonumber \\
\textrm{Li}_2(x)
&=&
-\textrm{Li}_2(y)
+\textrm{Li}_2(xy)
+\textrm{Li}_2\left[\frac{x(1-y)}{1-xy}\right]
+\textrm{Li}_2\left[\frac{y(1-x)}{1-xy}\right]
+\log\left(\frac{1-x}{1-xy}\right)
\log\left(\frac{1-y}{1-xy}\right),
\nonumber \\
\end{eqnarray}
and
\begin{eqnarray}
\textrm{Li}_3(x)
&=&
\textrm{Li}_3\left(\frac{1}{x}\right)
-\frac{1}{6}\log^3(-x)
-\frac{\pi^2}{6}\log(-x),
\quad
\textrm{for $x\notin(0,1)$}
\nonumber \\
\textrm{Li}_3(x)
&=&
-\textrm{Li}_3\left(\frac{x}{x-1}\right)
-\textrm{Li}_3(1-x)
+\frac{1}{6}\log^3(1-x)
-\frac{1}{2}\log x\log^2(1-x)
\nonumber \\
&&
+\frac{\pi^2}{6}\log(1-x)
+\zeta(3),
\quad
\textrm{for $x\notin(1,\infty)$},
\nonumber \\
\textrm{Li}_3(x)
&=&
\frac{1}{2}\textrm{Li}_3\left[\frac{x(1-y)^2}{y(1-x)^2}\right]
+\frac{1}{2}\textrm{Li}_3(xy)
+\frac{1}{2}\textrm{Li}_3\left(\frac{x}{y}\right)
-\textrm{Li}_3\left[\frac{x(1-y)}{y(1-x)}\right]
-\textrm{Li}_3\left[\frac{x(1-y)}{x-1}\right]
\nonumber \\
&&
-\textrm{Li}_3\left(\frac{1-y}{1-x}\right)
-\textrm{Li}_3\left[\frac{1-y}{y(x-1)}\right]
-\textrm{Li}_3(y)
+\zeta(3)
-\frac{1}{2}\log^2y\log\left(\frac{1-y}{1-x}\right)
\nonumber \\
&&
+\frac{\pi^2}{6}\log y
+\frac{1}{6}\log^3y,
\end{eqnarray}
\end{subequations}
where $\text{Li}_{3}(x)$ is the trilogarithm, which is defined by 
\begin{equation}
\text{Li}_{3}(x)=\int_0^{x}dz\,\frac{\text{Li}_{2}(z)}{z}.
\end{equation} 

\section{Phase-space integration
\label{app:phase-space-int-formula}}
In this appendix, we derive the phase-space-integration formulas
that are given in Eq.~(\ref{eq:k1-phase-int-table}).

First, let us consider
\begin{equation}
I_1\equiv\int\frac{d^Dk}{(2\pi)^D} 2\pi\delta(k^2)\theta(k^0)
\frac{1}{(2p\cdot k+M^2\pm i\varepsilon)^s},
\end{equation}
where $p$ is an arbitrary real four-vector and $M^2$ is a real number. 

In order to simplify the calculation, we initially choose a coordinate
frame in which $p_\perp=0$. Then, we rewrite the result of this
calculation in a form that is manifestly rotationally invariant.
In the frame in which $p_\perp=0$, $I_1$ is given, 
in light-cone coordinates, by
\begin{equation}
I_1
=
\frac{1}{(2\pi)^{D-1}}
\int d^{D-2}k_\perp
\int_{-\infty}^\infty dk^+
\int_{-\infty}^\infty dk^-
\frac{\delta(2k^+k^- - k_\perp^2)\theta(k^+ + k^-)}
{(2p^+k^- + 2p^-k^+ +M^2\pm i\varepsilon)^s}.
\end{equation}
Note that, because of the $\delta$ function and the $\theta$ function, 
we have $k^+\ge 0$ and $k^-\ge0$.

Using the $\delta$ function to integrate over $k^-$, we find that
\begin{eqnarray}
\label{eq:I-after-km-int}
I_1
&=&
\frac{1}{(2\pi)^{D-1}}
\int d^{D-2}k_\perp
\int_0^\infty \frac{dk^+}{2k^+}
\frac{1}
{\left(
2p^+\frac{k_\perp^2}{2k^+} + 2p^-k^+ +M^2
\pm i\varepsilon\right)^s}
\nonumber \\
&=&
\frac{1}{2(2\pi)^{D-1}}
\int_0^\infty d\lambda
\frac{1}{\lambda^{1-s}}
\int d^{D-2}k_\perp
\frac{1}
{\left[\textrm{sgn}(p^+)
\left(k_\perp^2
+2 \lambda^2 p^+p^-
\right)
+\lambda M^2
\pm i\varepsilon\right]^s}.
\end{eqnarray}
In the last line we have introduced the definitions
\begin{equation}
\textrm{sgn}(x) = 
\begin{cases}
1 & \textrm{for}~x>0,
\\
-1 & \textrm{for}~ x<0,
\end{cases}
\end{equation}
and $\lambda\equiv \frac{k^+}{|p^+|}$, where $p^+ = 
\textrm{sgn}(p^+)|p^+|$.

Next, let us perform the $k_\perp$ integration. 
Carrying out the angular integrations and making the change of variables
$z= |k_\perp|^2$, we find that
\begin{eqnarray}
\label{eq:kperp-new-int-table}
&&
\int d^{D-2}k_\perp
\frac{1}
{\left[\textrm{sgn}(p^+)
\left(k_\perp^2
+2\lambda^2 p^+ p^-
\right)
+\lambda M^2
\pm i\varepsilon\right]^s}
\nonumber \\
&=&
\frac{\Omega_{D-2}}{2}
\int_0^\infty dz
\frac{1}
{z^{2-\frac{D}{2}}
\left[\textrm{sgn}(p^+)
(z+2\lambda^2 p^+p^-)
+\lambda M^2
\pm i\varepsilon\right]^s}
\nonumber \\
&=&
\frac{\pi^{\frac{D}{2}-1}\Gamma(s-\frac{D}{2}+1)}{\Gamma(s)}
\frac{1}{\left[\textrm{sgn}(p^+)\pm i\varepsilon\right]^{\frac{D}{2}-1}
\left[
2\,\textrm{sgn}(p^+)p^+p^- \lambda^2+\lambda M^2\pm i\varepsilon
\right]^{s-\frac{D}{2}+1}
},
\end{eqnarray}
where in the last equality, we have carried out the $z$ integration
by making use of the integral formula in Eq.~(\ref{eq:ablam-general})
and have made use of the $n$-dimensional solid angle
\begin{equation}
\Omega_n = \frac{2\pi^{\frac{n}{2}}}{\Gamma(\frac{n}{2})}.
\end{equation}
Substituting this result into Eq.~(\ref{eq:I-after-km-int}), we have
\begin{equation}
I_1
=
\frac{1}{(4\pi)^{\frac{D}{2}}}
\frac{\Gamma(s-\frac{D}{2}+1)}{\Gamma(s)}
\frac{1}{\left[\textrm{sgn}(p^+)\pm i\varepsilon\right]^{\frac{D}{2}-1}}
\int_0^\infty d\lambda
\frac{1}{\lambda^{2-\frac{D}{2}}
\left[
2\,\textrm{sgn}(p^+)p^+p^- \lambda+M^2\pm i\varepsilon
\right]^{s-\frac{D}{2}+1}
}.
\end{equation}
We carry out the $\lambda$ integration by using Eq.~(\ref{eq:ablam-general})
once again. The result is
\begin{eqnarray}
I_1
&=&
\frac{1}{(4\pi)^{\frac{D}{2}}}
\frac{\Gamma(\frac{D}{2}-1)\Gamma(s-D+2)}{\Gamma(s)}
\frac{1}{\left[\textrm{sgn}(p^+)\pm i\varepsilon\right]^{\frac{D}{2}-1}
\left[2\,\textrm{sgn}(p^+)p^+p^-\pm i\varepsilon\right]
^{\frac{D}{2}-1}
{(M^2\pm i\varepsilon)^{s-D+2}}}\nonumber\\
&=&
\frac{1}{(4\pi)^{\frac{D}{2}}}
\frac{\Gamma(\frac{D}{2}-1)\Gamma(s-D+2)}{\Gamma(s)}
\frac{1}{\left(\sqrt{2}p^+\pm i\varepsilon\right)^{\frac{D}{2}-1}
\left(\sqrt{2}p^-\pm i\varepsilon\right)^{\frac{D}{2}-1}
(M^2\pm i\varepsilon)^{s-D+2}}.
\nonumber \\
\end{eqnarray}

Now we restore rotational invariance in our result by making the
replacements $p^+\to (p^0+|\bm{p}|)/\sqrt{2}$ and $p^-\to
(p^0-|\bm{p}|)/\sqrt{2}$, where $p^0$ is the temporal component of
$p$, and $\bm{p}$ is the vector of spatial components of $p$. (Note
that the right sides of these replacements are equal to the left sides
in the frame in which $p_\perp=0$.)
The result is 
\begin{eqnarray}
\label{eq:I1-ps-final}%
I_1&=&
\frac{1}{(4\pi)^{\frac{D}{2}}}
\frac{\Gamma(\frac{D}{2}-1)\Gamma(s-D+2)}{\Gamma(s)}
\frac{1}{\left(p^0+|\bm{p}|\pm i\varepsilon\right)^{\frac{D}{2}-1}
\left(p^0-|\bm{p}|\pm i\varepsilon\right)^{\frac{D}{2}-1}
(M^2\pm i\varepsilon)^{s-D+2}}.\nonumber\\
\end{eqnarray}
Clearly, this expression is rotationally invariant. It is also 
invariant under boosts because the magnitude of the first two denominator 
factors is $|p^2|^{\frac{D}{2}-1}$ and the signs of the arguments in the 
first two denominator factors are also boost invariant. 

Next, let us consider 
\begin{eqnarray}
q_\alpha
I_2^\alpha
\equiv
\int\frac{d^Dk}{(2\pi)^D} 2\pi\delta(k^2)\theta(k^0)
\frac{q\cdot k}{(2p\cdot k+M^2\pm i\varepsilon)^s},
\end{eqnarray}
where $q$ is an arbitrary four-vector.

Again, we initially choose a coordinate frame in which $p_\perp=0$
and, then, rewrite the result of the calculation in a form that is
manifestly rotationally invariant. In the frame in which $p_\perp=0$,
$q_\alpha I_2^\alpha$ is given, in light-cone coordinates, by
\begin{eqnarray}
q_\alpha
I_2^\alpha
&=&
\frac{1}{(2\pi)^{D-1}}
\int d^{D-2}k_\perp
\int_{-\infty}^\infty dk^+
\nonumber \\
&&
\times
\int_{-\infty}^\infty dk^-
\frac{\delta(2k^+k^- - k_\perp^2)\theta(k^+ + k^-)
(q^+ k^- + q^- k^+ - q_\perp\cdot k_\perp)}
{(2p^+k^- + 2p^-k^+ +M^2\pm i\varepsilon)^s}.
\end{eqnarray}

Using the $\delta$ function to carry out the $k_-$ integration and
performing the change of variables that is 
below Eq.~(\ref{eq:I-after-km-int}), we obtain
\begin{eqnarray}
\label{eq:phase-space-kperp-I2al}
q_\alpha I_2^\alpha
&=&
\frac{1}{2(2\pi)^{D-1}}
\int_0^\infty d\lambda
\frac{1}{\lambda^{1-s}}
\int d^{D-2}k_\perp
\frac{\frac{q^+k_\perp^2}{2|p^+|\lambda}
+\lambda q^-|p^+| }
{\left[\textrm{sgn}(p^+)
\left(k_\perp^2
+2\lambda^2 p^+p^-
\right)
+\lambda M^2
\pm i\varepsilon\right]^s},
\phantom{XX}
\end{eqnarray}
where we have dropped terms that are odd in $k_\perp$. 
The $k_\perp$ integration yields
\begin{eqnarray}
&&
\int d^{D-2}k_\perp
\frac{\frac{q^+k_\perp^2}{2|p^+|\lambda}
+\lambda q^-|p^+|}
{\left[\textrm{sgn}(p^+)
\left(k_\perp^2
+2\lambda^2 p^+p^-
\right)
+\lambda M^2
\pm i\varepsilon\right]^s}
\nonumber \\
&=&
\frac{\pi^{\frac{D}{2}-1}
\Gamma(\frac{D}{2})\Gamma(s-\frac{D}{2})}
{\Gamma(\frac{D}{2}-1)\Gamma(s)}
\frac{\frac{1}{2}\frac{q^+}{|p^+|\lambda}}{\left[\textrm{sgn}(p^+)\pm i\varepsilon\right]^{\frac{D}{2}}
\left[
2\,\textrm{sgn}(p^+)p^+p^- \lambda^2+\lambda M^2\pm i\varepsilon
\right]^{s-\frac{D}{2}}
}
\nonumber \\
&&
+
\frac{\pi^{\frac{D}{2}-1}\Gamma(s-\frac{D}{2}+1)}
{\Gamma(s)}
\frac{\lambda q^-|p^+|}
{\left[\textrm{sgn}(p^+)\pm i\varepsilon\right]^{\frac{D}{2}-1}
\left[
2\,\textrm{sgn}(p^+)p^+p^- \lambda^2+\lambda M^2\pm i\varepsilon
\right]^{s-\frac{D}{2}+1}
}.
\phantom{XX}
\end{eqnarray}
Substituting this result into Eq.~(\ref{eq:phase-space-kperp-I2al}) and
carrying out the remaining $\lambda$ integration by making use of
Eq.~(\ref{eq:ablam-general}), we find that
\begin{eqnarray}
q_\alpha I_2^\alpha
&=&
\frac{1}{(4\pi)^{\frac{D}{2}}}
\frac{\Gamma(\frac{D}{2})\Gamma(s-D+1)}
{\Gamma(s)
\left(M^2\pm i\varepsilon\right)^{s-D+1}}
\frac{\frac{q^+}{2|p^+|}}
{\left[\textrm{sgn}(p^+)\pm i\varepsilon\right]^{\frac{D}{2}}
\left[2\, \textrm{sgn}(p^+)p^+p^-\pm i\varepsilon\right]^{\frac{D}{2}-1}}
\nonumber \\
&&
+
\frac{1}{(4\pi)^{\frac{D}{2}}}
\frac{\Gamma(\frac{D}{2})\Gamma(s-D+1)}
{\Gamma(s)
\left(M^2\pm i\varepsilon\right)^{s-D+1}}
\frac{q^-|p^+|}
{\left[\textrm{sgn}(p^+)\pm i\varepsilon\right]^{\frac{D}{2}-1}
\left[2\,\textrm{sgn}(p^+)p^+p^-
\pm i\varepsilon\right]^{\frac{D}{2}}}\nonumber\\
&=&\frac{1}{(4\pi)^{\frac{D}{2}}}
\frac{\Gamma(\frac{D}{2})\Gamma(s-D+1)}
{\Gamma(s)
\left(M^2\pm i\varepsilon\right)^{s-D+1}}
\frac{q^+p^-+q^-p^+}
{\left[\textrm{sgn}(p^+)\pm i\varepsilon\right]^{\frac{D}{2}}
\left[2\,\textrm{sgn}(p^+)p^+p^-\pm i\varepsilon\right]^{\frac{D}{2}}}
\nonumber\\
&=&\frac{1}{(4\pi)^{\frac{D}{2}}}
\frac{\Gamma(\frac{D}{2})\Gamma(s-D+1)}
{\Gamma(s)
\left(M^2\pm i\varepsilon\right)^{s-D+1}}
\frac{q^+p^-+q^-p^+}
{
\left(\sqrt{2}p^+\pm i\varepsilon\right)^{\frac{D}{2}}
\left(\sqrt{2}p^-\pm i\varepsilon\right)^{\frac{D}{2}}
}.
\end{eqnarray}

Now, we restore rotational invariance in our result by making the 
replacements $p^+\to (p^0+|\bm{p}|)/\sqrt{2}$, $p^-\to 
(p^0-|\bm{p}|)/\sqrt{2}$, and $q^+p^- +q^-p^+\to q\cdot p$. The result is
\begin{eqnarray}
\label{eq:I2-ps-final}%
q_\alpha I_2^\alpha
&=&
\frac{1}{(4\pi)^{\frac{D}{2}}}
\frac{\Gamma(\frac{D}{2})\Gamma(s-D+1)}
{\Gamma(s)
\left(M^2\pm i\varepsilon\right)^{s-D+1}}
\frac{q\cdot p}
{\left(p^0+|\bm{p}|\pm i\varepsilon\right)^{\frac{D}{2}}
\left(p^0-|\bm{p}|\pm i\varepsilon\right)^{\frac{D}{2}}}.
\end{eqnarray}

\section{IR finiteness of integrations in
$\mathcal{B}_i$, $\mathcal{C}_i$, and $\mathcal{D}_i$}

In this appendix, we demonstrate the IR finiteness of the $k_1$
integration with $k_2$ fixed in $\mathcal{B}_{1a}+\mathcal{B}_2$,
$\mathcal{C}$ plus Hermitian conjugate, and 
$\mathcal{D}$ plus Hermitian conjugate.

\subsection{IR finiteness of the $k_1$ integration 
in $\mathcal{B}_{1a}+\mathcal{B}_2$ and the $k_2$ integration in 
$\mathcal{B}_{1a}+\mathcal{B}_3$
\label{app:IR-finiteness-of-B}}

We extract the $k_1$ integration in $\mathcal{B}_{1a}$ and 
$\mathcal{B}_{2}$ and extract the $k_2$ integration in 
$\mathcal{B}_{1b}$ and
$\mathcal{B}_{3}$, writing
\begin{eqnarray}
\mathcal{B}_{1a}^{P_1P_2}\big|_\textrm{Reg}
&=&
\frac{16g_s^4 \Lambda^2\Lambda\hspace{-0.1em}'^{\hspace{0.07em}4} cd}{(P_1^2)^{-2}}
\mu^{2\epsilon}
\int_{k_2} \textrm{PS}\,
\frac{1}
{(2P_2\cdot k_2)(2P_2\cdot k_2+\Lambda\hspace{-0.1em}'^{\hspace{0.07em}2})}
B_{1a}\big|_\textrm{Reg},
\nonumber \\
\mathcal{B}_{1b}^{P_1P_2}\big|_\textrm{Reg}
&=&
\frac{16g_s^4 \Lambda^2\Lambda\hspace{-0.1em}'^{\hspace{0.07em}4} cd}{(P_1^2)^{-2}}
\mu^{2\epsilon}
\int_{k_1} \textrm{PS}
\frac{1}
{(2P_1\cdot k_1)(2P_1\cdot k_1+\Lambda\hspace{-0.1em}'^{\hspace{0.07em}2})
}
B_{1b}\big|_\textrm{Reg},
\nonumber \\
\mathcal{B}_2^{P_1P_2}\big|_\textrm{Reg}
&=&
\frac{16ig_s^4 \Lambda^2\Lambda\hspace{-0.1em}'^{\hspace{0.07em}4} cd}{(P_1^2)^{-2}}
\mu^{2\epsilon}
\int_{k_2} \textrm{PS}\,
\frac{1}{(2P_2\cdot k_2)
(2P_2\cdot k_2+\Lambda\hspace{-0.1em}'^{\hspace{0.07em}2})
(2\ell\cdot k_2+\Lambda^2)} 
B_2\big|_\textrm{Reg},
\nonumber \\
\mathcal{B}_3^{P_1P_2}\big|_\textrm{Reg}
&=&
-\frac{16ig_s^4\Lambda^2\Lambda\hspace{-0.1em}'^{\hspace{0.07em}4} cd}{(P_1^2)^{-2}}
\mu^{2\epsilon}
\int_{k_1} \textrm{PS}
\frac{1}{\left(2P_1\cdot k_1\right)
(2P_1\cdot k_1+\Lambda\hspace{-0.1em}'^{\hspace{0.07em}2})
\left(2\ell\cdot k_1+\Lambda^2\right)}
B_3\big|_\textrm{Reg},
\phantom{XX}
\end{eqnarray}
where
\begin{eqnarray}
\label{eq:B1a1b23-UV}
B_{1a}\big|_\textrm{Reg}
&\equiv&
\mu^{2\epsilon}
\int_{k_1} \textrm{PS}
\frac{1}
{(2P_1\cdot k_1)(2P_1\cdot k_1+\Lambda\hspace{-0.1em}'^{\hspace{0.07em}2})(2\ell\cdot k_1)
\left[2\ell\cdot(-k_1+k_2)-i\varepsilon\right]
\left[2\ell\cdot (k_1+k_2)+\Lambda^2\right]},
\nonumber \\
B_{1b}\big|_\textrm{Reg}
&\equiv&
\mu^{2\epsilon}
\int_{k_2} \textrm{PS}\,
\frac{1}
{(2P_2\cdot k_2)(2P_2\cdot k_2+\Lambda\hspace{-0.1em}'^{\hspace{0.07em}2})(2\ell\cdot k_2)
\left[2\ell\cdot(k_1-k_2)+i\varepsilon\right]
\left[2\ell\cdot (k_1+k_2)+\Lambda^2\right]},
\nonumber \\
B_2\big|_\textrm{Reg}
&\equiv&
\mu^{2\epsilon}
\int\frac{d^Dk_1}{(2\pi)^D}
\frac{1}
{(2P_1\cdot k_1+i\varepsilon)
(2P_1\cdot k_1+\Lambda\hspace{-0.1em}'^{\hspace{0.07em}2})
(-2\ell\cdot k_1+i\varepsilon)
(k_1^2+i\varepsilon)
\left[
2\ell\cdot (-k_1+k_2)+i\varepsilon
\right]},
\nonumber \\
B_3\big|_\textrm{Reg}
&\equiv&
\mu^{2\epsilon}
\int\frac{d^Dk_2}{(2\pi)^D}\,
\frac{1}{
\left(2P_2\cdot k_2-i\varepsilon\right)
(2P_2\cdot k_2+\Lambda\hspace{-0.1em}'^{\hspace{0.07em}2})
\left(-2\ell\cdot k_2-i\varepsilon\right)
\left(k_2^2-i\varepsilon\right)
\left[2\ell\cdot(k_1-k_2)-i\varepsilon\right]
}.
\nonumber \\
\end{eqnarray}
We can obtain $\mathcal{B}_{1b}^{P_1P_2}\big|_\textrm{Reg}$ and
$\mathcal{B}_3^{P_1P_2}\big|_\textrm{Reg}$ from the expressions for
$\mathcal{B}_{1a}^{P_1P_2}\big|_\textrm{Reg}$ and
$\mathcal{B}_2^{P_1P_2}\big|_\textrm{Reg}$, respectively, by taking their
Hermitian conjugates and making the substitutions 
$k_2\leftrightarrow k_1$ and $P_1\leftrightarrow P_2$ in the 
integrands. Therefore, it follows that 
$\mathcal{B}_{1b}^{P_2P_1}\big|_\textrm{Reg}
+\mathcal{B}_3^{P_2P_1}\big|_\textrm{Reg}$ is the Hermitian conjugate
of $\mathcal{B}_{1a}^{P_1P_2}\big|_\textrm{Reg}
+\mathcal{B}_2^{P_1P_2}\big|_\textrm{Reg}$.

Applying Feynman parameters to Eq.~(\ref{eq:B1a1b23-UV}), we find that
\begin{eqnarray}
\label{eq:B1a-B2-UV-Feyn}
B_{1a}\big|_\textrm{Reg}
&=&
-
\mu^{2\epsilon}
\Gamma(5)
\int_0^\infty d\lambda_1
\int_0^1 dx
\int_0^1 dy
\int_0^1 dz
\nonumber \\
&&
\times
\int_{k_1} \textrm{PS}
\frac{\lambda_1 (1-y)}
{\left[2k_1\cdot (\ell+\lambda_1 P_1)+\lambda_1 x \Lambda\hspace{-0.1em}'^{\hspace{0.07em}2}
+(1-y)z(2\ell\cdot k_2+\Lambda^2)
-y(2\ell\cdot k_2)+i\varepsilon\right]^5},
\nonumber \\
B_2
\big|_\textrm{Reg}
&=&
\mu^{2\epsilon}
\Gamma(5)
\int_0^\infty d\lambda_1
\int_0^\infty d\lambda_2
\int_0^1 dx
\int_0^1 dy
\nonumber \\
&&
\times
\int\frac{d^dk_1}{(2\pi)^d}
\frac{\lambda_1 \lambda_2}
{
\left[k_1^2
+2k_1\cdot(\lambda_1 P_1-\lambda_2 \ell)
+\lambda_1 x\Lambda\hspace{-0.1em}'^{\hspace{0.07em}2}
+\lambda_2 y(2\ell\cdot k_2)
+i\varepsilon\right]^5
}.
\end{eqnarray}

Let us compute $B_{1a}\big|_\textrm{Reg}$ first. Performing 
the $k_1$ phase-space integration and the $z$ integration, 
we obtain
\begin{eqnarray}
\label{eq:B1a-two-terms}
B_{1a}\big|_\textrm{Reg}
&=&
-
\frac{\left(\tilde\mu^2P_1^2\right)^{\epsilon}}{(4\pi)^2P_1^2}
\frac{e^{\epsilon\gamma_{{}_{\textrm{E}}}}\Gamma(1-\epsilon)\Gamma(2+2\epsilon)}
{2\ell\cdot k_2+\Lambda^2}
\int_0^\infty d\lambda_1
\int_0^1 dx
\int_0^1 dy
\frac{1}
{\lambda_1^{-\epsilon}(\lambda_1+2c)^{1-\epsilon}}
\nonumber \\
&&
\times
\left\{
\frac{1}{\left[\lambda_1 x \Lambda\hspace{-0.1em}'^{\hspace{0.07em}2}
-y(2\ell\cdot k_2)+i\varepsilon\right]^{2+2\epsilon}}
-\frac{1}{\left[\lambda_1 x \Lambda\hspace{-0.1em}'^{\hspace{0.07em}2}
+(1-y)(2\ell\cdot k_2+\Lambda^2)
-y(2\ell\cdot k_2)+i\varepsilon\right]^{2+2\epsilon}}
\right\}.
\nonumber \\
\end{eqnarray}
Separating the contributions that correspond to the first and second 
terms in braces in Eq.~(\ref{eq:B1a-two-terms}), we have
\begin{eqnarray}
\mathcal{B}_{1a}^{P_1P_2}\big|_\textrm{Reg}
=
\mathcal{B}_{1aa}^{P_1P_2}\big|_\textrm{Reg}
+
\mathcal{B}_{1ab}^{P_1P_2}\big|_\textrm{Reg},
\end{eqnarray}
where
\begin{eqnarray}
\label{eq:B1aa-B1ab}%
\mathcal{B}_{1aa}^{P_1P_2}\big|_\textrm{Reg}
&\equiv&
-
\frac{16g_s^4 \Lambda^2\Lambda\hspace{-0.1em}'^{\hspace{0.07em}4} cd}{(4\pi)^2(P_1^2)^{-1}}
\mu^{2\epsilon}
\int_{k_2} \textrm{PS}\,
\frac{\left(\tilde\mu^2P_1^2\right)^{\epsilon}
e^{\epsilon\gamma_{{}_{\textrm{E}}}}\Gamma(1-\epsilon)\Gamma(2+2\epsilon)}
{(2P_2\cdot k_2)(2P_2\cdot k_2+\Lambda\hspace{-0.1em}'^{\hspace{0.07em}2})(2\ell\cdot k_2+\Lambda^2)}
\nonumber \\
&&
\times
\int_0^\infty d\lambda_1
\int_0^1 dx
\int_0^1 dy
\frac{1}
{\lambda_1^{-\epsilon}(\lambda_1+2c)^{1-\epsilon}
\left[\lambda_1 x \Lambda\hspace{-0.1em}'^{\hspace{0.07em}2}
-y(2\ell\cdot k_2)+i\varepsilon\right]^{2+2\epsilon}},
\nonumber \\
\mathcal{B}_{1ab}^{P_1P_2}\big|_\textrm{Reg}
&\equiv&
\frac{16g_s^4 \Lambda^2\Lambda\hspace{-0.1em}'^{\hspace{0.07em}4} cd}{(4\pi)^2(P_1^2)^{-1}}
\mu^{2\epsilon}
\int_{k_2} \textrm{PS}\,
\frac{\left(\tilde\mu^2P_1^2\right)^{\epsilon}
e^{\epsilon\gamma_{{}_{\textrm{E}}}}\Gamma(1-\epsilon)\Gamma(2+2\epsilon)}
{(2P_2\cdot k_2)(2P_2\cdot k_2+\Lambda\hspace{-0.1em}'^{\hspace{0.07em}2})(2\ell\cdot k_2+\Lambda^2)}
\nonumber \\
&&
\!\!\!\!\!\!\!\!\!\!
\times
\int_0^\infty d\lambda_1
\int_0^1 dx
\int_0^1 dy
\frac{1}
{\lambda_1^{-\epsilon}(\lambda_1+2c)^{1-\epsilon}
\left[\lambda_1 x \Lambda\hspace{-0.1em}'^{\hspace{0.07em}2}
+(1-y)(2\ell\cdot k_2+\Lambda^2)
-y(2\ell\cdot k_2)+i\varepsilon\right]^{2+2\epsilon}}.
\nonumber \\
\end{eqnarray}

Let us consider the $\lambda_1$ integration in 
$\mathcal{B}_{1aa}^{P_1P_2}$. We can rotate the contour of integration 
counterclockwise by an angle of $\pi$ without encountering any 
singularities. Therefore, by Cauchy's theorem, we have 
\begin{eqnarray}
\mathcal{B}_{1aa}^{P_1P_2}\big|_\textrm{Reg}
&=&
-
\frac{16g_s^4 \Lambda^2\Lambda\hspace{-0.1em}'^{\hspace{0.07em}4} cd}{(4\pi)^2(P_1^2)^{-1}}
\mu^{2\epsilon}
\int_{k_2} \textrm{PS}\,
\frac{\left(\tilde\mu^2P_1^2\right)^{\epsilon}
e^{\epsilon\gamma_{{}_{\textrm{E}}}}\Gamma(1-\epsilon)\Gamma(2+2\epsilon)}
{(2P_2\cdot k_2)(2P_2\cdot k_2+\Lambda\hspace{-0.1em}'^{\hspace{0.07em}2})(2\ell\cdot k_2+\Lambda^2)}
\nonumber \\
&&
\times
\int_0^\infty d\lambda_1
\int_0^1 dx
\int_0^1 dy
\frac{1}
{\lambda_1^{-\epsilon}
\left(\lambda_1-2c-i\varepsilon\right)^{1-\epsilon}
\left[\lambda_1 x \Lambda\hspace{-0.1em}'^{\hspace{0.07em}2}
+y(2\ell\cdot k_2)\right]^{2+2\epsilon}}.
\end{eqnarray}

Next, let us consider $B_2
\big|_\textrm{Reg}$ in Eq.~(\ref{eq:B1a-B2-UV-Feyn}).
Carrying out the $k_1$ virtual integration and $\lambda_1$ 
integration, we obtain
\begin{equation}
B_2\big|_\textrm{Reg}
=
-
\frac{i(\tilde\mu^2P_1^2)^{\epsilon}}{(4\pi)^2(P_1^2)}
\int_0^\infty d\lambda_1
\int_0^1 dx
\int_0^1 dy
\frac{e^{\epsilon\gamma_{{}_{\textrm{E}}}}\Gamma(1-\epsilon)\Gamma(2+2\epsilon)
e^{2i\pi\epsilon}}
{\lambda_1^{-\epsilon}
(\lambda_1-2c-i\varepsilon)^{1-\epsilon}
\left[\lambda_1 x\Lambda\hspace{-0.1em}'^{\hspace{0.07em}2}
+y (2\ell\cdot k_2)
\right]^{2+2\epsilon}
},
\end{equation}
where, in the result, we have made a change of variables 
$\lambda_2\to 1/\lambda_1$.
Then, $\mathcal{B}_2^{P_1P_2}\big|_\textrm{Reg}$ becomes
\begin{eqnarray}
\mathcal{B}_2^{P_1P_2}\big|_\textrm{Reg}
&=&
\frac{16g_s^4 \Lambda^2\Lambda\hspace{-0.1em}'^{\hspace{0.07em}4} cd}{(4\pi)^2(P_1^2)^{-1}}
\mu^{2\epsilon}
\int_{k_2} \textrm{PS}\,
\frac{(\tilde\mu^2P_1^2)^{\epsilon}
e^{\epsilon\gamma_{{}_{\textrm{E}}}}\Gamma(1-\epsilon)\Gamma(2+2\epsilon)
e^{2i\pi\epsilon}}{(2P_2\cdot k_2)
(2P_2\cdot k_2+\Lambda\hspace{-0.1em}'^{\hspace{0.07em}2})
(2\ell\cdot k_2+\Lambda^2)}
\nonumber \\
&&
\times
\int_0^\infty d\lambda_1
\int_0^1 dx
\int_0^1 dy
\frac{1}{\lambda_1^{-\epsilon}
(\lambda_1-2c-i\varepsilon)^{1-\epsilon}
\left[\lambda_1 x\Lambda\hspace{-0.1em}'^{\hspace{0.07em}2}
+y (2\ell\cdot k_2)
\right]^{2+2\epsilon}
}.
\end{eqnarray}

Therefore, we find that 
\begin{eqnarray}
\mathcal{B}_{1aa}^{P_1P_2}
+
\mathcal{B}_2^{P_1P_2}
\big|_\textrm{Reg}
&=&
-
\frac{16g_s^4 \Lambda^2\Lambda\hspace{-0.1em}'^{\hspace{0.07em}4} cd}{(4\pi)^2(P_1^2)^{-1}}
\mu^{2\epsilon}
\int_{k_2} \textrm{PS}\,
\frac{\left(\tilde\mu^2P_1^2\right)^{\epsilon}
e^{\epsilon\gamma_{{}_{\textrm{E}}}}\Gamma(1-\epsilon)\Gamma(2+2\epsilon)
\left(
1-e^{2i\pi\epsilon}
\right)}
{(2P_2\cdot k_2)(2P_2\cdot k_2+\Lambda\hspace{-0.1em}'^{\hspace{0.07em}2})(2\ell\cdot k_2+\Lambda^2)}
\nonumber \\
&&
\times
\int_0^\infty d\lambda_1
\int_0^1 dx
\int_0^1 dy
\frac{1}
{\lambda_1^{-\epsilon}
\left(\lambda_1-2c-i\varepsilon\right)^{1-\epsilon}
\left[\lambda_1 x \Lambda\hspace{-0.1em}'^{\hspace{0.07em}2}
+y(2\ell\cdot k_2)\right]^{2+2\epsilon}}.
\nonumber \\
\end{eqnarray}
Carrying out the $x$ and $y$ integrations, we obtain
\begin{eqnarray}
\mathcal{B}_{1aa}^{P_1P_2}
+
\mathcal{B}_2^{P_1P_2}
\big|_\textrm{Reg}
&=&
-
\frac{16g_s^4 \Lambda^2\Lambda\hspace{-0.1em}'^{\hspace{0.07em}2} cd}{(4\pi)^2(P_1^2)^{-1}}
\times
\mu^{2\epsilon}
\int_{k_2} \textrm{PS}\,
\frac{\left(\tilde\mu^2P_1^2\right)^{\epsilon}
e^{\epsilon\gamma_{{}_{\textrm{E}}}}\Gamma(1-\epsilon)\Gamma(1+2\epsilon)
\times
\left(
1-e^{2i\pi\epsilon}
\right)}
{(2P_2\cdot k_2)(2P_2\cdot k_2+\Lambda\hspace{-0.1em}'^{\hspace{0.07em}2})(2\ell\cdot k_2+\Lambda^2)
(2\ell\cdot k_2)}
\nonumber \\
&&
\times
\frac{1}{2\epsilon}
\int_0^\infty d\lambda_1
\frac{(\lambda_1\Lambda\hspace{-0.1em}'^{\hspace{0.07em}2}+2\ell\cdot k_2)^{-2\epsilon}
-(2\ell\cdot k_2)^{-2\epsilon}
-(\lambda_1\Lambda\hspace{-0.1em}'^{\hspace{0.07em}2})^{-2\epsilon}}
{\lambda_1^{1-\epsilon}
\left(\lambda_1-2c-i\varepsilon\right)^{1-\epsilon}}.
\end{eqnarray}
Note that the first two terms in the numerator cancel as $\lambda_1\to
0$, and so they result in a single IR pole. The $\lambda_1$ integration
of the remaining numerator term can be carried out by making use of the
integration formula in Eq.~(\ref{eq:ablam-general}), and it results in a
real double pole. 
Since
\begin{equation}
1-e^{2i\pi\epsilon} = 
-2i\pi\epsilon +2\pi^2\epsilon^2+O(\epsilon^3),
\end{equation}
we can conclude that the IR double pole from the $k_1$ integration in
$\mathcal{B}_2^{P_1P_2}\big|_\textrm{Reg}$ is canceled by the IR double pole 
from the $k_1$ integration in
$\mathcal{B}_{1aa}^{P_1P_2}\big|_\textrm{Reg}$.
The residual imaginary pole will be canceled by
the corresponding pole in  the Hermitian-conjugate contribution
$\mathcal{B}_{1b}^{P_2P_1}\big|_\textrm{Reg}
+\mathcal{B}_3^{P_2P_1}\big|_\textrm{Reg}$.

The remaining contribution to $\mathcal{B}^{P_1P_2}$ is
$\mathcal{B}_{1ab}^{P_1P_2}\big|_\textrm{Reg}$
[Eq.~(\ref{eq:B1aa-B1ab})]. Let us consider the parameter integrations
of $\mathcal{B}_{1ab}^{P_1P_2}$. Carrying out the $y$ integration, we
obtain
\begin{eqnarray}
\label{eq:B1ab-param}
&&
\int_0^\infty d\lambda_1
\int_0^1 dx
\int_0^1 dy
\frac{1}
{\lambda_1^{-\epsilon}(\lambda_1+2c)^{1-\epsilon}
\left[\lambda_1 x \Lambda\hspace{-0.1em}'^{\hspace{0.07em}2}
+(1-y)(2\ell\cdot k_2+\Lambda^2)
-y(2\ell\cdot k_2)+i\varepsilon\right]^{2+2\epsilon}}
\nonumber \\
&=&
-
\frac{1}{(1+2\epsilon)(4\ell\cdot k_2+\Lambda^2)}
\int_0^{\infty} d\lambda_1
\int_0^1 dx
\frac{1}
{\lambda_1^{-\epsilon}
\left(\lambda_1-2c-i\varepsilon\right)^{1-\epsilon}
\left(\lambda_1 x \Lambda\hspace{-0.1em}'^{\hspace{0.07em}2}+2\ell\cdot k_2\right)^{1+2\epsilon}}
\nonumber \\
&&
-
\frac{1}{(1+2\epsilon)(4\ell\cdot k_2+\Lambda^2)}
\int_0^\infty d\lambda_1
\int_0^1 dx
\frac{1}
{\lambda_1^{-\epsilon}(\lambda_1+2c)^{1-\epsilon}
\left(\lambda_1 x \Lambda\hspace{-0.1em}'^{\hspace{0.07em}2}+2\ell\cdot 
k_2+\Lambda^2\right)^{1+2\epsilon}}.
\end{eqnarray}
In the last line, we have rotated the  $\lambda_1$ contour of integration 
counterclockwise by an angle of $\pi$, using the fact that one does not 
encounter any singularities in carrying out that rotation.
Then, we find that
\begin{eqnarray}
\mathcal{B}_{1ab}^{P_1P_2}\big|_\textrm{Reg}
&=&
-
\frac{16g_s^4 \Lambda^2\Lambda\hspace{-0.1em}'^{\hspace{0.07em}4} cd}{(4\pi)^2(P_1^2)^{-1}}
\mu^{2\epsilon}
\int_{k_2} \textrm{PS}\,
\frac{\left(\tilde\mu^2P_1^2\right)^{\epsilon}
e^{\epsilon\gamma_{{}_{\textrm{E}}}}\Gamma(1-\epsilon)\Gamma(1+2\epsilon)}
{(2P_2\cdot k_2)(2P_2\cdot k_2+\Lambda\hspace{-0.1em}'^{\hspace{0.07em}2})(2\ell\cdot k_2+\Lambda^2)
(4\ell\cdot k_2+\Lambda^2)}
\nonumber \\
&&
\times
\bigg\{
\int_0^{\infty} d\lambda_1
\int_0^1 dx
\frac{1}
{\lambda_1^{-\epsilon}
\left(\lambda_1-2c-i\varepsilon\right)^{1-\epsilon}
\left(\lambda_1 x \Lambda\hspace{-0.1em}'^{\hspace{0.07em}2}+2\ell\cdot k_2\right)^{1+2\epsilon}}
\nonumber \\
&&
\quad
+
\int_0^\infty d\lambda_1
\int_0^1 dx
\frac{1}
{\lambda_1^{-\epsilon}(\lambda_1+2c)^{1-\epsilon}
\left(\lambda_1 x \Lambda\hspace{-0.1em}'^{\hspace{0.07em}2}+2\ell\cdot k_2+\Lambda^2\right)^{1+2\epsilon}}
\bigg\}.
\end{eqnarray}
Carrying out the $x$ integration, we obtain
\begin{eqnarray}
\label{eq:B1ab-final}
\mathcal{B}_{1ab}^{P_1P_2}\big|_\textrm{Reg}
&=&
-
\frac{16g_s^4 \Lambda^2\Lambda\hspace{-0.1em}'^{\hspace{0.07em}2} cd}{(4\pi)^2(P_1^2)^{-1}}
\mu^{2\epsilon}
\int_{k_2} \textrm{PS}\,
\frac{\left(\tilde\mu^2P_1^2\right)^{\epsilon}
e^{\epsilon\gamma_{{}_{\textrm{E}}}}\Gamma(1-\epsilon)\Gamma(1+2\epsilon)}
{(2P_2\cdot k_2)(2P_2\cdot k_2+\Lambda\hspace{-0.1em}'^{\hspace{0.07em}2})(2\ell\cdot k_2+\Lambda^2)
(4\ell\cdot k_2+\Lambda^2)}
\nonumber \\
&&
\times
\bigg\{
\frac{1}{2\epsilon}
\int_0^{\infty} d\lambda_1
\frac{1}
{\lambda_1^{1-\epsilon}
\left(\lambda_1-2c-i\varepsilon\right)^{1-\epsilon}}
\left[
\frac{1}{\left(2\ell\cdot k_2\right)^{2\epsilon}}
-
\frac{1}{\left(\lambda_1 \Lambda\hspace{-0.1em}'^{\hspace{0.07em}2}+2\ell\cdot k_2\right)^{2\epsilon}}
\right]
\nonumber \\
&&
\quad
+
\frac{1}{2\epsilon}
\int_0^\infty d\lambda_1
\frac{1}
{\lambda_1^{1-\epsilon}(\lambda_1+2c)^{1-\epsilon}}
\left[
\frac{1}
{\left(2\ell\cdot k_2+\Lambda^2\right)^{2\epsilon}}
-\frac{1}
{\left(\lambda_1 \Lambda\hspace{-0.1em}'^{\hspace{0.07em}2}+2\ell\cdot k_2+\Lambda^2\right)^{2\epsilon}}
\right]
\bigg\}.
\nonumber \\
\end{eqnarray}
The $\lambda_1$ integration does not produce a pole in $\epsilon$.
Therefore, we can expand the integrand in powers of $\epsilon$. The
order-$\epsilon^0$ term vanishes and the order-$\epsilon^1$ term cancels
the $1/\epsilon$ in Eq.~(\ref{eq:B1ab-final}). Hence, we conclude that
the $k_1$ integration with $k_2$ fixed in
$\mathcal{B}_{1ab}^{P_1P_2}\big|_\textrm{Reg}$ is IR finite.

Therefore, we conclude that the $k_1$ integration with $k_2$ fixed
in $\mathcal{B}_{1a}+\mathcal{B}_{2}$ is IR finite, from which it
follows that the $k_2$ integration with $k_1$ fixed in
$\mathcal{B}_{1b}+\mathcal{B}_{3}$ is IR finite.

\subsection{IR finiteness of the $k_1$ integration 
in $\mathcal{C}_{1}+\mathcal{C}_2$
\label{app:IR-finiteness-C12}}

We extract the $k_1$ integrations in $\mathcal{C}_{1}^{P_1P_2}$ and 
$\mathcal{C}_{2}^{P_1P_2}$, writing
\begin{eqnarray}
\label{eq:Ci-P1P2-UV-organized}
\mathcal{C}_1^{P_1P_2}\big|_\textrm{Reg}
&=&
-
\frac{16i
g_s^4
\mu^{2\epsilon}
\Lambda^2
\Lambda\hspace{-0.1em}'^{\hspace{0.07em}2}
ac}
{(P_1^2)^{-2}}
\int_{k_2} \textrm{PS}\,
\frac{1}
{(2P_1\cdot k_2)
(2P_2\cdot k_2)
}
C_1\big|_\textrm{Reg},
\nonumber \\
\mathcal{C}_2^{P_1P_2}\big|_\textrm{Reg}
&=&
\frac{16g_s^4
\mu^{2\epsilon}\Lambda^2\Lambda\hspace{-0.1em}'^{\hspace{0.07em}2}
ac}
{(P_1^2)^{-2}}
\int_{k_2} \textrm{PS}\,
\frac{1}
{(2P_1\cdot k_2)
(2P_2\cdot k_2)
(2\ell\cdot k_2+\Lambda^2)}
C_2\big|_\textrm{Reg},
\end{eqnarray}
where
\begin{eqnarray}
\label{eq:C1C2-UV}
C_1\big|_\textrm{Reg}
&\equiv&
\mu^{2\epsilon}
\int_{k_1} \textrm{PS}
\frac{1}{\left[2P_1\cdot (k_1+k_2)\right]
\left(2\ell\cdot k_1\right)\left[2\ell\cdot (k_1+k_2)+\Lambda^2\right]
\left[2P_1\cdot (k_1+k_2) + \Lambda\hspace{-0.1em}'^{\hspace{0.07em}2}\right]},
\nonumber \\
C_2\big|_\textrm{Reg}
&\equiv&
\mu^{2\epsilon}
\int\frac{d^Dk_1}{(2\pi)^D}
\frac{1}
{\left[2P_1\cdot (k_1+k_2)+i\varepsilon\right]
\left[2\ell\cdot (-k_1)+i\varepsilon\right]
\left(k_1^2+i\varepsilon\right)
\left[2P_1\cdot (k_1+k_2) + \Lambda\hspace{-0.1em}'^{\hspace{0.07em}2}\right]}.
\nonumber \\
\end{eqnarray}
Applying Feynman parameters to Eq.~(\ref{eq:C1C2-UV}), we find that
\begin{eqnarray}
C_1\big|_\textrm{Reg}
&=&
\mu^{2\epsilon}
\Gamma(4)
\int_0^{\infty}d\lambda_1 
\int_0^{\infty}d\lambda_2
\int_0^1 dx
\nonumber \\
&&
\times
\int_{k_1} \textrm{PS}
\frac{\lambda_1}
{\left[
2k_1\cdot \left(\ell+ \lambda_1 P_1+\lambda_2\ell \right)
+2\lambda_1 P_1\cdot k_2
+x\lambda_1 \Lambda\hspace{-0.1em}'^{\hspace{0.07em}2}
+2\ell\cdot k_2+\Lambda^2
\right]^4},
\nonumber \\
C_2\big|_\textrm{Reg}
&=&
\mu^{2\epsilon}
\Gamma(4)
\int_0^\infty d\lambda_1
\int_0^\infty d\lambda_2
\int_0^1 dx
\nonumber \\
&&
\times
\int\frac{d^Dk_1}{(2\pi)^D}
\frac{\lambda_1}
{\left[
k_1^2+2k_1\cdot(\lambda_1P_1-\lambda_2\ell)+2\lambda_1 P_1\cdot k_2
+\lambda_1 x \Lambda\hspace{-0.1em}'^{\hspace{0.07em}2}
+i\varepsilon
\right]^4}.
\end{eqnarray}

Let us compute $C_1\big|_\textrm{Reg}$ first.
Performing the $k_1$ phase-space integration
and $\lambda_2$ and $x$ parameter integrations, we obtain
\begin{eqnarray}
C_1\big|_\textrm{Reg}
&=&
-
\frac{1}{2c}
\frac{\left(\frac{\tilde\mu^2}{P_1^2}\right)^{\epsilon}}
{(4\pi)^{2}(P_1^2)^2\Lambda\hspace{-0.1em}'^{\hspace{0.07em}2}}
\frac{e^{\epsilon\gamma_{{}_{\textrm{E}}}}\Gamma(1-\epsilon)\Gamma(1+2\epsilon)}
{\epsilon_\textrm{IR}}
\int_0^{\infty}d\lambda_1\, 
\frac{1}
{\lambda_1^{1-\epsilon}
\left(\lambda_1+2c\right)^{-\epsilon}}
\nonumber\\
&&\times
\left[
\frac{1}
{\left(
\lambda_1\frac{2P_1\cdot k_2}{P_1^2}
+\frac{2\ell\cdot k_2+\Lambda^2}{P_1^2}
\right)^{1+2\epsilon}}
-
\frac{1}
{\left(
\lambda_1\frac{2P_1\cdot k_2+\Lambda\hspace{-0.1em}'^{\hspace{0.07em}2}}{P_1^2}
+\frac{2\ell\cdot k_2+\Lambda^2}{P_1^2}
\right)^{1+2\epsilon}}
\right].
\end{eqnarray}
Since the $\lambda_1$ integration converges at both $0$ and $\infty$,
we can expand the integrand as a series in $\epsilon$ and carry out
the $\lambda_1$ integration term by term to obtain
\begin{eqnarray}
\label{eq:C1-UV-finite}%
C_1\big|_\textrm{Reg}
&=&
-
\frac{1}{2c}
\frac{1}
{(4\pi)^{2}P_1^2\Lambda\hspace{-0.1em}'^{\hspace{0.07em}2}}
\frac{\frac{1}{\epsilon_\textrm{IR}}
\log\left(\frac{2P_1\cdot k_2+\Lambda\hspace{-0.1em}'^{\hspace{0.07em}2}}{2P_1\cdot k_2}\right)
+O(\epsilon^0)}
{2\ell\cdot k_2+\Lambda^2}.
\end{eqnarray}

Next, let us consider $C_2\big|_\textrm{Reg}$. Carrying out the virtual
$k_1$ integration and the $\lambda_1$ integration, using
Eq.~(\ref{eq:ablam-general}),  we obtain
\begin{equation}
C_2\big|_\textrm{Reg}
=
-
\frac{i\left(\frac{\tilde\mu^2}{P_1^2}\right)^{\epsilon}}
{(4\pi)^{2}(P_1^2)^2}
\frac{e^{\epsilon\gamma_{{}_{\textrm{E}}}}
\Gamma(1-\epsilon)\Gamma(2+2\epsilon)}{e^{-2i\pi\epsilon}\epsilon_\textrm{IR}}
\int_0^\infty d\lambda_1
\int_0^1 dx
\frac{1}
{\lambda_1^{-2\epsilon}\left(
\lambda_1\frac{2 P_1\cdot k_2+x\Lambda\hspace{-0.1em}'^{\hspace{0.07em}2}}{P_1^2}
+2c
\right)^{2+2\epsilon}},
\end{equation}
where, in the result, we have made a change of variables 
$\lambda_2\to 1/\lambda_1$. 
Then, carrying out the remaining $\lambda_1$ and $x$ integrations, we obtain
\begin{eqnarray}
\label{eq:C2-UV-finite}%
C_2\big|_\textrm{Reg}
&=&
-
\frac{i}{2c}
\frac{\left(\frac{\tilde\mu^2}{P_1^2}\right)^{\epsilon}}
{(4\pi)^{2}(P_1^2)^2}
\frac{e^{\epsilon\gamma_{{}_{\textrm{E}}}}
\Gamma(1-\epsilon)\Gamma(1+2\epsilon)}{e^{-2i\pi\epsilon}\epsilon_\textrm{IR}}
\int_0^1 dx
\frac{1}{\left(\frac{2 P_1\cdot k_2+x\Lambda\hspace{-0.1em}'^{\hspace{0.07em}2}}{P_1^2}\right)^{1+2\epsilon}}
\nonumber \\
&=&
-
\frac{i}{2c}
\frac{\left(\frac{\tilde\mu^2}{P_1^2}\right)^{\epsilon}}
{(4\pi)^{2}P_1^2 \Lambda\hspace{-0.1em}'^{\hspace{0.07em}2}}
\frac{e^{\epsilon\gamma_{{}_{\textrm{E}}}}
\Gamma(1-\epsilon)\Gamma(1+2\epsilon)}{e^{-2i\pi\epsilon}\epsilon_\textrm{IR}}
\frac{1}{2\epsilon}
\left[
\left(\frac{2P_1\cdot k_2}{P_1^2}\right)^{-2\epsilon}
-
\left(\frac{2P_1\cdot k_2+\Lambda\hspace{-0.1em}'^{\hspace{0.07em}2}}{P_1^2}\right)^{-2\epsilon}
\right]
\nonumber \\
&=&
-
\frac{i}{2c}
\frac{1}
{(4\pi)^{2}P_1^2\Lambda\hspace{-0.1em}'^{\hspace{0.07em}2}}
\left[
\frac{1}{\epsilon_\textrm{IR}}
\log\left(\frac{2P_1\cdot k_2+\Lambda\hspace{-0.1em}'^{\hspace{0.07em}2}}{2P_1\cdot k_2}\right)
+O(\epsilon^0)
\right].
\end{eqnarray}

Inserting the results in Eqs.~(\ref{eq:C1-UV-finite}) and
(\ref{eq:C2-UV-finite})  into Eq.~(\ref{eq:Ci-P1P2-UV-organized}), we
obtain
\begin{eqnarray}
\label{eq:C-UV-finite}%
\mathcal{C}_1^{P_1P_2}\big|_\textrm{Reg}
&=&
\frac{8i
g_s^4
\mu^{2\epsilon}
\Lambda^2
a}
{(4\pi)^{2}(P_1^2)^{-1}}
\int_{k_2} \textrm{PS}\,
\frac{\frac{1}{\epsilon_\textrm{IR}}
\log\left(\frac{2P_1\cdot k_2+\Lambda\hspace{-0.1em}'^{\hspace{0.07em}2}}{2P_1\cdot k_2}\right)
+O(\epsilon^0)}
{(2P_1\cdot k_2)
(2P_2\cdot k_2)
(2\ell\cdot k_2+\Lambda^2)
},
\nonumber \\
\mathcal{C}_2^{P_1P_2}\big|_\textrm{Reg}
&=&
-
\frac{8ig_s^4
\mu^{2\epsilon}\Lambda^2
a}
{(4\pi)^{2}(P_1^2)^{-1}}
\int_{k_2} \textrm{PS}\,
\frac{\frac{1}{\epsilon_\textrm{IR}}
\log\left(\frac{2P_1\cdot k_2+\Lambda\hspace{-0.1em}'^{\hspace{0.07em}2}}{2P_1\cdot k_2}\right)
+O(\epsilon^0)}
{(2P_1\cdot k_2)
(2P_2\cdot k_2)
(2\ell\cdot k_2+\Lambda^2)}.
\end{eqnarray}
Hence, the $k_1$ integration in $\mathcal{C}_{1}+\mathcal{C}_2$ is IR
finite.

\subsection{IR finiteness of the $k_1$ integration 
in the $\mathcal{D}_1+\mathcal{D}_2$
\label{app:IR-finiteness-D12}}

We extract the $k_1$ integrations in
$\mathcal{D}_1^{P_1P_2}\big|_\textrm{Reg}$ and
$\mathcal{D}_2^{P_1P_2}\big|_\textrm{Reg}$, writing
\begin{eqnarray}
\label{eq:Di-P1P2-UV-organized}
\mathcal{D}_1^{P_1P_2}\big|_\textrm{Reg}
&=&
\frac{16i
g_s^4
\mu^{2\epsilon}
\Lambda^2
\Lambda\hspace{-0.1em}'^{\hspace{0.07em}2}
ad}
{(P_1^2)^{-2}}
\int_{k_2} \textrm{PS}\,
\frac{1}
{(2P_1\cdot k_2)
(2P_2\cdot k_2)}
D_1\big|_\textrm{Reg},
\nonumber \\
\mathcal{D}_2^{P_1P_2}\big|_\textrm{Reg}
&=&
-
\frac{16g_s^4
\mu^{2\epsilon}\Lambda^2\Lambda\hspace{-0.1em}'^{\hspace{0.07em}2}
ad}
{(P_1^2)^{-2}}
\int_{k_2} \textrm{PS}\,
\frac{1}
{(2P_1\cdot k_2)
(2P_2\cdot k_2)
(2\ell\cdot k_2+\Lambda^2)}
D_2\big|_\textrm{Reg},
\end{eqnarray}
where
\begin{eqnarray}
\label{eq:D12-UV-k1-int}
D_1\big|_\textrm{Reg}
&\equiv&
\mu^{2\epsilon}
\int_{k_1} \textrm{PS}\,
\frac{1}
{\left(2P_2\cdot k_1\right)
\left(2\ell\cdot k_1\right)
\left[2\ell\cdot(k_1+k_2)+\Lambda^2\right]
\left(2P_2\cdot k_1+\Lambda\hspace{-0.1em}'^{\hspace{0.07em}2}\right)},
\nonumber \\
D_2\big|_\textrm{Reg}
&\equiv&
\mu^{2\epsilon}
\int\frac{d^Dk_1}{(2\pi)^D}
\frac{1}
{\left(2P_2\cdot k_1+i\varepsilon\right)
\left[2\ell\cdot (-k_1)+i\varepsilon\right]
\left(k_1^2+i\varepsilon\right)
\left(2P_2\cdot k_1 +\Lambda\hspace{-0.1em}'^{\hspace{0.07em}2}\right)}.
\phantom{X}
\end{eqnarray}
Applying Feynman parameters to Eq.~(\ref{eq:D12-UV-k1-int}), we find 
that 
\begin{eqnarray}
D_1\big|_\textrm{Reg}
&=&
\mu^{2\epsilon}\Gamma(4)
\int_0^1 dx\,
\int_0^{\infty}d\lambda_1 
\int_0^{\infty}d\lambda_2
\nonumber \\
&&
\times
\int_{k_1} \textrm{PS}
\frac{\lambda_1}
{
\left[
2k_1\cdot \left(\ell+ \lambda_1 P_2+\lambda_2\ell \right)
+x\lambda_1 \Lambda\hspace{-0.1em}'^{\hspace{0.07em}2}+2\ell\cdot k_2+\Lambda^2
\right]^4},
\nonumber \\
D_2\big|_\textrm{Reg}
&=&
\mu^{2\epsilon}
\Gamma(4)
\int_0^1 dx\,
\int_0^\infty d\lambda_1
\int_0^\infty d\lambda_2
\nonumber \\
&&
\times
\int\frac{d^Dk_1}{(2\pi)^D}
\frac{\lambda_1}
{\left[
k_1^2+2k_1\cdot(\lambda_1P_2-\lambda_2\ell)
+\lambda_1 x \Lambda\hspace{-0.1em}'^{\hspace{0.07em}2}
+i\varepsilon
\right]^4}.
\end{eqnarray}

Let us compute $D_1\big|_\textrm{Reg}$ first. 
Carrying out the $k_1$ phase-space integration 
and $\lambda_2$ and $x$ parameter integrations, we obtain
\begin{eqnarray}
\label{eq:D1-UV-before-lam1-int}
D_1\big|_\textrm{Reg}
&=&
-
\frac{1}{2d}
\frac{\left(\frac{\tilde\mu^2}{P_1^2}\right)^\epsilon}
{(4\pi)^2(P_1^2)^2\Lambda\hspace{-0.1em}'^{\hspace{0.07em}2}}
\frac{e^{\epsilon\gamma_{{}_{\textrm{E}}}}\Gamma(1-\epsilon)\Gamma(1+2\epsilon)}
{\epsilon_\textrm{IR}}
\nonumber \\
&&
\times
\int_0^{\infty}d\lambda_1 
\frac{1}{\lambda_1^{1-\epsilon}
(\lambda_1 +2d)^{-\epsilon}}
\left[
\frac{1}{\left(\frac{2\ell\cdot k_2+\Lambda^2}{P_1^2}\right)^{1+2\epsilon}}
-
\frac{1}
{\left(\frac{2\ell\cdot k_2+\Lambda^2}{P_1^2}
+\lambda_1 \frac{\Lambda\hspace{-0.1em}'^{\hspace{0.07em}2}}{P_1^2}\right)^{1+2\epsilon}}
\right].
\end{eqnarray}
The two terms in brackets cancel as $\lambda_1\to0$, and so there is no
divergence in the $\lambda_1$ integration from that limit. However, a
divergence remains in the first term from the limit $\lambda_1\to\infty$.
This divergence will eventually be canceled by the
corresponding divergence in $D_2\big|_\textrm{Reg}$. We separate the
divergences in the first term by inserting
\begin{equation}
1=\frac{2d}{\lambda_1+2d}+\frac{\lambda_1}{\lambda_1+2d},
\label{eq:D-trick}%
\end{equation}
where the first term on the right side of Eq.~(\ref{eq:D-trick}) yields
an IR pole, and the second term gives a UV pole.
Then, we can write Eq.~(\ref{eq:D1-UV-before-lam1-int}) as follows:
\begin{eqnarray}
\label{eq:D1-UV-rearranged}%
D_1\big|_\textrm{Reg}
&=&
-
\frac{1}{2d}
\frac{\left(\frac{\tilde\mu^2}{P_1^2}\right)^\epsilon}
{(4\pi)^2(P_1^2)^2\Lambda\hspace{-0.1em}'^{\hspace{0.07em}2}}
\frac{e^{\epsilon\gamma_{{}_{\textrm{E}}}}\Gamma(1-\epsilon)\Gamma(1+2\epsilon)}
{\epsilon_\textrm{IR}}
\nonumber \\
&&
\times\Bigg\{
\int_0^{\infty}d\lambda_1 
\frac{1}{\lambda_1^{1-\epsilon}
(\lambda_1 +2d)^{-\epsilon}}
\Bigg[
\frac{2d}{\lambda_1+2d}
\frac{1}{\left(\frac{2\ell\cdot k_2+\Lambda^2}{P_1^2}\right)^{1+2\epsilon}}
-
\frac{1}
{\left(\frac{2\ell\cdot k_2+\Lambda^2}{P_1^2}
+\lambda_1 \frac{\Lambda\hspace{-0.1em}'^{\hspace{0.07em}2}}{P_1^2}\right)^{1+2\epsilon}}
\Bigg]
\nonumber \\
&&
\quad
+\frac{1}{\left(\frac{2\ell\cdot k_2+\Lambda^2}{P_1^2}\right)^{1+2\epsilon}}
\int_0^{\infty}d\lambda_1 
\frac{1}{\lambda_1^{-\epsilon}
(\lambda_1 +2d)^{1-\epsilon}}\Bigg\}.
\nonumber \\
\end{eqnarray}

Now the $\lambda_1$ integration of the first term in
Eq.~(\ref{eq:D1-UV-rearranged}) converges in the UV and the IR, and so,
for it, we can expand the integrand in a series in $\epsilon$. The
integral of the second term in Eq.~(\ref{eq:D1-UV-rearranged}) can be
expressed as a beta function and gives an IR pole. Then, expanding
$D_1\big|_\textrm{Reg}$ in a series in $\epsilon$, we obtain
\begin{eqnarray}
\label{eq:D1-expanded}%
D_1\big|_\textrm{Reg}
&=&
\frac{1}{2d}
\frac{1}
{(4\pi)^2P_1^2\Lambda\hspace{-0.1em}'^{\hspace{0.07em}2}}
\frac{\frac{1}{\epsilon_\textrm{IR}}
\log\left(\frac{2\ell\cdot k_2+\Lambda^2}{\Lambda\hspace{-0.1em}'^{\hspace{0.07em}2}(2d)}\right)
+O(\epsilon^0)}
{(2\ell\cdot k_2+\Lambda^2)}
\nonumber \\
&&
+
\frac{1}{2d}
\frac{1}
{(4\pi)^2P_1^2\Lambda\hspace{-0.1em}'^{\hspace{0.07em}2}}
\frac{\frac{1}{2\epsilon_\textrm{IR}^2}
+\frac{1}{2\epsilon_\textrm{IR}}
\left[
\log\left(\frac{\tilde\mu^2}{P_1^2}\right)
-2\log\left(\frac{2\ell\cdot k_2+\Lambda^2}{P_1^2(2d)}\right)
\right]
+O(\epsilon^0)
}{(2\ell\cdot k_2+\Lambda^2)}
\nonumber \\
&=&
\frac{1}{2d}
\frac{1}
{(4\pi)^2P_1^2\Lambda\hspace{-0.1em}'^{\hspace{0.07em}2}}
\frac{1}{(2\ell\cdot k_2+\Lambda^2)}
\left[\frac{1}{2\epsilon_\textrm{IR}^2}
+\frac{1}{2\epsilon_\textrm{IR}}
\log\left(\frac{\tilde\mu^2P_1^2}{\Lambda\hspace{-0.1em}'^{\hspace{0.07em}4}}\right)
+O(\epsilon^0)\right].
\end{eqnarray}

Next, let us consider $D_2\big|_\textrm{Reg}$. Carrying out the virtual
$k_1$ integration and the $\lambda_1$ integration using
Eq.~(\ref{eq:ablam-general}), we obtain
\begin{eqnarray}
D_2\big|_\textrm{Reg}
&=&
-
\frac{i\left(\frac{\tilde\mu^2}{P_1^2}\right)^{\epsilon}}
{(4\pi)^{2}(P_1^2)^2}
\frac{e^{\epsilon\gamma_{{}_{\textrm{E}}}}\Gamma(1-\epsilon)\Gamma(2+2\epsilon)}
{e^{-2i\pi\epsilon}\epsilon_\textrm{IR}}
\int_0^1 dx\,
\int_0^\infty d\lambda_1
\frac{1}
{\lambda_1^{-2\epsilon}
\left(2d+\lambda_1 x\frac{\Lambda\hspace{-0.1em}'^{\hspace{0.07em}2}}{P_1^2}\right)^{2+2\epsilon}},
\phantom{XX}
\end{eqnarray}
where, in the result, we have made a change of variables 
$\lambda_2\to 1/\lambda_1$. 
Then, carrying out the $x$ integration, we find that
\begin{equation}
\label{eq:D2-UV-IR}%
D_2\big|_\textrm{Reg}
=
-
\frac{i\left(\frac{\tilde\mu^2}{P_1^2}\right)^{\epsilon}}
{(4\pi)^{2}P_1^2\Lambda\hspace{-0.1em}'^{\hspace{0.07em}2}}
\frac{e^{\epsilon\gamma_{{}_{\textrm{E}}}}\Gamma(1-\epsilon)\Gamma(1+2\epsilon)}
{e^{-2i\pi\epsilon}\epsilon_\textrm{IR}}
\int_0^\infty d\lambda_1
\frac{1}
{\lambda_1^{1-2\epsilon}}
\left[
\frac{1}{\left(2d\right)^{1+2\epsilon}}
-\frac{1}{\left(2d+\lambda_1 \frac{\Lambda\hspace{-0.1em}'^{\hspace{0.07em}2}}{P_1^2}\right)^{1+2\epsilon}}
\right].
\end{equation}
As in the case of $D_1\big|_\textrm{Reg}$, the divergence in
Eq.~(\ref{eq:D2-UV-IR}) that appears in the integration over $\lambda_1$ as
$\lambda_1\to0$  cancels between the two terms in brackets. However,
there remains a divergence that appears as $\lambda_1\to\infty$. We
separate the UV and IR divergences in the first term by inserting
\begin{equation} 
1=\frac{1}{\lambda_1+1}+\frac{\lambda_{1}}{\lambda_1+1}
\end{equation} 
to obtain
\begin{eqnarray}
\label{eq:D2-separated}%
D_2\big|_\textrm{Reg}
&=&
-
\frac{i\left(\frac{\tilde\mu^2}{P_1^2}\right)^{\epsilon}}
{(4\pi)^{2}P_1^2\Lambda\hspace{-0.1em}'^{\hspace{0.07em}2}}
\frac{e^{\epsilon\gamma_{{}_{\textrm{E}}}}\Gamma(1-\epsilon)\Gamma(1+2\epsilon)}
{e^{-2i\pi\epsilon}\epsilon_\textrm{IR}}
\Bigg\{
\int_0^\infty d\lambda_1
\frac{1}
{\lambda_1^{1-2\epsilon}}
\Bigg[
\frac{1}{\lambda_1+1}
\frac{1}{\left(2d\right)^{1+2\epsilon}}
-\frac{1}{\left(2d+\lambda_1 \frac{\Lambda\hspace{-0.1em}'^{\hspace{0.07em}2}}{P_1^2}\right)^{1+2\epsilon}}
\Bigg]
\nonumber \\
&&
+\int_0^\infty d\lambda_1
\frac{1}
{\lambda_1^{-2\epsilon}(\lambda_1+1)}
\frac{1}{\left(2d\right)^{1+2\epsilon}}
\bigg\}.
\end{eqnarray}
In Eq.~(\ref{eq:D2-separated}),
the integration in the first term over $\lambda_1$ is finite, and so, for 
this term, we can expand the integrand in a series in $\epsilon$. The 
integration in the second term over $\lambda_1$ can be expressed as a beta 
function and yields an IR pole.

Then, expanding $D_2\big|_\textrm{Reg}$ as a series in $\epsilon$, we
obtain
\begin{eqnarray}
\label{eq:D2-expanded}%
D_2\big|_\textrm{Reg}
&=&
-
\frac{i}{2d}
\frac{1}
{(4\pi)^{2}P_1^2\Lambda\hspace{-0.1em}'^{\hspace{0.07em}2}}
\left[
\frac{1}{\epsilon_\textrm{IR}}\log\left(\frac{\Lambda\hspace{-0.1em}'^{\hspace{0.07em}2}}{P_1^2(2d)}\right)
+O(\epsilon^0)
\right]
\nonumber \\
&&
+\frac{i}{2d}\frac{1}
{(4\pi)^{2}P_1^2\Lambda\hspace{-0.1em}'^{\hspace{0.07em}2}}
\left[
\frac{1}{2\epsilon_\textrm{IR}^2}
+\frac{2i\pi-2\log(2d)+\log\left(\frac{\tilde\mu^2}{P_1^2}\right)}
{2\epsilon_\textrm{IR}}
+O(\epsilon^0)
\right]
\nonumber \\
&=&
\frac{i}{2d}\frac{1}
{(4\pi)^{2}P_1^2\Lambda\hspace{-0.1em}'^{\hspace{0.07em}2}}
\left[
\frac{1}{2\epsilon_\textrm{IR}^2}
+\frac{2i\pi+\log\left(\frac{\tilde\mu^2P_1^2}{\Lambda\hspace{-0.1em}'^{\hspace{0.07em}4}}\right)}
{2\epsilon_\textrm{IR}}
+O(\epsilon^0)
\right].
\end{eqnarray}

Inserting the results in Eqs.~(\ref{eq:D1-expanded}) and
(\ref{eq:D2-expanded}) into Eq.~(\ref{eq:Di-P1P2-UV-organized}), we find
that
\begin{eqnarray}
\mathcal{D}_1^{P_1P_2}\big|_\textrm{Reg}
&=&
\frac{8i
g_s^4
\mu^{2\epsilon}
\Lambda^2
a}
{(4\pi)^2(P_1^2)^{-1}}
\int_{k_2} \textrm{PS}\,
\frac{\frac{1}{2\epsilon_\textrm{IR}^2}
+\frac{1}{2\epsilon_\textrm{IR}}
\log\left(\frac{\tilde\mu^2P_1^2}{\Lambda\hspace{-0.1em}'^{\hspace{0.07em}4}}\right)
+O(\epsilon^0)}
{(2P_1\cdot k_2)
(2P_2\cdot k_2)
(2\ell\cdot k_2+\Lambda^2)},
\nonumber \\
\mathcal{D}_2^{P_1P_2}\big|_\textrm{Reg}
&=&
-
\frac{8ig_s^4
\mu^{2\epsilon}\Lambda^2
a}
{(4\pi)^{2}(P_1^2)^{-1}}
\int_{k_2} \textrm{PS}\,
\frac{\frac{1}{2\epsilon_\textrm{IR}^2}
+\frac{1}
{2\epsilon_\textrm{IR}}
\left[2i\pi+\log\left(\frac{\tilde\mu^2P_1^2}{\Lambda\hspace{-0.1em}'^{\hspace{0.07em}4}}\right)\right]
+O(\epsilon^0)}
{(2P_1\cdot k_2)
(2P_2\cdot k_2)
(2\ell\cdot k_2+\Lambda^2)}.
\end{eqnarray}
We see that the real double and single poles cancel between 
$\mathcal{D}_1$ and $\mathcal{D}_2$, while the single imaginary pole in 
$\mathcal{D}_2$ cancels when we add the Hermitian-conjugate contribution.
Hence, the $k_1$ integration with $k_2$ fixed in $\mathcal{D}_1+\mathcal{D}_2$ 
plus Hermitian-conjugate is IR finite.

\section{NQS Calculations\label{app:NQS-calc}}

In this appendix, we correct some signs and typographical errors and
supply color factors in the expressions for diagrams IV, V, and VI in
Ref.~\cite{Nayak:2005rt}. We mark these changes relative to the
expressions in Ref.~\cite{Nayak:2005rt} with double brackets.  We also
complete the calculations of the integrals in the expression for diagram
V.

\subsection{Diagram IV}
The contribution of diagram IV is given by
\begin{eqnarray}
\label{Eq:IVA-after-k1m-k2m-int-rescale}
IV^{(k_2^0)}
&=&
[[-\frac{f_{abc}^2}{4N_c}]]
\times
4\left(\frac{\alpha_s}{\pi}\right)^2
2^{2\epsilon}\pi^{3\epsilon-1}\Gamma(1+\epsilon)
\int_0^\Lambda \frac{dk_1^+}{(k_1^+)^{1+4\epsilon}}
\int d^{2-2\epsilon}\kappa_1
\nonumber \\
&&
\times
\int_{-\infty}^\infty dy
\frac{
\frac{1}{2}q_\perp^2
(1 + \kappa_1^2)
+q_3^2
}
{(1 + \kappa_1^2)^2
\left[(y-1)(y+\kappa_1^2)
+i\varepsilon
\right]^{1+\epsilon}},
\end{eqnarray} 
which agrees with Eq.~(51) of Ref.~\cite{Nayak:2005rt} up to the missing
color factor $\frac{f_{abc}^2}{4N_c}$ and an overall sign. Performing the
remaining integrations, except for the $k_1^+$ integration, we obtain
\begin{eqnarray}
\label{eq:NQS-final-IV}
IV^{(k_2^0)}
&=&
-
\left(\frac{\alpha_s}{\pi}\right)^2
\frac{f_{abc}^2}{N_c}
(2\pi)^{2\epsilon}
\Gamma(1+\epsilon)
\left[
-i\pi-\pi^2\epsilon + O(\epsilon^2)
\right]
\int_0^\Lambda \frac{dk_1^+}{(k_1^+)^{1+4\epsilon}}
\nonumber \\
&&
\times
\left[\bm{q}^2
-\epsilon\left(2q_\perp^2+\gamma_{{}_{\textrm{E}}}\bm{q}^2\right)
+O(\epsilon^2)\right].
\end{eqnarray}

\subsection{Diagram V}
\label{app:diagram-V}%
The contribution of diagram V is given by
\begin{eqnarray}
\label{eq:VA-after-k2m-int}
V^{(k_2^0)}
&=&
[[-\frac{f_{abc}^2}{4N_c}]]
\left(\frac{\alpha_s}{\pi}\right)^2
\frac{8}{(4\pi^2)^{1-2\epsilon}}
\int_0^\Lambda \frac{dk_1^+}{2k_1^+}
\int_{-\infty}^\infty dk_2^+
\int d^{2-2\epsilon}k_{1\perp}
\int d^{2-2\epsilon}k_{2\perp}
\frac{1}{(k_1^+ + \frac{k_{1\perp}^2}{2k_1^+})^3}
\nonumber \\
&&
\times
\frac{
-2k_{[[1]]\perp}^2 q_3^2
-q_\perp^2(k_1^+ + \frac{k_{1\perp}^2}{2k_1^+})^2
[[+]]2(q_\perp\cdot k_{1\perp})^2}{
\left[2(k_1^+-k_2^+)(k_2^++\frac{k_{1\perp}^2}{2k_1^+})
-(k_{2\perp}-k_{1\perp})^2
-i\varepsilon
\right]
}
\nonumber \\
&&
\times
\left\{
\frac{1}{
\left(k_1^+ - k_2^+-i\varepsilon\right)
(k_1^+ + \frac{k_{1\perp}^2}{2k_1^+})}
+
\frac{2}{
\left[2(k_1^+-k_2^+)(k_2^++\frac{k_{1\perp}^2}{2k_1^+})
-(k_{2\perp}-k_{1\perp})^2
-i\varepsilon
\right]}
\right\},
\nonumber \\
\end{eqnarray} 
which agrees with Eq.~(54) of Ref.~\cite{Nayak:2005rt} up to the
corrections that are marked with the double brackets. Using the
rescaling of integration variables in Ref.~\cite{Nayak:2005rt}
\begin{equation}
k_2^+
=(k_1^+) y,
\quad
k_{i\perp}
=(\sqrt{2}k_1^+) \kappa_i,
\end{equation}
we can rewrite Eq.~(\ref{eq:VA-after-k2m-int}) as follows:
\begin{eqnarray}
\label{eq:V-k2-NQS}
V^{(k_2^0)}
&=&
[[\frac{f_{abc}^2}{4N_c}]]
\times
8
\left(
\frac{\alpha_s}{\pi}
\right)^2
\frac{2^{-2\epsilon}}{(4\pi^{[[2]]})^{1-2\epsilon}}
\int_0^\Lambda \frac{dk_1^+}{(k_1^+)^{1+4\epsilon}}
\int d^{2-2\epsilon}\kappa_1
\nonumber \\
&&
\times
\frac{4q_3^2 \kappa_1^2
+q_\perp^2(1+\kappa_1^2)^{[[2]]}
-4(q_\perp\cdot \kappa_{1})^2}{(1 +\kappa_1^2)^3}
J_V(\kappa_1)
,
\end{eqnarray} 
which agrees with Eq.~(55) of Ref.~\cite{Nayak:2005rt}, up to the
corrections that are marked with double brackets. $J_V(\kappa_1)$ is
defined by
\begin{eqnarray}
J_V(\kappa_1)
&=&
-\int_{-\infty}^{\infty}
dy
\int d^{2-2\epsilon} \kappa_2
\bigg\{
\frac{1}{(1+\kappa_1^2)(1-y-i\varepsilon)
\left[
(y-1)(y+\kappa_1^2)+(\kappa_2-\kappa_1)^2+i\varepsilon
\right]
}
\nonumber \\
&&
\quad\quad\quad\quad\quad\quad\quad\quad\quad
-\frac{1}{[(y-1)(y+\kappa_1^2)+(\kappa_2-\kappa_1)^2+i\varepsilon]^2}
\bigg\}.
\label{eq:nVa-JV}
\end{eqnarray}
Performing the $\kappa_2$ integration, we can write 
Eq.~(\ref{eq:nVa-JV}) as 
\begin{eqnarray}
J_V(\kappa_1)
&=&
-\pi^{1-\epsilon}\Gamma(1+\epsilon)
\int_{-\infty}^{\infty}
dy
\bigg\{
\frac{1}{\epsilon_\textrm{UV}}
\frac{1}{(1+\kappa_1^2)(1-y-i\varepsilon)
\left[(y-1)(y+\kappa_1^2)+i\varepsilon\right]^{\epsilon}
}
\nonumber \\
&&
\quad\quad\quad\quad\quad\quad\quad\quad\quad\quad
-\frac{1}{[(y-1)(y+\kappa_1^2)+i\varepsilon]^{1+\epsilon}}
\bigg\},
\end{eqnarray}
which agrees with Eq.~(57) of Ref.~\cite{Nayak:2005rt}.
Performing the remaining $y$ and $\kappa_1$ integrations, 
we obtain 
\begin{eqnarray}
\label{eq:NQS-V-final}
V^{(k_2^0)}
&=&
-
\left(\frac{\alpha_s}{\pi}\right)^2
\frac{f_{abc}^2}{N_c}
(2\pi)^{2\epsilon}\Gamma(1+\epsilon)
\left[
-i\pi-\pi^2\epsilon + O(\epsilon^2)
\right]
\int_0 \frac{dk_1^+}{(k_1^+)^{1+4\epsilon}}
\nonumber\\
&&\times\left(
\frac{1}{2\epsilon_\textrm{UV}}
+1
\right)
\left[
-
\frac{2}{3}\bm{q}^2
+O(\epsilon)
\right].
\end{eqnarray}

In Ref.~\cite{Nayak:2005rt}, the complete factor that multiplies the 
$k_1^+$ integral in Eq.~(\ref{eq:NQS-V-final})
was considered to be UV in nature and, therefore, to be a contribution
to an SDC. Consequently, the entire contribution in
Eq.~(\ref{eq:NQS-V-final}) was taken to be a one-loop contribution to an
SDC times the one-loop contribution to the LDME, and it was dropped in
Ref.~\cite{Nayak:2005rt}. As we will now show, the real contribution
that comes from the product of the UV pole and the term $-\pi^2\epsilon$
in Eq.~(\ref{eq:NQS-V-final}) should actually be considered to be IR in
nature. We show this by completing the $k_2$ integration in $V$ before
carrying out the $k_1$ integration.

First, we rewrite Eq.~(\ref{eq:VA-after-k2m-int}) as
\begin{eqnarray}
\label{eq:VA-for-NQS52}
V^{(k_2^0)}
&=&
-
\frac{2^{1/2}g_s^4\frac{f_{abc}^2}{N_c}}{(2\pi)^{2D-2}}
\int^\Lambda d^{D}k_1
\,\delta_+(k_1^2)
\left\{
\sqrt{2}
\left[
2(q_\perp\cdot k_{1\perp})^2
-4 k_1^- k_1^+ q_3^2
-q_\perp^2(k_1^+ + k_1^-)^2
\right]
\right\}
\nonumber \\
&&
\times
\left[
\frac{K_2^a}{(k_1^+ + k_1^-)^3} +
\frac{K_2^b}{(k_1^+ + k_1^-)^4}
\right],
\end{eqnarray} 
where $\delta_+(k_1^2) \equiv \delta(k_1^2)\theta(k_1^+  + k_1^-)$,
and $K_2^a$ and $K_2^b$ are
\begin{eqnarray}
K_2^a
&=&
\int_{-\infty}^\infty dk_2^+\int d^{2-2\epsilon}k_{2\perp}
\frac{2(k_1^+ - k_2^+)}{
\left(k_1^+ - k_2^+-i\varepsilon\right)
\left[2(k_1^+-k_2^+)(k_2^++k_1^-)
-(k_{2\perp}-k_{1\perp})^2
-i\varepsilon
\right]^2},
\nonumber \\
K_2^b
&=&
\int_{-\infty}^\infty dk_2^+\int d^{2-2\epsilon}k_{2\perp}
\frac{1}{
\left(k_1^+ - k_2^+-i\varepsilon\right)
\left[2(k_1^+-k_2^+)(k_2^++k_1^-)
-(k_{2\perp}-k_{1\perp})^2
-i\varepsilon
\right]}.
\nonumber \\
\end{eqnarray}
Making the changes of variables 
$k_2^+\to k_2^+ +k_1^+$ and $k_{2\perp}\to k_{2\perp}+k_{1\perp}$ and 
carrying out the $k_{2\perp}$ integration, we obtain
\begin{eqnarray}
K_2^a
&=&
\frac{\pi\Gamma(1+\epsilon)}{(2\pi)^\epsilon} 
\int_{-\infty}^\infty dk_2^+
\frac{1}{
\left[k_2^+(k_2^++k_1^++k_1^-)+i\varepsilon
\right]^{1+\epsilon}},
\nonumber \\
K_2^b
&=&
\frac{\pi\Gamma(\epsilon_\textrm{UV})}{(2\pi)^\epsilon} 
\int_{-\infty}^\infty dk_2^+
\frac{1}{
\left(k_2^++i\varepsilon\right)
\left[k_2^+(k_2^++k_1^++k_1^-)
+i\varepsilon
\right]^{\epsilon}}.
\end{eqnarray}
We can carry out the $k_2^+$ integration by making use of the following 
formulas:
\begin{subequations}%
\label{eq:K2a-K2b}%
\begin{eqnarray}
\int_{-\infty}^{\infty}dy
\frac{1}{[y(y+k_1^++k_1^-)+i\varepsilon]^{1+\epsilon}}
&=&
\frac{1}{(k_1^++k_1^-)^{1+2\epsilon}}
\left[
\frac{2\Gamma(-\epsilon)\Gamma(1+2\epsilon)}{\Gamma(1+\epsilon)}
-\frac{\Gamma^2(-\epsilon)}{e^{i\pi\epsilon}\Gamma(-2\epsilon)}
\right]
\nonumber \\
&=&
\frac{2}{(k_1^++k_1^-)^{1+2\epsilon}}
\left[
-i\pi-\pi^2\epsilon + O(\epsilon^2)
\right],
\\
\int_{-\infty}^{\infty}dy
\frac{1}{(y+i\varepsilon)
\left[y(y+k_1^++k_1^-)+i\varepsilon\right]^{\epsilon}}
&=&
\frac{1}{(k_1^++k_1^-)^{2\epsilon}}
\left[
\frac{\Gamma(-\epsilon)\Gamma(1+2\epsilon)}{\Gamma(1+\epsilon)}
-
\frac{\Gamma(-\epsilon)\Gamma(1-\epsilon)}
{e^{i\pi\epsilon}\Gamma(1-2\epsilon)}
\right]
\nonumber \\
&=&
\frac{1}{(k_1^++k_1^-)^{2\epsilon}}
\left[
-i\pi-\pi^2\epsilon+O(\epsilon^2)
\right].
\end{eqnarray}
\end{subequations}%
Then, we find that
\begin{eqnarray}
\label{eq:K2a-K2b-2}
K_2^a
&=&
\frac{\pi\Gamma(1+\epsilon)}{(2\pi)^\epsilon} 
\frac{2}{(k_1^++k_1^-)^{1+2\epsilon}}
\left[
-i\pi-\pi^2\epsilon + O(\epsilon^2)
\right],
\nonumber \\
K_2^b
&=& 
\frac{\pi\Gamma(\epsilon_\textrm{UV})}{(2\pi)^\epsilon} 
\frac{1}{(k_1^++k_1^-)^{2\epsilon}}
\left[
-i\pi-\pi^2\epsilon+O(\epsilon^2)
\right].
\end{eqnarray}
We can see that these expressions have an IR sensitivity 
because of the denominator factors $(k_1^+ + k_1^-)^{1+2\epsilon}$ and
$(k_1^+ + k_1^-)^{2\epsilon}$, which 
become singular when $k_1$ goes to zero, as happens for the IR 
divergence in the $k_1$ integral. These factors from the $k_2$ 
integration affect the strength of the IR pole in the $k_1$ integration. 
The only part of Eqs.~(\ref{eq:K2a-K2b-2}) that can be considered to be UV 
in nature is the pure UV pole in the expression for $K_2^b$, for which 
the denominator factor $(k_1^+ + k_1^-)^{2\epsilon}$ is set to unity. 
However, this pure UV pole cancels when we add the Hermitian-conjugate
contribution.
All of the other contributions in Eqs.~(\ref{eq:K2a-K2b}) are IR in 
nature. This is true, in particular, of the only real contribution, which 
comes from the product of the UV pole and the 
$-\pi^2\epsilon$ term in $K_2^b$.

Having demonstrated the IR nature of the real part of the result, we
complete the calculation by carrying out the $k_1$ integration. Using
our results from the $k_2$ integration, we have
\begin{eqnarray}
\label{eq:VA-for-NQS52-2}
V^{(k_2^0)}
&=&
-
\frac{g_s^4\frac{f_{abc}^2}{N_c}}{(2\pi)^{4-3\epsilon}\pi}
\Gamma(1+\epsilon)
\left(
1
+
\frac{1}{2\epsilon_\textrm{UV}}
\right)
\left[
-i\pi-\pi^2\epsilon+O(\epsilon^2)
\right]
\nonumber \\
&&
\times
\int^\Lambda d^{D}k_1
\,\delta_+(k_1^2)
\frac{\left[
2(q_\perp\cdot k_{1\perp})^2
-4 k_1^- k_1^+ q_3^2
-q_\perp^2(k_1^+ + k_1^-)^2
\right]}{(k_1^+ + k_1^-)^{4+2\epsilon}}.
\end{eqnarray} 
We carry out the $k_1$ integration by making use of the
$\delta_+$ function, make the change of variables $k_{1\perp}\to
\sqrt{2}k_1^+\kappa_1$, and carry out the $\kappa_1$ integration. The
result is
\begin{eqnarray}
\label{eq:VA-for-NQS52-3}
V^{(k_2^0)}
&=&
-
\frac{16g_s^4\frac{f_{abc}^2}{N_c}}{(4\pi)^{4}}
(2\pi)^{2\epsilon}
\Gamma(1+\epsilon)
\left(
1
+
\frac{1}{2\epsilon_\textrm{UV}}
\right)
\left[
-i\pi-\pi^2\epsilon+O(\epsilon^2)
\right]
\int_0^\Lambda \frac{dk_1^+}{(k_1^+)^{1+4\epsilon}}
\left[-
\frac{2}{3}\bm{q}^2
+O(\epsilon)\right],
\nonumber \\
\end{eqnarray} 
which is in agreement with Eq.~(\ref{eq:NQS-V-final}).

\subsection{Diagram VI}
The contribution of diagram VI is given by
\begin{eqnarray}
VI^{(k_2^0,k_2^0-k_1^0)}
&=&
-[[\frac{f_{abc}^2}{4N_c}]]
\times
4
\left(\frac{\alpha_s}{\pi}\right)^2
2^{2\epsilon}\pi^{3\epsilon-1}
[[\Gamma(1+\epsilon)]]
\int_0^\Lambda \frac{dk_1^+}{(k_1^+)^{1+4\epsilon}}
\int d^{2-2\epsilon}\kappa_1
\frac{q_3^2(1 - \kappa_1^2)}{
\left(1 + \kappa_1^2\right)^3}
\nonumber \\
&&
\times
\int_{-\infty}^\infty dy\,
(1- y)
\Bigg\{
\frac{1}{
\left[
[[]]
(y-1)^2[[+]]i\varepsilon\right]^{1+\epsilon}}
-
\frac{1}{
\left[
[[]](y-1)\left(y+\kappa_1^2\right)
[[+]]i\varepsilon\right]^{1+\epsilon}}
\Bigg\},
\nonumber \\
\end{eqnarray}
which agrees with Eq.~(61) of Ref.~\cite{Nayak:2005rt}, up to the
corrections that are marked with double brackets. Here, the empty
double brackets $[[]]$ indicate factors of $2$ that were present in
Eq.~(61) of Ref.~\cite{Nayak:2005rt} and have been removed. Carrying out
the remaining $y$ and $\kappa_1$ integrations, we find that
\begin{eqnarray}
\label{eq:VIA-Eq-58-final-q3}
VI
&=&
-
\left(\frac{\alpha_s}{\pi}\right)^2
\frac{f_{abc}^2}{N_c}
(2\pi)^{2\epsilon}
\Gamma(1+\epsilon)
\left[
-i\pi-\pi^2\epsilon
+O(\epsilon^2)
\right]
\int_0^\Lambda \frac{dk_1^+}{(k_1^+)^{1+4\epsilon}}
\times
\left[-2\epsilon q_3^2+O(\epsilon^2)\right].\nonumber\\
\end{eqnarray}

\section{Source of the UV double poles in an NQS-style 
calculation\label{app:source-disc}}

In this section, we demonstrate how the UV double poles in $\mathcal{C}$
and $\mathcal{D}$ arise in an NQS-style calculation. As we have
mentioned, the source of the double poles is a mismatch between a
real-gluon contribution, which is subject to a phase-space UV
regulator, and the contribution from the residue of a pole in the
propagator of the corresponding virtual gluon, which is unregulated in
the UV. 
Following NQS, we impose a UV
cutoff $\Lambda$ on each phase-space integration over a plus light-cone
momentum component. While this cutoff is different from our standard UV
phase-space regulator, we expect that this difference will not affect
the coefficients of the leading (double) UV poles.

\subsection{Diagram IV}

Let us first consider the contribution of diagram IV in
Ref.~\cite{Nayak:2005rt}. In terms of light-cone variables, the
real-virtual contribution is \cite{Nayak:2005rt}
\begin{eqnarray}
IVA
&=&
\frac{4ig_s^4 \frac{f_{abc}^2}{N_c}}{(2\pi)^{2D-1}}
\int_0^\Lambda \frac{dk_1^+}{2k_1^+}
\int_{-\infty}^\infty dk_2^+
\int_{-\infty}^\infty dk_2^-
\int d^{2-2\epsilon}k_{1\perp}
\int d^{2-2\epsilon}k_{2\perp}
\nonumber \\
&&
\times
\frac{
\frac{1}{2}q_\perp^2
(k_1^+ + \frac{k_{1\perp}^2}{2k_1^+})
+q_3^2k_1^+
+\frac{1}{\sqrt{2}}q_3q_\perp\cdot k_{1\perp}}
{(k_1^+ + \frac{k_{1\perp}^2}{2k_1^+})^2
(k_2^+ + k_2^- -i\varepsilon)^2
\left(k_1^+ - k_2^+-i\varepsilon\right)}
\nonumber \\
&&
\times
\frac{1}{\left[
(2k_2^+ k_2^--k_{2\perp}^2)
-2(k_1^+ k_2^-+\frac{k_{1\perp}^2}{2k_1^+}k_2^+
-k_{1\perp}\cdot k_{2\perp})
-i\varepsilon\right]}.
\end{eqnarray} 
Note that the last denominator can be rewritten as
\begin{eqnarray}
&&
\left[
(2k_2^+ k_2^--k_{2\perp}^2)
-2\left(k_1^+ k_2^-+\frac{k_{1\perp}^2}{2k_1^+}k_2^+
-k_{1\perp}\cdot k_{2\perp}\right)
-i\varepsilon\right]
\nonumber \\
&=&
2(k_2^+ - k_1^+)
\left[k_2^-
-\frac{k_{1\perp}^2}{2k_1^+}
-\frac{\left(k_{2\perp}-k_{1\perp}\right)^2+i\varepsilon}{2(k_2^+ -k_1^+)}
\right].
\end{eqnarray}
We close the $k_2^-$ contour in the upper half-plane
and pick up the residue of the pole at $(k_2-k_1)^2=0$. The location of 
the pole is 
\begin{equation}
\label{eq:k2m-k2-k1-sq-pole}
k^-_{2[(k_2-k_1)^2]}
=
\frac{k_{1\perp}^2}{2k_1^+}
+\frac{\left(k_{2\perp}-k_{1\perp}\right)^2+i\varepsilon}{2(k_2^+ 
-k_1^+)},
\end{equation}
and the residue is nonzero only when $k_2^+ > k_1^+$. The residue is 
\begin{eqnarray}
IVA^{(k_2-k_1)^2}
&=&
\frac{g_s^4 \frac{f_{abc}^2}{N_c}}{(2\pi)^{6-4\epsilon}}
\int_0^\Lambda \frac{dk_1^+}{k_1^+}
\int_{k_1^+}^\infty dk_2^+
\int d^{2-2\epsilon}k_{1\perp}
\int d^{2-2\epsilon}k_{2\perp}
\nonumber \\
&&
\times
\frac{\frac{1}{2}q_\perp^2
(k_1^+ + \frac{k_{1\perp}^2}{2k_1^+})
+q_3^2k_1^+
+\frac{1}{\sqrt{2}}q_3q_\perp\cdot k_{1\perp}}
{(k_1^+ + \frac{k_{1\perp}^2}{2k_1^+})^2
\left[k_2^+ + \frac{k_{1\perp}^2}{2k_1^+}
+\frac{\left(k_{2\perp}-k_{1\perp}\right)^2}{2(k_2^+ -k_1^+)}\right]^2
(k_2^+ - k_1^+)^2}.
\end{eqnarray} 
Carrying out the $k_2^+$, $k_{1\perp}$, and $k_{2\perp}$ integrations, 
we obtain
\begin{eqnarray}
IVA^{(k_2-k_1)^2}
&=&
-
\left(\frac{\alpha_s}{\pi}\right)^2
(2\pi)^{2\epsilon}
\left(\frac{f_{abc}^2}{4N_c}\right)
\left[-2\bm{q}^2+O(\epsilon)\right]
\Gamma(1+2\epsilon)\Gamma(-\epsilon_\textrm{IR})
\int_0^\Lambda \frac{dk_1^+}{(k_1^+)^{1+4\epsilon}}.
\phantom{XX}
\end{eqnarray} 
We note that the poles in this expression are purely IR in origin. That
is, the integrations in $IVA^{(k_2-k_1)^2}$ are UV convergent. The 
real-real contribution $IVB$ is precisely the negative of the
real-virtual contribution $IVA^{(k_2-k_1)^2}$, except that the $k_2$
integration in $IVB$ is cut off by a UV regulator. Since the
integrations in both $IVA^{(k_2-k_1)^2}$ and $IVB$ are UV convergent,
they cancel, up to terms that are suppressed by inverse powers of the
UV-regulator scale $\Lambda$. Hence, the diagrams IV do not yield any UV
double poles.

\subsection{Diagram V}

Next, let us consider the contribution of diagram V in
Ref.~\cite{Nayak:2005rt}. In terms of light-cone variables the
real-virtual contribution is \cite{Nayak:2005rt}
\begin{eqnarray}
VA
&=&
\frac{i2^{5/2}g_s^4\frac{f_{abc}^2}{4N_c}}{(2\pi)^{2D-1}}
\int_0^\Lambda \frac{dk_1^+}{2k_1^+}
\int_{-\infty}^\infty dk_2^+
\int_{-\infty}^\infty dk_2^-
\int d^{2-2\epsilon}k_{1\perp}
\int d^{2-2\epsilon}k_{2\perp}
\nonumber \\
&&
\times
\frac{\sqrt{2}
\left[
2(q_\perp\cdot k_{1\perp})^2
-2k_{1\perp}^2 q_3^2
-q_\perp^2(k_1^+ + \frac{k_{1\perp}^2}{2k_1^+})^2
\right]}{
\left(k_1^+ - k_2^+-i\varepsilon\right)
\left[
(2k_2^+ k_2^--k_{2\perp}^2)
-2(k_1^+ k_2^-+\frac{k_{1\perp}^2}{2k_1^+}k_2^+
-k_{1\perp}\cdot k_{2\perp})
-i\varepsilon\right]
}
\nonumber \\
&&
\times
\left[
\frac{1}{(k_2^+ + k_2^- -i\varepsilon)^2
(k_1^+ + \frac{k_{1\perp}^2}{2k_1^+})^3}
+
\frac{1}{(k_2^+ + k_2^- -i\varepsilon)
(k_1^+ + \frac{k_{1\perp}^2}{2k_1^+})^4}
\right].
\end{eqnarray} 
Again, we close the $k_2^-$ contour in the upper half-plane and pick up 
the residue of the pole at $(k_2-k_1)^2=0$. The location of the pole is 
given in Eq.~(\ref{eq:k2m-k2-k1-sq-pole}), and the residue is nonzero only 
when $k_2^+ > k_1^+$. The residue is 
\begin{eqnarray}
\label{eq:VA-before-rescaling}
VA^{(k_2-k_1)^2}
&=&
\frac{2^3g_s^4}{(2\pi)^{2D-2}}
\left(\frac{f_{abc}^2}{4N_c}\right)
\int_0^\Lambda \frac{dk_1^+}{2k_1^+}
\int_{k_1^+}^\infty dk_2^+
\int d^{2-2\epsilon}k_{1\perp}
\int d^{2-2\epsilon}k_{2\perp}
\nonumber \\
&&
\times
\frac{
2(q_\perp\cdot k_{1\perp})^2
-2k_{1\perp}^2 q_3^2
-q_\perp^2(k_1^+ + \frac{k_{1\perp}^2}{2k_1^+})^2}{
2(k_2^+ - k_1^+)^2
}
\nonumber \\
&&
\times
\left\{
\frac{1}{
\left[k_2^+ + \frac{k_{1\perp}^2}{2k_1^+}
+\frac{\left(k_{2\perp}-k_{1\perp}\right)^2}{2(k_2^+ -k_1^+)}\right]^2
(k_1^+ + \frac{k_{1\perp}^2}{2k_1^+})^3}
+
\frac{1}{\left[k_2^+ + \frac{k_{1\perp}^2}{2k_1^+}
+\frac{\left(k_{2\perp}-k_{1\perp}\right)^2}{2(k_2^+ -k_1^+)}\right]
(k_1^+ + \frac{k_{1\perp}^2}{2k_1^+})^4}
\right\}.
\nonumber \\
\end{eqnarray}
After the changes of variables
\begin{eqnarray}
k_2^+ &\to& k_2^+ + k_1^+,\nonumber\\
k_{2\perp} &\to& k_{2\perp} + k_{1\perp},\nonumber\\
k_{i\perp}&\to&(\sqrt{2}k_i^+)\kappa_i,
\end{eqnarray}
for $i=1$ and $2$,
we can perform the $\kappa_2$ integration to obtain
\begin{eqnarray}
\label{eq:VA-k1-k2-almost-final}%
&&
VA^{(k_2-k_1)^2}
\nonumber \\
&=&
\left(\frac{\alpha_s}{\pi}\right)^2
2^{1+2\epsilon}\pi^{-1+3\epsilon}
\left(\frac{f_{abc}^2}{4N_c}\right)
\int_0^\Lambda dk_1^+(k_1^+)^{-2\epsilon}
\int d^{2-2\epsilon}\kappa_1
\frac{\left[4(q_\perp\cdot \kappa_1)^2
-4\kappa_1^2 q_3^2
-q_\perp^2(1 + \kappa_1^2)^2
\right]}{(1+\kappa_1^2)^3}
\nonumber \\
&&
\times
\int_{0}^\infty \frac{dk_2^+}{(k_2^+)^{1+\epsilon}}
\left[
\frac{\Gamma(1+\epsilon)}
{
\left(
k_2^+ + k_1^+ + k_1^+\kappa_1^2
\right)^{1+\epsilon}}
+
\frac{\Gamma(\epsilon_\textrm{UV})}
{(k_1^+)(1+\kappa_1^2)
\left(
k_2^+ +k_1^+ + k_1^+\kappa_1^2
\right)^{\epsilon}}
\right].
\end{eqnarray}

The contribution of the corresponding real diagram is 
\begin{eqnarray}
\label{eq:VB-almost-final}%
&&
VB
\nonumber \\
&=&
-
\left(\frac{\alpha_s}{\pi}\right)^2
2^{1+2\epsilon}\pi^{-1+3\epsilon}
\left(\frac{f_{abc}^2}{4N_c}\right)
\int_0^\Lambda dk_1^+(k_1^+)^{-2\epsilon}
\int d^{2-2\epsilon}\kappa_1
\frac{\left[4(q_\perp\cdot \kappa_1)^2
-4\kappa_1^2 q_3^2
-q_\perp^2(1 + \kappa_1^2)^2
\right]}{(1+\kappa_1^2)^3}
\nonumber \\
&&
\times
\int_{0}^\Lambda \frac{dk_2^+}{(k_2^+)^{1+\epsilon}}
\left[
\frac{\Gamma(1+\epsilon)}
{
\left(
k_2^+ +k_1^+ + k_1^+\kappa_1^2
\right)^{1+\epsilon}}
+
\frac{\Gamma(\epsilon_\textrm{UV})}
{(k_1^+)(1+\kappa_1^2)
\left(
k_2^+ +k_1^++  k_1^+\kappa_1^2
\right)^{\epsilon}}
\right].
\end{eqnarray}

Performing the remaining integrations in Eqs.~(\ref{eq:VA-k1-k2-almost-final}) 
and (\ref{eq:VB-almost-final}), we obtain 
\begin{equation}
\label{eq:VA-VB-before-kappa-1-int-new2}
2\textrm{Re}
\left(VA^{(k_2-k_1)^2}
+VB\right)
=
-
2
\left[\frac{1}{\epsilon_\textrm{UV}^2}+O(\epsilon^{-1})\right]
\frac{(N_c^2-1)\alpha_s^2}{\pi^2\epsilon_\textrm{IR}}
\left(
-
\frac{\bm{q}^2}{12}
\right)
+O(1/\Lambda^2),
\end{equation}
which contains a UV double pole. This UV-double-pole contribution 
accounts for the UV double poles that we find in $\mathcal{C}$ and 
$\mathcal{D}$ in Eq.~(\ref{eq:expanded-C-D-E}).

\subsection{Diagram VI}

Finally, let us consider the contribution of diagram VI in
Ref.~\cite{Nayak:2005rt}. In terms of light-cone variables the
real-virtual contribution is \cite{Nayak:2005rt}
\begin{eqnarray}
VIA
&=&
\frac{i2^{5/2}g_s^4}{(2\pi)^{2D-1}}
\left(\frac{f_{abc}^2}{4N_c}\right)
\int_0^\Lambda \frac{dk_1^+}{2k_1^+}
\int_{-\infty}^\infty dk_2^+
\int d^{2-2\epsilon}k_{1\perp}
\int d^{2-2\epsilon}k_{2\perp}
\frac{1}{(k_1^+ + \frac{k_{1\perp}^2}{2k_1^+})^2
\left(k_1^+ - k_2^+ -i\varepsilon\right)}
\nonumber \\
&&
\times
\int_{-\infty}^\infty dk_2^-
\frac{2\sqrt{2}q_3^2(k_1^+ - \frac{k_{1\perp}^2}{2k_1^+})
(k_1^+ - k_2^+)
-2\sqrt{2}(q_\perp\cdot k_{1\perp})
\left[
q_\perp\cdot(k_{1\perp}-k_{2\perp})
\right]}
{\left[
(2k_2^+ k_2^--k_{2\perp}^2)
-2(k_1^+ k_2^-+\frac{k_{1\perp}^2}{2k_1^+}k_2^+
-k_{1\perp}\cdot k_{2\perp})
-i\varepsilon\right]}
\nonumber \\
&&
\!\!\!\!\!\!\!\!
\times
\left[
\frac{1}{(k_2^+ - k_1^+ + k_2^- - \frac{k_{1\perp}^2}{2k_1^+} 
-i\varepsilon)^2
(k_2^+ + k_2^--i\varepsilon)}
+
\frac{1}{(k_2^+ - k_1^+ + k_2^- - \frac{k_{1\perp}^2}{2k_1^+} 
-i\varepsilon)
(k_2^+ + k_2^--i\varepsilon)^2}
\right].
\nonumber \\
\end{eqnarray}
Again, we close the $k_2^-$ contour in the upper half-plane and pick up 
the residue of the pole at $(k_2-k_1)^2=0$. The location of the pole is 
given in Eq.~(\ref{eq:k2m-k2-k1-sq-pole}), and the residue is nonzero only 
when $k_2^+ > k_1^+$. The residue is 
\begin{eqnarray}
VIA^{(k_2-k_1)^2}
&=&
-
2^{4\epsilon}\pi^{-4+4\epsilon}\alpha_s^2
\left(\frac{f_{abc}^2}{4N_c}\right)
\int_0^\Lambda \frac{dk_1^+}{k_1^+}
\int_{k_1^+}^\infty dk_2^+
\int d^{2-2\epsilon}k_{1\perp}
\int d^{2-2\epsilon}k_{2\perp}
\nonumber \\
&&
\times
\frac{q_3^2(k_1^+ - \frac{k_{1\perp}^2}{2k_1^+})
(k_1^+ - k_2^+)
-(q_\perp\cdot k_{1\perp})
\left[
q_\perp\cdot(k_{1\perp}-k_{2\perp})
\right]}
{(k_1^+ + \frac{k_{1\perp}^2}{2k_1^+})^2
\left(k_1^+ - k_2^+\right)(k_2^+-k_1^+)}
\nonumber \\
&&
\times
\Bigg\{
\frac{1}{
\left[
k_2^+ - k_1^+ +
\frac{\left(k_{2\perp}-k_{1\perp}\right)^2}{2(k_2^+ -k_1^+)}
\right]^2
\left[
k_2^+ + \frac{k_{1\perp}^2}{2k_1^+}
+\frac{(k_{2\perp}-k_{1\perp})^2}{2(k_2^+ - k_1^+)}
\right]}
\nonumber \\
&&
\quad
+
\frac{1}{
\left[
k_2^+ - k_1^+ +
\frac{\left(k_{2\perp}-k_{1\perp}\right)^2}{2(k_2^+ -k_1^+)}
\right]
\left[
k_2^+ + \frac{k_{1\perp}^2}{2k_1^+}
+\frac{(k_{2\perp}-k_{1\perp})^2}{2(k_2^+ - k_1^+)}
\right]^2}
\Bigg\}.
\end{eqnarray}
Making the changes of variables
\begin{equation}
k_2^+ = (k_1^+)y,
\quad
k_i^\perp = (\sqrt{2}k_1^+)\kappa_i,
\end{equation}
we obtain
\begin{eqnarray}
VIA^{(k_2-k_1)^2}
&=&
-
2^{2+2\epsilon}\pi^{-4+4\epsilon}\alpha_s^2
\left(\frac{f_{abc}^2}{4N_c}\right)
\int_0^\Lambda \frac{dk_1^+}{(k_1^+)^{1+4\epsilon}}
\int_{1}^\infty dy
\int d^{2-2\epsilon}\kappa_1
\int d^{2-2\epsilon}\kappa_2
\frac{q_3^2(y-1)^2}
{(1 + \kappa_1^2)}
\nonumber \\
&&
\times
\Bigg\{
\frac{1}{
\left[
(y-1)^2 +
\kappa_2^2
\right]^2
\left[
(y-1)
(y + \kappa_1^2)
+\kappa_2^2
\right]}
+
\frac{1}{
\left[
(y-1)^2 +\kappa_2^2\right]
\left[
(y-1)
(y + \kappa_1^2)
+\kappa_2^2
\right]^2}
\Bigg\}.
\nonumber \\
\end{eqnarray}
It can be seen easily that the $\kappa_1$ and $\kappa_2$ integrations cannot 
give UV poles, and so the diagram VI does not contribute any double UV 
poles. 



\end{document}